\documentclass[pra,preprint,superscriptaddress,floatfix,showkeys,showpacs]{revtex4-1}

\usepackage{amsmath,amsfonts,amssymb}
\usepackage{graphicx,color}
\allowdisplaybreaks

\def\beq{\begin{equation}}
\def\eeq{\end{equation}}
\def\beqn{\begin{eqnarray}}
\def\eeqn{\end{eqnarray}}

\def\r {{\bf r}}

\def\r {{\bf r}}

\begin{document}

\title{Solvable model of a generic driven mixture of trapped Bose-Einstein condensates
and properties of a many-boson Floquet state at the limit of an infinite number of particles}
\author{Ofir E. Alon}
\email{ofir@research.haifa.ac.il}
\affiliation{Department of Mathematics, University of Haifa, Haifa 3498838, Israel}
\affiliation{Haifa Research Center for Theoretical Physics and Astrophysics, University of Haifa,
Haifa 3498838, Israel}

\begin{abstract}
A solvable model of a periodically-driven trapped mixture of Bose-Einstein condensates,
consisting of
$N_1$ interacting bosons of mass $m_1$ driven by a force of amplitude $f_{L,1}$
and $N_2$ interacting bosons of mass $m_2$ driven by a force of amplitude $f_{L,2}$,
is presented.
The model generalizes the harmonic-interaction model for mixtures to the time-dependent domain.
The resulting many-particle ground Floquet wavefunction and quasienergy,
as well as the time-dependent densities and reduced density matrices,
are prescribed explicitly and analyzed at the many-body and mean-field levels of theory
for finite systems and at the limit of an infinite number of particles.
We prove that the time-dependent densities per particle are
given at the limit of an infinite number of particles by their respective mean-field quantities,
and that the time-dependent reduced one-particle and two-particle density matrices per particle of the driven mixture 
are $100\%$ condensed.
Interestingly, the quasienergy per particle {\it does not} coincide with the mean-field value at this limit,
unless the relative center-of-mass coordinate 
of the two Bose-Einstein condensates is not activated by the driving forces $f_{L,1}$ and $f_{L,2}$.
As an application, we investigate the imprinting of
angular momentum and its fluctuations
when steering a Bose-Einstein condensate
by an interacting bosonic impurity,
and the resulting modes of rotations.
Whereas the expectation values per particle of the angular-momentum operator 
for the many-body and mean-field solutions
coincide at the limit of an infinite number of particles,
the respective fluctuations can differ substantially.
The results are analyzed in terms of the transformation properties
of the angular-momentum operator under translations and boosts and
the interactions between the particles.
Implications are briefly discussed.
\end{abstract}


\maketitle 

\section{Introduction}\label{Intro}

Quantum dynamics is central to nuclear, atomic, molecular, and optical
physics \cite{QDb1,QDb2,QDb3,QDb4,QDb5,QDb6,QDb7,QDb8}.
In order to theoretically describe the dynamics of a quantum many-particle system it is generally required to solve and
analyze the time-dependent Schr\"odinger equation.
Usually, a numerical treatment of a faithful representation to the
time-dependent system under investigation is the only available tool,
whereas the existence of solvable time-dependent many-particle models is scarcer.

A particularly exciting arena for quantum dynamics across many disciplines is that in which the system's
Hamiltonian is periodic in time
\cite{Floquet,FL1,FL2,FL3,FL4,FL5,FL6,FL7,FL8,FL9,FL10,FL11,FL12,FL13,FL14,FL15,
FL16,FL17,FL18,FL19,FL20,FL21,FL22,FL23,FL24,FL25,FL26,FL27,FL28,FL29,FL30,
FL31,FL32,FL33,FL34,FL35,FL36,FL37,FL38,FL39,FL40,FL41,FL42,FL43,FL44,FL45,FL46}.
The time-periodic Hamiltonian defines the Floquet Hamiltonian and governs the Floquet eigenvalue equation
in an extended Hilbert space \cite{FL2,FL3},
in analogy to the static Hamiltonian which governs the Schr\"odinger eigenvlaue equation in the standard Hilbert space
of a time-independent system.

In the field of Bose-Einstein condensates \cite{QDb4,QDb6},
the relation between the many-body and mean-field treatments
of bosonic systems at the limit of an infinite number of particles
has drawn much attention \cite{IN1,IN2,IN3,IN4,IN5,IN6,IN7,IN8,IN8p5,IN9,IN10,IN11,IN12,IN13,IN14,IN15}.
At the infinite-particle-number limit the interaction parameter (i.e., the product of the number of bosons times the interaction strength)
is held fixed while the number of particles is increased to infinity.
Under well defined conditions it has been proved for the ground state of trapped bosons
that the many-body reduced one-particle and two-particle density matrices \cite{RDMb1,RDMb2}
per particle are $100\%$ condensed and coincide with the respective Gross-Pitaevskii results \cite{IN2,IN3}.
Similarly, the many-body energy per particle as well as the one-particle and two-particle densities per particle
coincide in this limit with the Gross-Pitaevskii results.
Analogous results exist for the dynamics of interacting bosons with time-independent Hamiltonians
at a definite propagation time \cite{IN4,IN5}.
Generalizations for trapped mixtures of two kinds of identical bosons cover
the many-body energy per particle, intraspecies and interspecies densities per particle
and reduced density matrices per particle, and show
coincidence in the limit of an infinite number of particles with the
Gross-Pitaevskii results, 
$100\%$ condensation of each of the species,
and separability of interspecies quantities in the mixture \cite{IN8p5,IN9,IN10}.

However,
differences between many-body and mean-field results of fully condensed bosonic systems
can emerge in other quantities,
in particular variances of operators,
such as the position, momentum, and angular-momentum operators \cite{IN6,IN7,IN13,IN15},
and stem from the differences between the many-body and mean-field wavefunctions themselves \cite{IN8,IN11}.
The differences in the variances of observables
are instrumental in defining and quantifying correlations at the infinite-particle-number limit.
For instance,
distinguishing between $100\%$ condensed trapped bosons,
either in the ground state or undergoing dynamics in two spatial dimensions,
in which the anisotropy along the $x$ and $y$ directions
of their position variance per particle and density per particle is a like, 
and $100\%$ condensed bosons
in which this anisotropy is opposite \cite{IN12,IN14}.

In this work we present a solvable model of a periodically-driven trapped mixture of Bose-Einstein condensates.
Mixtures of Bose-Einstein condensates have been a vivid topic \cite{MX1,MX2,MX3,MX4,MX5,MX6,MX7,MX8,MX9,MX10,MX11,MX12,MX13,MX14,
MX15,MX16,MX17,MX18,MX19,MX20,MX21,MX22,MX23,MX24,MX25,MX26,MX27,MX28}.
Here, we introduce the driven harmonic-interaction model for mixtures,
that is, 
$N_1$ bosons of mass $m_1$ interacting by harmonic interparticle interaction of strength $\lambda_1$
and driven by a force of amplitude $f_{L,1}$,
and
$N_2$ bosons of mass $m_2$ interacting by harmonic interparticle interaction of strength $\lambda_2$
and driven by a force of amplitude $f_{L,1}$.
All bosons are trapped in an harmonic potential of frequency $\omega$
and, furthermore, bosons of species $1$ and species $2$ interact with each other
by harmonic interparticle interaction of strength $\lambda_{12}$.
The model generalizes the harmonic-interaction model for mixtures to the time-dependent domain,
and builds on a long-standing interest in harmonic-interaction models \cite{IN10,HM1,HM2,HM3,HM4,HM5,HM6,HM7,HM8,HM9,HM10,HM11,HM12,HM13,HM14,HM15,HM16,HM17}.

Besides solving the model and discussing properties,
our main purpose is to investigate the infinite-particle-limit of a Floquet many-boson state,
namely,
the connection between the many-body and mean-field time-dependent reduced density matrices,
time-dependent densities, quasienergies,
and eventually variances.
In this way, we expand literature studies on the ground state and dynamics with time-independent Hamiltonians
of bosonic systems to the realm of Floquet bosonic systems.
Both the many-body and mean-field solutions of the trapped driven mixture are
required and derived in closed form, making the proofs of properties explicit.
As an application,    
the expectation value and variance of the many-boson angular momentum operator  
when a Bose-Einstein condensate is steered by interacting bosons consisting of a different species
is explored, within our model, fully analytically.
The modes of rotations are investigated as a function of the steering frequency and interspecies interaction,
and the differences between the many-body and mean-field variances explored.

The structure of the paper is as follows.
In Sec.~\ref{DRIV_SING} we consider the driven single-species harmonic-interaction model,
solve it at the many-body and mean-field levels, and derive working tools.
In Sec.~\ref{HIM_MIX} we present the driven harmonic-interaction model for mixtures,
solve it at the many-body and mean-field levels of theory,
discuss its properties,
and analyze the limit of an infinite number of particles.
In Sec.~\ref{ANG_MIX} we analyze the imprinting of angular momentum
when one of the species is steered and embedded in a second species which is not steered,
and discuss the many-body and mean-field angular-momentum variances 
at the infinite-particle-number limit.
In Sec.~\ref{SUM_OUT} a summary and outlook are given.
Finally, appendices \ref{TD_GPEs}-\ref{VAR_TRANS_BST}
collect supplemental results, details of derivations, and further tools.

\section{The driven harmonic-interaction model}\label{DRIV_SING}

Consider the driven classical harmonic oscillator in one spatial dimension.
The classical equation of motion and its particular solution are given by
\beq\label{CM_FHO}
\frac{d^2x(t)}{dt^2} + \omega^2 x(t) = \frac{f_L \cos(\omega_L t)}{m}, \qquad
x(t) = a \cos(\omega_L t), \qquad a = \frac{f_L}{m(\omega^2-\omega_L^2)}.
\eeq
Here and hereafter the amplitude of the driving force satisfies $f_L \ne 0$,
the driving frequency $\omega_L>0$,
the trapping frequency $w>0$,
and for convenience the off-resonance condition $\omega_L \ne \omega$ holds.
For $\omega_L < \omega$ the amplitude of oscillations $a$ is in phase with $f_L$,
whereas for $\omega_L > \omega$ and the high-frequency limit 
$a$ has the opposite phase to $f_L$.
The particular solution $x(t)$ 
appears in the driven quantum harmonic oscillator problem and
in its many-boson generalization,
the driven harmonic-interaction model.

The driven quantum harmonic oscillator \cite{dis,hng}
obeys the time-dependent Schr\"odinger equation
\beqn
& & \hat h(x,t) \psi(x,t) = i \frac{\partial \psi(x,t)}{\partial t}, \qquad 
\hat h(x,t) = -\frac{1}{2m}\frac{\partial^2}{\partial x^2} + \frac{1}{2} m \omega^2 x^2 - x f_L \cos(\omega_L t) = \nonumber \\
& & = -\frac{1}{2m}\frac{\partial^2}{\partial x^2} +  \frac{1}{2} m \omega^2 
\left[x - \frac{f_L}{m(\omega^2-\omega_L^2)} \cos(\omega_L t)\right]^2 -
\frac{f_L^2\omega^2}{2m(\omega^2-\omega_L^2)^2} \cos^2(\omega_L t) + \nonumber \\
& & + \frac{\omega_L^2}{\omega^2-\omega_L^2} x f_L \cos(\omega_L t) = 
 -\frac{1}{2m}\frac{\partial^2}{\partial x^2} + \frac{1}{2} m \omega^2 \left[x - x(t)\right]^2
 - \frac{1}{2}m\omega^2 x^2(t) + m\omega_L^2 x(t) x, \
\eeqn
where, throughout this work, $\hbar=1$.
It has a set of Floquet solutions corresponding to the eigenfucntions of the non-driven system.
The solution corresponding to the ground state of the harmonic oscillator,
hereafter termed for brevity the ground Floquet solution, is given by
\beqn\label{1P_Floquet}
& & \psi(x,t) = \left(\frac{m\omega}{\pi}\right)^{\frac{1}{4}} e^{-i\varepsilon(t)}
 e^{-\frac{m\omega}{2}\left[x - \frac{f_L}{m(\omega^2-\omega_L^2)} \cos(\omega_L t)\right]^2}
 e^{-i \frac{f_L \omega_L}{(\omega^2-\omega_L^2)}\sin(\omega_L t)x} = \nonumber \\
& & = \left(\frac{m\omega}{\pi}\right)^{\frac{1}{4}} e^{-i\varepsilon(t)} 
e^{-\frac{m\omega}{2}\left[x - x(t)\right]^2}
 e^{+i m\dot{x}(t) x}, \
\eeqn
where
\beqn\label{1P_Phase}
& & \varepsilon(t) = \frac{\omega}{2}t 
+ \int^t dt' \left[\frac{1}{2}m\dot{x}^2(t) - \frac{1}{2}m \omega^2 x^2(t)\right] = \frac{\omega}{2}t + 
\int^t dt' \Bigg[\frac{f_L^2\omega_L^2}{2m(\omega^2-\omega_L^2)^2} \sin^2(\omega_L t') - \nonumber \\
& & - \frac{f_L^2\omega^2}{2m(\omega^2-\omega_L^2)^2} \cos^2(\omega_L t') \Bigg] 
= \varepsilon_F t -
\frac{f_L^2(\omega^2+\omega_L^2)}{8m\omega_L(\omega^2-\omega_L^2)^2} \sin(2\omega_L t) \
\eeqn
is the time-dependent phase and \cite{hng}
\beq\label{1P_Ef}
\varepsilon_F = \frac{\omega}{2} - \frac{f_L^2}{4m(\omega^2-\omega_L^2)}
\eeq
is an eigenvalue of the Floquet Hamiltonian, 
$\left[\hat h(x,t) - i\frac{\partial}{\partial t}\right] \bar{\psi}(x,t) = \varepsilon_F \bar{\psi}(x,t)$,
and called quasienergy.
The Floquet quasienergy state $\bar{\psi}(x,t)$ is the periodic part of
$\psi(x,t)=e^{-i\varepsilon_F t} \bar{\psi}(x,t)$.
There are three contributions to the time-dependent phase (\ref{1P_Phase}) and hence to the quasienergy (\ref{1P_Ef}):
the first is the energy of the stationary problem,
the second originates, upon acting with the kinetic-energy operator, 
from the spatial-dependent phase of the time-dependent wavefunction,
and the third results from completing-the-square in the time-dependent Hamiltonian.
The latter contribution and its renormalization by interaction
in particular would play an interesting role in the driven
many-boson problem.
In the present work we study interacting many-boson generalizations of the driven harmonic oscillator,
explicitly time-dependent harmonic-interaction models,
and analyze their ground Floquet solutions at the many-body and mean-field levels of theory.

Consider now the single-species harmonic-interaction model \cite{HM4} plus a driving term,
\beq\label{HIM_driv_crtz}
\hat H(x_1,\ldots,x_N,t) = \sum_{j=1}^N \left[ -\frac{1}{2m}\frac{\partial^2}{\partial x_j^2} + \frac{1}{2}m\omega^2 x_j^2 
- x_j f_L\cos(\omega_L t) \right] + \lambda \sum_{1\le j <k}^N \left(x_j-x_k\right)^2, 
\eeq
describing $N$ harmonically-interacting trapped bosons driven by a time-periodic force of amplitude $f_L$.
Using the Jacobi coordinates
\beq\label{Jac_cor}
Q_k = \frac{1}{\sqrt{k(k+1)}} \sum_{j=1}^{k} (x_{k+1}-x_j), \qquad 1 \le k \le N-1,  \qquad
Q_N = \frac{1}{\sqrt{N}}  \sum_{j=1}^{N} x_j
\eeq
the time-dependent Hamiltonian is readily diagonalized,
\beqn\label{HIM_driv_Jac}
& & \hat H(Q_1,\ldots,Q_N,t) = \nonumber \\
& & 
= \sum_{j=1}^{N-1} \left(-\frac{1}{2m}\frac{\partial^2}{\partial Q_j^2} + \frac{1}{2}m\Omega^2 Q_j^2\right) + \left[-\frac{1}{2m}\frac{\partial^2}{\partial Q_N^2} + \frac{1}{2}m\omega^2 Q_N^2
 - Q_N \sqrt{N}f_L\cos(\omega_L t)\right], \
\eeqn
and describes a single driven center-of-mass oscillator 
and $N-1$ time-independent relative-coordinates' degrees-of-freedom.
The frequency $\Omega=\sqrt{\omega^2+\frac{2\lambda N}{m}}$
of the relative coordinates is dressed by the interparticle interaction.
The (driven) system is bound provided
$\lambda > -\frac{m\omega^2}{2N}$, i.e.,
the interaction is either attractive ($\lambda > 0$)
or not too repulsive ($\lambda < 0$).

The normalized ground Floquet solution of the time-dependent many-boson Schr\"odinger equation
$\hat H(x_1,\ldots,x_N,t)\Psi(x_1,\ldots,x_N,t)=i\frac{\Psi(x_1,\ldots,x_N,t)}{\partial t}$
is given in terms of the Jacoby coordinates by
\beqn\label{HIM_driv_Jac_WF}
& & 
\Psi(Q_1,\ldots,Q_N,t) = \left(\frac{m\Omega}{\pi}\right)^{\frac{N-1}{4}} \left(\frac{m\omega}{\pi}\right)^{\frac{1}{4}}
e^{-i\mathcal{E}(t)}
e^{-\frac{1}{2}m\left\{\Omega\sum_{j=1}^{N-1} Q^2_j +\omega\left[Q_N-Q_N(t)\right]^2\right\}} e^{+im\dot{Q}_N{(t)}Q_N},
\nonumber \\
& & 
\qquad Q_N(t) = \frac{\sqrt{N}f_L}{m(\omega^2-\omega_L^2)} \cos(\omega t) = \sqrt{N} x(t), \
\eeqn
where
\beqn\label{HIM_driv_phase}
& & \mathcal{E}(t) = 
\frac{(N-1)\Omega+\omega}{2}t + 
\int^t dt' \left[\frac{1}{2}m\dot{Q}_N^2(t) - \frac{1}{2}m \omega^2 Q_N^2(t)\right] = \nonumber \\
& &  =
\frac{(N-1)\Omega+\omega}{2}t + 
\int^t dt' \left[\frac{Nf_L^2\omega_L^2}{2m(\omega^2-\omega_L^2)^2} \sin^2(\omega_L t') -
\frac{Nf_L^2\omega^2}{2m(\omega^2-\omega_L^2)^2} \cos^2(\omega_L t') \right] = \nonumber \\
& &  = \mathcal{E}_F t -
\frac{Nf_L^2(\omega^2+\omega_L^2)}{8m\omega_L(\omega^2-\omega_L^2)^2} \sin(2\omega_L t) \
\eeqn
is the time-dependent phase,
and
\beq\label{MB_Ef}
\mathcal{E}_F=\frac{(N-1)\Omega+\omega}{2} - \frac{Nf_L^2}{4m(\omega^2-\omega_L^2)}
\eeq
the many-boson quasienergy.
Comparison of the many-boson (\ref{MB_Ef}) and single-particle (\ref{1P_Ef})
quasienergies shows that they have a similar structure.
The later emanates from the fact that for both systems the
driving force is coupled to the center-of-mass
only, the frequency of which being the trapping frequency $\omega$.
The many-boson quasienergy state $\bar{\Psi}(Q_1,\ldots,Q_N,t)$ 
is an eigenfunction of the many-body Floquet Hamiltonain,
\beq\label{MB_F_Ham}
\left[\hat H(Q_1,\ldots,Q_N,t) - i\frac{\partial}{\partial t}\right] \bar{\Psi}(Q_1,\ldots,Q_N,t)
= \mathcal{E}_F \bar{\Psi}(Q_1,\ldots,Q_N,t),
\eeq
where $\Psi(Q_1,\ldots,Q_N,t)=e^{-i \mathcal{E}_F t} \bar{\Psi}(Q_1,\ldots,Q_N,t)$.

To proceed we transform from the Jacoby coordinates, for which the wavefunction is separable,
to the laboratory frame and get
\beqn\label{HIM_driv_crtz_WF}
& & 
\Psi(x_1,\ldots,x_N,t) = \left(\frac{m\Omega}{\pi}\right)^{\frac{N-1}{4}} \left(\frac{m\omega}{\pi}\right)^{\frac{1}{4}}
e^{-i\mathcal{E}(t)} e^{+imN\dot{x}(t) x(t)} \times \nonumber \\
& & \times
e^{-\frac{\alpha}{2}\sum_{j=1}^N \left[x_j-x(t)\right]^2 - \beta \sum_{1\le j < k}^N \left[x_j-x(t)\right]\left[x_k-x(t)\right]} 
e^{+i m \dot{x}(t) \sum_{j=1}^N\left[x_j-x(t)\right]}, \nonumber \\
& &
\alpha
= m\Omega + \beta, \qquad
\beta = m \frac{\omega-\Omega}{N}. \
\eeqn
In obtaining the wavefunction in the laboratory frame we made use of the relations
that $Q_N-Q_N(t) = \frac{1}{\sqrt{N}} \sum_{j=1}^{N} [x_j-x(t)]$
and that the relative Jacoby coordinates are invariant
under the translation of all coordinates by $x(t)$.

The determination of the time-dependent reduced density matrices follows in essentially the same way as
for the undriven system \cite{HM4}.
Importantly, since the driving is coupled to the center-of-mass coordinate only,
all coordinates are translated by the same amplitude, the classical amplitude $x(t)$,
and the coherence properties of the driven system, like the eigenvalues of the time-dependent reduced one-particle density matrix,
are exactly those of the undriven system, in particular these eigenvalues are time independent.
Explicitly,
using the normalized time-dependent wavefunction
and from the $N$-particle density matrix,
\beqn\label{N_RDM}
& &
\Psi(x_1,\ldots,x_N,t)\Psi^\ast(x'_1,\ldots,x'_N,t) = 
 \left(\frac{m\Omega}{\pi}\right)^{\frac{N-1}{2}} \left(\frac{m\omega}{\pi}\right)^{\frac{1}{2}} \times \nonumber \\
& &
\times 
e^{-\frac{\alpha}{2}\sum_{j=1}^N \left\{\left[x_j-x(t)\right]^2 + \left[x'_j-x(t)\right]^2\right\}
- \beta \sum_{1\le j < k}^N \left\{\left[x_j-x(t)\right]\left[x_k-x(t)\right] + \left[x'_j-x(t)\right]\left[x'_k-x(t)\right] \right\}} 
\times \nonumber \\
& & 
\times
e^{+i m \dot{x}(t) \sum_{j=1}^N \left\{\left[x_j-x(t)\right] - \left[x'_j-x(t)\right]\right\}}, \
\eeqn
the reduced one-particle density matrix
$\rho^{(1)}(x,x',t) = N \int dx_2 \cdots dx_N \Psi(x,x_2,\ldots,x_N,t)\times$ $\Psi^\ast(x',x_2,\ldots,x_N,t)$ is defined.
Preforming the integrations consecutively over $x_j, x'_j=x_j, j=N,\ldots,2$, for which the spatial-dependent phase factor falls
and while changing the variable $x_j-x(t) \to x_j$,
the final result reads
\beqn\label{1_RDM}
& &
\rho^{(1)}(x,x',t) = N \left(\frac{\alpha+C_1}{\pi}\right)^{\frac{1}{2}} e^{-\frac{\alpha}{2}\left\{\left[x-x(t)\right]^2 + 
\left[x'-x(t)\right]^2 \right\}} 
e^{- \frac{1}{4} C_1 \left\{\left[x-x(t)\right]+\left[x'-x(t)\right]\right\}^2} \times \nonumber \\
& &
\times e^{+i m \dot{x}(t) \left\{\left[x-x(t)\right]-\left[x'-x(t)\right]\right\}},
\qquad
 C_1 = - \beta^2 \frac{N-1}{(\alpha-\beta)+(N-1)\beta}.
\eeqn
The diagonal of the time-dependent reduced one-particle density matrix is the time-dependent density and given by
\beq\label{1_density}
\rho^{(1)}(x,t) = N \left(\frac{\alpha+C_1}{\pi}\right)^{\frac{1}{2}} e^{-(\alpha+C_1)\left[x-x(t)\right]^2} =
N \left(\frac{\alpha+C_1}{\pi}\right)^{\frac{1}{2}} e^{-(\alpha+C_1)\left[x-\frac{f_L}{m(\omega^2-\omega_L^2)}\cos(\omega_Lt)\right]^2}. 
\eeq
For the respective two-particle quantities see appendix \ref{TD_2RDMs}.

So far, we have presented the solution of the driven harmonic-oscillator model at the many-body level of theory.
The similar structure of the time-dependent reduced one-particle density matrix per particle of the ground Floquet state (\ref{1_RDM})
and the static reduced one-particle density matrix per particle of the ground state \cite{HM4}
already tells one that in the limit of an infinite number of particles the former, like the latter, is $100\%$ condensed.
To discuss properties of the driven system at the infinite-particle-number limit,
we now derive the solution of the mean-field level of theory.
This would allow us to compare explicitly the time-dependent reduced density matrices per particle
and quasienergies per particle,
and prescribe the connection 
between the many-body and mean-field levels of theory
for driven Bose-Einstein condensates,
at least within the driven harmonic-interaction model.

Starting from the Gross-Pitaevskii mean-field ansatz for the time-dependent many-boson wavefunction,
\beq\label{MF_ansatz}
 \Phi(x_1,\ldots,x_N,t) = \prod_{j=1}^N \phi(x_j,t), 
\eeq
and using the time-dependent variational principle, in either of its forms,
the Gross-Pitaevskii mean-field equation for the driven system takes on the form
\beq\label{TDGP}
\left[-\frac{1}{2m}\frac{\partial^2}{\partial x^2} + \frac{1}{2} m \omega^2 x^2 - x f_L \cos(\omega_L t)
+ \Lambda \int dx' |\phi(x',t)|^2(x-x')^2\right]\phi(x,t)
=i\frac{\partial \phi(x,t)}{\partial t},
\eeq
where $\Lambda=\lambda(N-1)$ is the mean-field interaction parameter.
Solving the
Gross-Pitaevskii equation (\ref{TDGP}) combines the approach used
in the treatment of the non-interacting driven oscillator 
and that of the undriven static mean-field system,
and is detailed in appendix \ref{TD_GPEs}.
The final result can be cast as
\beqn\label{TDGP_SOL_1}
& & \phi(x,t) = \left(\frac{m\Omega_{GP}}{\pi}\right)^{\frac{1}{4}} e^{-i\mu(t)}
 e^{-\frac{m\Omega_{GP}}{2}\left[x - \frac{f_L}{m(\omega^2-\omega_L^2)} \cos(\omega_L t)\right]^2}
 e^{-i \frac{f_L \omega_L}{(\omega^2-\omega_L^2)}\sin(\omega_L t)x} = \nonumber \\
& & = \left(\frac{m\Omega_{GP}}{\pi}\right)^{\frac{1}{4}} e^{-i\mu(t)} 
e^{-\frac{m\Omega_{GP}}{2}\left[x - x(t)\right]^2}
 e^{+i m\dot{x}(t) x}, \
\eeqn
where $\Omega_{GP} = \sqrt{\omega^2+\frac{2\Lambda}{m}}$ is the mean-field interaction-dressed frequency
and
\beq\label{MF_mu}
\mu(t) = \mu_F t -\frac{f_L^2(\omega^2+\omega_L^2)}{8m\omega_L(\omega^2-\omega_L^2)^2} \sin(2\omega_L t),
\qquad
\mu_F = \frac{\Omega_{GP}}{2}+\frac{\Lambda}{2\Omega_{GP}} - \frac{f_L^2}{4m(\omega^2-\omega_L^2)}
\eeq
is a time-dependent phase.
Here, $\mu_F$ can in analogy to the static case be called the quasichemical-potential
and is an eigenvalue of the mean-field Floquet Hamiltonian,
\beq\label{MF_Floquet_ham}
\left[-\frac{1}{2m}\frac{\partial^2}{\partial x^2} + \frac{1}{2} m \omega^2 x^2 - x f_L \cos(\omega_L t)
+ \Lambda \int dx' |\bar{\phi}(x',t)|^2(x-x')^2 - i\frac{\partial}{\partial t} \right] \bar{\phi}(x,t)
= \mu_F \bar{\phi}(x,t).
\eeq
Analogously, $\bar{\phi}(x,t)$,
which may be called the mean-field quasichemical-potential state, 
is the periodic part of $\phi(x,t) = e^{-i\mu_F t}\bar{\phi}(x,t)$.
Since $|\phi(x,t)|^2 = |\bar{\phi}(x,t)|^2$,
the mean-field Floquet Hamiltonian (\ref{MF_Floquet_ham}) 
can be defined solely in terms of $\bar{\phi}(x,t)$.
Of course, the Gross-Pitaevskii solution $\phi(x,t)$ boils down to $\psi(x,t)$ for non-interacting particles, $\Lambda=0$.

The determination of the mean-field wavefunction (\ref{MF_ansatz}) 
allows one to prescribe the reduced density matrices.
Thus, the time-dependent reduced one-particle density matrix at the mean-field level reads
\beqn\label{1_density_MF}
& & \rho^{(1)}_{MF}(x,x',t) = N \rho^{(1)}_{GP}(x,x',t), \nonumber \\
& &  \rho^{(1)}_{GP}(x,x',t) = \left(\frac{m\Omega_{GP}}{\pi}\right)^{\frac{1}{2}} 
e^{-\frac{m\Omega_{GP}}{2}\left\{\left[x - x(t)\right]^2 + \left[x' - x(t)\right]^2\right\}}
 e^{+i m \dot{x}(t) \left\{\left[x-x(t)\right]-\left[x'-x(t)\right]\right\}} \
\eeqn
and its diagonal, the density, is
$\rho^{(1)}_{MF}(x,t) = N \rho^{(1)}_{GP}(x,t)$,
$\rho^{(1)}_{GP}(x,t) = \left(\frac{m\Omega_{GP}}{\pi}\right)^{\frac{1}{2}} 
e^{-m\Omega_{GP}\left[x - x(t)\right]^2}$.

Next, let $\bar{\Phi}(x_1,\ldots,x_N,t)=\prod_{j=1}^{N} \bar{\phi}_j(x_j,t)$
be the periodic part of the mean-field many-boson wavefunction (\ref{MF_ansatz}),
i.e., $\Phi(x_1,\ldots,x_N,t) = e^{-iN\mu_F t} \bar{\Phi}(x_1,\ldots,x_N,t)$. 
Then, 
the mean-field Floquet energy per particle is defined by sandwiching the many-boson Floquet Hamiltonian,
\beq\label{MF_QE_HIM_Single}
\varepsilon_F^{GP}=\frac{\langle\bar{\Phi}|\hat H - i\frac{\partial}{\partial t}|\bar{\Phi}\rangle}{N} = 
\frac{\Omega_{GP}}{2} - \frac{f_L^2}{4m(\omega^2-\omega_L^2)},
\eeq
in analogy to a Bose-Einstein condensate in its ground state 
for which the mean-field energy per particle is
obtained by sandwiching the static many-boson Hamiltonian
with the static mean-field wavefunction.
We are now in the position to examine the relations
between the many-body and mean-field solutions of the driven interacting
bosons at the limit of an infinite number of particles.

Thus, the following relations readily hold:
\beq\label{RDM_1_limit}
 \lim_{N \to \infty} \frac{\rho^{(1)}(x,x',t)}{N} = \rho^{(1)}_{GP}(x,x',t)
\eeq
for the time-dependent reduced one-particle density matrix per particle,
$\lim_{N \to \infty} \frac{\rho^{(1)}(x,t)}{N} = \rho^{(1)}_{GP}(x,t)$ for the density per particle,
and
\beq\label{QE_limit}
 \lim_{N \to \infty} \frac{\mathcal{E}_F}{N} = \varepsilon_F^{GP}
\eeq
for the quasienergy per particle,
thus generalizing literature results
for the ground state of trapped interacting bosons
at the limit of an infinite number of particles.

\section{The driven harmonic-interaction model for mixtures}\label{HIM_MIX}

Consider a driven mixture of two distinguishable 
kinds of identical bosons which we denote as species $1$ and species $2$.
The bosons are trapped in a one-dimensional harmonic potential of frequency $\omega$ 
and interact via harmonic particle-particle interactions.
In the present work we only deal with the ground Floquet solution of the driven mixture.
We solve for the case of a generic driven mixture, i.e.,
a mixture comprising
$N_1$ bosons of species $1$ and mass $m_1$ 
and $N_2$ bosons of species $2$ and mass $m_2$.
The total number of bosons is denoted by $N=N_1+N_2$.
The species are driven independently by forces of amplitude $f_{L,1}$ and $f_{L,2}$, respectively.
Furthermore, the two intraspecies interaction strengths are denoted by $\lambda_1$ and $\lambda_2$,
and the interspecies interaction strength by $\lambda_{12}$.
Positive values of $\lambda_1$, $\lambda_2$, and $\lambda_{12}$ mean 
attractive particle-particle interactions
whereas negative values imply repulsive interactions.
In what follows we solve the driven harmonic-interaction-model for mixtures at the many-body and mean-field levels,
obtain the respective quasienergies,
construct the time-dependent intraspecies and intraspecies reduced density matrices and densities,
discuss properties of these quantities as a function of the interactions between the bosons and strengths of the driving forces,
and establish connections between the many-body and mean-field Floquet quantities at the infinite-particle-number limit. 

\subsection{Many-body solution and time-dependent reduced density matrices}\label{subsec1}

The Hamiltonian of the driven mixture in the laboratory frame
is then that of the harmonic-interaction model for mixtures \cite{IN10}
plus two driving terms and reads
\beqn\label{HAM_MIX}
& & \hat H(x_{1,1},\ldots,x_{1,N_1},x_{2,1},\ldots,x_{2,N_2},t) = \nonumber \\
& & = \sum_{j=1}^{N_1} \left[ -\frac{1}{2m_1} \frac{\partial^2}{\partial x_{1,j}^2} + 
\frac{1}{2} m_1\omega^2 x_{1,j}^2 - x_{1,j} f_{L,1} \cos(\omega_L t) \right] + \lambda_1 \sum_{1 \le j < k}^{N_1} (x_{1,j}-x_{1,k})^2 +
\nonumber \\
& & + \sum_{j=1}^{N_2} \left[ -\frac{1}{2m_2} \frac{\partial^2}{\partial x_{2,j}^2} + 
\frac{1}{2} m_2\omega^2 x_{2,j}^2 - x_{2,j} f_{L,2} \cos(\omega_L t) \right] + \lambda_2 \sum_{1 \le j < k}^{N_2} (x_{2,j}-x_{2,k})^2 +
\nonumber \\
& & + 
\lambda_{12} \sum_{j=1}^{N_1} \sum_{k=1}^{N_2} (x_{1,j}-x_{2,k})^2. \  
\eeqn
Here, the coordinates $x_{1,j}$ with subscript $1$ stand for species $1$ bosons
and $x_{2,k}$ with subscript $2$ for species $2$ bosons.
In the present section we solve the one-dimensional driven mixture
where the amplitudes of the two driving forces $f_{L,1}$ and $f_{L,2}$ are general.
In section \ref{ANG_MIX} we examine an application in two spatial dimensions
with $f_{L,2}=0$, i.e., when only species $1$ in the interacting mixture is driven.

Using the Jacoby coordinates \cite{HM5}
\beqn\label{MIX_COOR}
& & Q_k = \frac{1}{\sqrt{k(k+1)}} \sum_{j=1}^{k} (x_{1,k+1}-x_{1,j}), \qquad 1 \le k \le N_1-1,  \nonumber \\
& & Q_{N_1-1+k} = \frac{1}{\sqrt{k(k+1)}} \sum_{j=1}^{k} (x_{2,k+1}-x_{2,j}), \qquad 1 \le k \le N_2-1,  \nonumber \\
& & Q_{N-1} = \sqrt{\frac{N_2}{N_1}} \sum_{j=1}^{N_1} x_{1,j} - \sqrt{\frac{N_1}{N_2}} \sum_{j=1}^{N_2} x_{2,j}, \nonumber \\
& & Q_N = \frac{m_1}{M} \sum_{j=1}^{N_1} x_{1,j} + \frac{m_2}{M} \sum_{j=1}^{N_2} x_{2,j}, \
\eeqn
 the Hamiltonian (\ref{HAM_MIX}) transforms to the diagonal form
\beqn\label{HAM_DIAG}
& & \hat H(Q_1,\ldots,Q_N,t) = 
\sum_{k=1}^{N_1-1} \left( -\frac{1}{2m_1} \frac{\partial^2}{\partial Q_k^2} +
\frac{1}{2} m_1\Omega_1^2 Q_k^2 \right) + \sum_{k=N_1}^{N-2} \left( -\frac{1}{2m_2} \frac{\partial^2}{\partial Q_k^2} + \frac{1}{2} m_2\Omega_2^2 Q_k^2 \right) + \nonumber \\
& & +
\left[-\frac{1}{2M_{12}} \frac{\partial^2}{\partial Q_{N-1}^2} + \frac{1}{2} M_{12} \Omega_{12}^2 Q_{N-1}^2
- M_{12} \sqrt{N_1N_2} \left(\frac{f_{L,1}}{m_1} - \frac{f_{L,2}}{m_2}\right)  \cos(\omega_L t) Q_{N-1}\right] + \nonumber \\
& & +
\left[-\frac{1}{2M} \frac{\partial^2}{\partial Q_N^2} + \frac{1}{2} M \omega^2 Q_N^2
- \left(N_1 f_{L,1} + N_2 f_{L,2} \right) \cos(\omega_L t) Q_N\right], \
\eeqn
where the masses $M_{12} = \frac{m_1m_2}{M}$, $M=N_1m_1+N_2m_2$,
and inverse relations 
$\sum_{j=1}^{N_1} x_{1,j} =  \sqrt{N_1N_2} \frac{m_2}{M} Q_{N-1} + N_1 Q_N$,
$\sum_{j=1}^{N_1} x_{2,j} =  -\sqrt{N_1N_2} \frac{m_1}{M} Q_{N-1} + N_2 Q_N$
are used.
The transformed Hamiltonian of the mixture (\ref{HAM_DIAG}) is that of $N$ uncoupled harmonic oscillators,
two driven oscillators built on the center-of-mass $Q_N$ and relative center-of-mass $Q_{N-1}$ coordinates 
and $N-2$ time-independent oscillators for the relative coordinates of the individual species.
The frequencies of the relative coordinates are dressed by the interactions,
\beqn\label{MIX_FREQ}
& & \Omega_1 = \sqrt{\omega^2 + \frac{2}{m_1}(N_1\lambda_1+N_2\lambda_{12})}, \quad
 \Omega_2 = \sqrt{\omega^2 + \frac{2}{m_2}(N_2\lambda_2+N_1\lambda_{12})}, \nonumber \\
& & \Omega_{12} =
 \sqrt{\omega^2 + \frac{2\lambda_{12}}{M_{12}}}, \
\eeqn
and have to be positive for the driven mixture to be bound,
implying the following conditions on the interspecies
$\lambda_{12} > -\frac{M_{12}\omega^2}{2}$,
and
intraspecies
$\lambda_1 > - m_1 \frac{N_2}{N_1} \lambda_{12} - \frac{m_1\omega^2}{2N_1}$
and
$\lambda_2 > - m_2 \frac{N_1}{N_2} \lambda_{12} - \frac{m_2\omega^2}{2N_2}$
interaction strengths.

We can now proceed and 
prescribe the normalized ground Floquet wavefunction of (\ref{HAM_DIAG}),
\beqn\label{WAVE_FUN_1}
& & \Psi(Q_1,\ldots,Q_N,t) = 
\left(\frac{m_1\Omega_1}{\pi}\right)^{\frac{N_1-1}{4}}
\left(\frac{m_2\Omega_2}{\pi}\right)^{\frac{N_2-1}{4}}
\left(\frac{M_{12}\Omega_{12}}{\pi}\right)^{\frac{1}{4}}
\left(\frac{M\omega}{\pi}\right)^{\frac{1}{4}}
\times \nonumber \\
& & \times
e^{-i\mathcal{E}(t)}
e^{-\frac{1}{2} \left\{ m_1\Omega_1 \sum_{k=1}^{N_1-1} Q_k^2 + m_2\Omega_2 \sum_{k=N_1}^{N-2} Q_k^2 
+ M_{12}\Omega_{12} \left[Q_{N-1}-Q_{N-1}(t)\right]^2 + M\omega \left[Q_N-Q_N(t)\right]^2 \right\}} 
\times \nonumber \\
& & \times
e^{+i \left[M_{12} \dot Q_{N-1}(t) Q_{N-1} + M \dot Q_N(t) Q_N \right]} =
e^{-i\mathcal{E}_F t} \bar\Psi(Q_1,\dots,Q_N,t), \
\eeqn
where
\beqn\label{WAVE_FUN_1.1}
& & 
Q_{N-1}(t) = 
\frac{\sqrt{N_1N_2}\left(\frac{f_{L,1}}{m_1}-\frac{f_{L,2}}{m_2}\right)}{\Omega^2_{12}-\omega^2_L}\cos(\omega_L t), \quad
Q_N(t) = \frac{1}{M}\frac{N_1f_{L,1}+N_2f_{L,2}}{\omega^2-\omega^2_L}\cos(\omega_L t) \
\eeqn
are the time-dependent oscillating amplitudes of the relative center-of-mass and center-of-mass coordinates, respectively,
\beqn\label{WAVE_FUN_1.2}
& & 
\mathcal{E}(t) = \frac{(N_1-1) \Omega_1 + (N_2-1) \Omega_2 + \Omega_{12} + \omega}{2}t +
 \\
& &
+ \int^t dt' \left\{\left[\frac{1}{2}M_{12} \dot Q^2_{N-1}(t) - \frac{1}{2} M_{12} \Omega^2_{12} Q^2_{N-1}(t)\right] + 
\left[\frac{1}{2}M \dot Q^2_N(t) - \frac{1}{2} M \omega^2 Q^2_N(t)\right] \right\} = \nonumber \\
& & 
= \mathcal{E}_F t - \nonumber \\
& & -
\frac{1}{8\omega_LM}\left[\frac{m_1m_2N_1N_2\left(\frac{f_{L,1}}{m_1}-\frac{f_{L,2}}{m_2}\right)^2(\Omega^2_{12} + \omega^2_L)}
{(\Omega^2_{12} - \omega^2_L)^2} 
+ \frac{\left(N_1 f_{L,1} + N_2 f_{L,2} \right)^2(\omega^2 + \omega^2_L)}{(\omega^2 - \omega^2_L)^2}
\right] \sin(2\omega_L t) \nonumber \
\eeqn
is the time-dependent phase,
and
\beqn\label{E_F_MIX}
& & \mathcal{E}_F =  \frac{(N_1-1) \Omega_1 + (N_2-1) \Omega_2 + \Omega_{12} + \omega}{2} - \nonumber \\
& & 
- \frac{1}{4M} \left[\frac{m_1m_2N_1N_2\left(\frac{f_{L,1}}{m_1}-\frac{f_{L,2}}{m_2}\right)^2}{\Omega^2_{12} - \omega^2_L} 
+ \frac{\left(N_1 f_{L,1} + N_2 f_{L,2} \right)^2}{\omega^2 - \omega^2_L}
\right] \
\eeqn
is the many-particle quasienergy.
The quasienergy for the driven mixture (\ref{E_F_MIX})
is, of course, more intricate than the above quasienergy of the driven Bose-Einstein condensate alone, see (\ref{MB_Ef}).
In particular, the appearance of the interspecies interaction-dependent resonance is appealing.
For simplicity in what follows,
we assume the off-resonance condition $\omega_L \ne \Omega_{12}$ with the relative center-of-mass frequency,
in addition to the above-used
off-resonance condition $\omega_L \ne \omega$ with the center-of-mass frequency.

It is instructive to analyze the solution of the driven mixture.
We begin with the poles.
There are two poles in the quasienergy,
the first is associated with driving the center-of-mass coordinate $Q_N$ of the mixture,
and the second with driving the relative center-of-mass coordinate $Q_{N-1}$.
In the absence of interspecies interaction the two poles coincide, $\Omega_{12}=\omega$, 
and the solution boils down to two independent driven Bose-Einstein condensates.
Let us move to the zeros.
For general forces $f_{L,1}$ and $f_{L,2}$ both the $Q_N$ and $Q_{N-1}$ oscillations are activated.
There are, however, particular relations between the forces and parameters of the mixture for which 
either of the oscillations is not activated,
i.e., there is zero contribution to the quasienergy.
Explicitly, for the relation $\frac{f_{L,1}}{m_1}=\frac{f_{L,2}}{m_2}$ between the forces and masses the relative center-of-mass
coordinate is not activated,
whereas for the relation $f_{L,1}N_1+f_{L,2}N_2=0$ between the forces and numbers of particles the center-of-mass
coordinate in not activated.
Importantly, both modes of oscillations are (non-trivially) activated if only species $1$ is driven, i.e., when $f_{L,2}=0$,
due to the interspecies interaction.
We shall be investigating this particular case in two spatial dimensions in Sec.~\ref{ANG_MIX},
focusing on angular-momentum matters.

To move from the Jacobi-coordinate to the laboratory-frame representation for the driven mixture
one must express the time-dependent oscillation amplitudes $Q_{N-1}(t)$ and $Q_N(t)$ in the former representation 
in terms of
corresponding amplitudes for each of the species in the latter.
Since each species is made of identical particles,
only two such amplitudes are required
which we shall denote by $x_1(t)$ and $x_2(t)$ for species $1$ and $2$, respectively.
Recall that the Jacoby coordinates of the relative motions, $Q_j, j=1,\ldots,N-2$, are invariant under translations
of each species.
Using the definition of the relative center-of-mass and center-of-mass Jacobi coordinates (\ref{MIX_COOR}),
we require $Q_{N-1}-Q_{N-1}(t) = \sqrt{\frac{N_2}{N_1}} \sum_{j=1}^{N_1} \left[x_{1,j}-x_1(t)\right] - 
\sqrt{\frac{N_1}{N_2}} \sum_{j=1}^{N_2} \left[x_{2,j}-x_2(t)\right]$ and
$Q_N - Q_N(t) = \frac{m_1}{M} \sum_{j=1}^{N_1} \left[x_{1,j}-x_1(t)\right] + 
\frac{m_2}{M} \sum_{j=1}^{N_2} \left[x_{2,j}-x_2(t)\right]$
or
$\sqrt{N_1N_2} \left[x_1(t)-x_2(t)\right] = Q_{N-1}(t)$ and 
$\frac{1}{M}\left[m_1N_1x(t)+m_2N_2y(t)\right]=Q_N(t)$.
The solution to these conditions are the translations
\beqn\label{CRTZ_TRNS}
& & x_1(t) = Q_N(t) + \frac{m_2}{M} \sqrt{\frac{N_2}{N_1}}Q_{N-1}(t) = 
\frac{1}{M}\left[\frac{N_1f_{L,1}+N_2f_{L,2}}{\omega^2-\omega^2_L} +
\frac{m_2N_2\left(\frac{f_{L,1}}{m_1}-\frac{f_{L,2}}{m_2}\right)}{\Omega^2_{12}-\omega^2_L}
\right]\cos(\omega_L t),
\nonumber \\
& & x_2(t) =  Q_N(t) - \frac{m_1}{M} \sqrt{\frac{N_1}{N_2}}Q_{N-1}(t) = 
\frac{1}{M}\left[\frac{N_1f_{L,1}+N_2f_{L,2}}{\omega^2-\omega^2_L} -
\frac{m_1N_1\left(\frac{f_{L,1}}{m_1}-\frac{f_{L,2}}{m_2}\right)}{\Omega^2_{12}-\omega^2_L}
\right]\cos(\omega_L t),  \nonumber \\ \
\eeqn
which are the relations between the individual species'
center-of-mass coordinates
and the respective Jacobi coordinates \cite{HM17}. 
We see that, in the laboratory frame, both contributing poles (\ref{WAVE_FUN_1.1}) are intermixed and
the time-dependent translation of each species is more intricate.

We can now translate the Floquet wavefunction (\ref{WAVE_FUN_1}) from the Jacoby to Cartesian coordinates,
\beqn\label{WAVE_FUN_1_CRTZ}
& & \Psi(x_{1,1},\ldots,x_{1,N_1},x_{2,1},\ldots,x_{2,N_2},t) = 
\left(\frac{m_1\Omega_1}{\pi}\right)^{\frac{N_1-1}{4}}
\left(\frac{m_2\Omega_2}{\pi}\right)^{\frac{N_2-1}{4}}
\left(\frac{M_{12}\Omega_{12}}{\pi}\right)^{\frac{1}{4}}
\left(\frac{M\omega}{\pi}\right)^{\frac{1}{4}}
\times \nonumber \\
& & \times
e^{-i\mathcal{E}(t)}
e^{+i\left[m_1N_1 \dot x_1(t) x_1(t)+m_2N_2 \dot x_2(t) x_2(t)\right]}
e^{-\frac{\alpha_1}{2} \sum_{j=1}^{N_1}\left[x_{1,j} - x_1(t)\right]^2 - 
\beta_1 \sum_{1 \le j < k}^{N_1} \left[x_{1,j} - x_1(t)\right] \left[x_{1,k} - x_1(t)\right]} \times \nonumber \\
& & \times e^{-\frac{\alpha_1}{2} \sum_{j=1}^{N_2}\left[x_{2,j} - x_2(t)\right]^2 - 
\beta_2 \sum_{1 \le j < k}^{N_2} \left[x_{2,j} - x_2(t)\right] \left[x_{2,k} - x_2(t)\right]} \times 
e^{+\gamma \sum_{j=1}^{N_1} \sum_{k=1}^{N_2} \left[x_{1,j} - x_1(t)\right]\left[x_{2,k} - x_2(t)\right]} \times  
\nonumber \\
& & \times
e^{+i \left\{m_1 \dot x_1(t)  \sum_{j=1}^{N_1}\left[x_{1,j} - x_1(t)\right] + 
m_2 \dot x_2(t) \sum_{j=1}^{N_2}\left[x_{2,j} - x_2(t)\right]\right\}},
\
\eeqn
where the parameters are
\beqn\label{WAVE_FUN_3}
& & \alpha_1 = m_1\Omega_1+\beta_1, \qquad 
\beta_1 = m_1 \left[- \Omega_1 \frac{1}{N_1} 
+ (m_2N_2\Omega_{12} + m_1N_1\omega) \frac{1}{MN_1} \right], \nonumber \\
& & \alpha_2 = m_2\Omega_2+\beta_2, \qquad
\beta_2 = m_2 \left[- \Omega_2 \frac{1}{N_2} 
+ (m_1N_1\Omega_{12} + m_2N_2\omega) \frac{1}{MN_2} \right], \nonumber \\
& & \gamma = M_{12}(\Omega_{12}-\omega) \
\eeqn
and $M_{12} \dot Q_{N-1}(t) Q_{N-1} + M \dot Q_N(t) Q_N = m_1 \dot x_1(t) \sum_{j=1}^{N_1}x_{1,j} + 
m_2 \dot x_2(t)\sum_{j=1}^{N_2}x_{2,j}$ is used.

From the time-dependent wavefunction (\ref{WAVE_FUN_1_CRTZ},\ref{WAVE_FUN_3})
the $(N_1+N_2)$-particle density matrix of the driven mixture is defined and
given by
\beqn\label{N1N2_RDM}
& & \Psi(x_{1,1},\ldots,x_{1,N_1},x_{2,1},\ldots,x_{2,N_2},t)\Psi^\ast(x'_{1,1},\ldots,x'_{1,N_1},x'_{2,1},\ldots,x'_{2,N_2},t) =\nonumber \\
& & = \left(\frac{m_1\Omega_1}{\pi}\right)^{\frac{N_1-1}{2}}
\left(\frac{m_2\Omega_2}{\pi}\right)^{\frac{N_2-1}{2}}
\left(\frac{M_{12}\Omega_{12}}{\pi}\right)^{\frac{1}{2}}
\left(\frac{M\omega}{\pi}\right)^{\frac{1}{2}} \times \nonumber \\
& & \times e^{-\frac{\alpha_1}{2} \sum_{j=1}^{N_1}\left\{\left[x_{1,j} - x_1(t)\right]^2 + \left[x'_{1,j} - x_1(t)\right]^2\right\} - 
\beta_1 \sum_{1 \le j < k}^{N_1} \left\{\left[x_{1,j} - x_1(t)\right] \left[x_{1,k} - x_1(t)\right] +
\left[x'_{1,j} - x_1(t)\right] \left[x'_{1,k} - x_1(t)\right] \right\}} \times \nonumber \\
& & \times e^{-\frac{\alpha_1}{2} \sum_{j=1}^{N_2} \left\{ \left[x_{2,j} - x_2(t)\right]^2 + \left[x'_{2,j} - x_2(t)\right]^2 \right\} - 
\beta_2 \sum_{1 \le j < k}^{N_2} \left\{\left[x_{2,j} - x_2(t)\right] \left[x_{2,k} - x_2(t)\right] +
\left[x'_{2,j} - x_2(t)\right] \left[x'_{2,k} - x_2(t)\right] \right\}} \times \nonumber \\
& & \times
e^{+\gamma \sum_{j=1}^{N_1} \sum_{k=1}^{N_2} \left\{\left[x_{1,j} - x_1(t)\right]\left[x_{2,k} - x_2(t)\right] +
\left[x'_{1,j} - x_1(t)\right]\left[x'_{2,k} - x_2(t)\right] \right\}} \times \nonumber \\
& & \times
e^{+i m_1 \dot x_1(t) \sum_{j=1}^{N_1} \left\{ \left[x_{1,j} - x_1(t)\right] - \left[x'_{1,j} - x_1(t)\right] \right\}}
e^{+i m_2 \dot x_2(t) \sum_{j=1}^{N_2} \left\{ \left[x_{2,j} - x_2(t)\right] - \left[x'_{2,j} - x_2(t)\right] \right\}}.
\eeqn
This allows us to compute the reduced density matrices of the driven mixture.

It turns out that the determination of the time-dependent reduced density matrices of the driven mixture 
follows in practically the same way as for the undriven mixtures \cite{IN10},
although this is not a straightforward result.
The driving forces are coupled to the relative center-of-mass $Q_{N-1}$ and center-of-mass $Q_N$ coordinates only.
Consequently, as shown above,
all coordinates of species $1$ are translated by the same amplitude $x_1(t)$,
and likewise all coordinates of species $2$ are translated by $x_2(t)$.
Preforming the integrations consecutively over
$x_{1,j}, x'_{1,j}=x_{1,j}, j=N_1,\ldots,2,1$ and
$x_{2,k}, x'_{2,k}=x_{2,k}, k=N_2,\ldots,2,1$, 
for which the spatial-dependent phase factors are canceled
and while changing the variables
$x_{1,j}-x_1(t) \to x_{1,j}$ and
$x_{2,k}-x_2(t) \to x_{2,k}$,
one finds the same recurrence relations \cite{IN10} determining the coefficients 
of the now time-dependent
reduced density matrices, see below.
As a result of this,
the coherence properties of the driven mixture described by the ground Floquet wavefunction
(\ref{WAVE_FUN_1_CRTZ},\ref{WAVE_FUN_3}),
like the eigenvalues of the time-dependent 
intraspecies reduced one-particle density matrices and interspecies reduced two-particle density matrix,
are exactly those of the undriven mixtures, in particular these eigenvalues are time independent.
Thus, the time-dependent
intraspecies reduced one-particle density matrices are
explicitly given as
\beqn\label{1_RDM_1_2}
& &
\rho_1^{(1)}(x_1,x'_1,t) = N_1 \left(\frac{\alpha+C_{1,0}}{\pi}\right)^{\frac{1}{2}} e^{-\frac{\alpha_1}{2}\left\{\left[x_1-x_1(t)\right]^2 + 
\left[x'_1-x_1(t)\right]^2 \right\}} \times \nonumber \\
& & \times e^{- \frac{1}{4} C_{1,0} \left\{\left[x_1-x_1(t)\right]+\left[x'_1-x_1(t)\right]\right\}^2}
e^{+i m_1 \dot x_1 (t) \left\{\left[x_1-x_1(t)\right]-\left[x'_1-x_1(t)\right]\right\}}, \nonumber \\
& & C_{1,0} =
\frac{(\alpha_1-\beta_1)C_{N_1,0}-(N_1-1)(C_{N_1,0}+\beta_1)\beta_1}{(\alpha_1-\beta_1)+(N_1-1)(C_{N_1,0}+\beta_1)},
\qquad C_{N_1,0} = - \gamma^2 \frac{N_2}{(\alpha_2-\beta_2)+N_2\beta_2}, \nonumber \\
& &
\rho_2^{(1)}(x_2,x'_2,t) = N_2 \left(\frac{\alpha+C'_{0,1}}{\pi}\right)^{\frac{1}{2}} e^{-\frac{\alpha_2}{2}\left\{\left[x_2-x_2(t)\right]^2 + 
\left[x'_2-x_2(t)\right]^2 \right\}} \times \nonumber \\
& & e^{- \frac{1}{4} C'_{0,1} \left\{\left[x_2-x_2(t)\right]+\left[x'_2-x_2(t)\right]\right\}^2}
e^{+i m_2 \dot x_2 (t) \left\{\left[x_2-x_2(t)\right]-\left[x'_2-x_2(t)\right]\right\}}, \\
& & C'_{0,1} =  
\frac{(\alpha_2-\beta_2)C'_{0,N_2}-(N_2-1)(C'_{0,N_2}+\beta_2)\beta_2}{(\alpha_2-\beta_2)+(N_2-1)(C'_{0,N_2}+\beta_2)}, 
\qquad C'_{0,N_2} = - \gamma^2 \frac{N_1}{(\alpha_1-\beta_1)+N_1\beta_1}. \nonumber \
\eeqn
The diagonals of $\rho_1^{(1)}(x_1,x'_1,t)$ and $\rho_2^{(1)}(x_2,x'_2,t)$
are the intraspecies time-dependent densities and given explicitly by
\beqn\label{1_density_1_2}
& & \rho_1^{(1)}(x_1,t) = N_1 \left(\frac{\alpha_1+C_{1,0}}{\pi}\right)^{\frac{1}{2}}
e^{-(\alpha_1+C_{1,0})\left[x_1-x_1(t)\right]^2} = \nonumber \\
& & 
= N_1 \left(\frac{\alpha_1+C_{1,0}}{\pi}\right)^{\frac{1}{2}}
e^{-(\alpha_1+C_{1,0})\left\{x-\frac{1}{M}\left[\frac{N_1f_{L,1}+N_2f_{L,2}}{\omega^2-\omega^2_L} +
\frac{m_2N_2\left(\frac{f_{L,1}}{m_1}-\frac{f_{L,2}}{m_2}\right)}{\Omega^2_{12}-\omega^2_L}
\right]\cos(\omega_L t)\right\}^2}, \nonumber \\
& & \rho_2^{(1)}(x_2,t) = N_2 \left(\frac{\alpha_2+C'_{0,1}}{\pi}\right)^{\frac{1}{2}}
e^{-(\alpha_2+C'_{0,1})\left[x_2-x_2(t)\right]^2} = \nonumber \\
& & 
= N_2 \left(\frac{\alpha_2+C'_{0,1}}{\pi}\right)^{\frac{1}{2}}
e^{-(\alpha_2+C'_{0,1})\left\{x-\frac{1}{M}\left[\frac{N_1f_{L,1}+N_2f_{L,2}}{\omega^2-\omega^2_L} -
\frac{m_1N_1\left(\frac{f_{L,1}}{m_1}-\frac{f_{L,2}}{m_2}\right)}{\Omega^2_{12}-\omega^2_L}
\right]\cos(\omega_L t)\right\}^2}. \
\eeqn
We see that the two species generally oscillate with different amplitudes due to the intermixing of the two poles,
with the specific cases of a single-pole contribution when only the
center-of-mass (same-amplitude motion; for $\frac{f_{L,1}}{m_1}=\frac{f_{L,2}}{m_2}$)
or relative center-of-mass (anti-phase motion; for $f_{L,1}N_1+f_{L,2}N_2=0$) coordinate is activated.

Finally, the time-dependent interspecies reduced two-particle density matrix of\break\hfill the mixture,
$\rho_{12}^{(2)}(x_1,x'_1,x_2,x'_2,t) = N_1N_2 \int dx_{1,2}\cdots dx_{1,N_1}dx_{2,2}\cdots dx_{2,N_2} \times$\break\hfill
$\times \Psi(x_1,x_{1,2},\ldots,x_{1,N_1},x_2,x_{2,2},\ldots,x_{2,N_2},t) 
\Psi^\ast(x'_1,x_{1,2},\ldots,x_{1,N_1},x'_2,x_{2,2},\ldots,x_{2,N_2},t)$,
which is the lowest-order interspecies quantity, 
is given by
\beqn\label{2_RDM_12}
& & \rho_{12}^{(2)}(x_1,x'_1,x_2,x'_2,t) = N_1N_2
\left[\frac{(\alpha_1+C_{1,1})(\alpha_2+C'_{1,1})-D_{1,1}^2}{\pi^2}\right]^\frac{1}{2} \times \nonumber \\
& & \times e^{-\frac{\alpha_1}{2} \left\{\left[x_1 - x_1(t)\right]^2 + \left[x'_1 - x'_1(t)\right]^2\right\}} 
e^{-\frac{\alpha_2}{2} \left\{\left[x_2 - x_2(t)\right]^2 + \left[x'_2 - x'_2(t)\right]^2\right\}} 
e^{-\frac{1}{4}C_{1,1} \left\{\left[x_1 - x_1(t)\right] + \left[x'_1 - x'_1(t)\right]\right\}^2} \times \nonumber \\
& & e^{-\frac{1}{4}C'_{1,1} \left\{\left[x_2 - x_2(t)\right] + \left[x'_2 - x'_2(t)\right]\right\}^2} 
e^{+\frac{1}{2}D_{1,1} \left\{\left[x_1 - x_1(t)\right] + \left[x'_1 - x'_1(t)\right]\right\}
\left\{\left[x_2 - x_2(t)\right]^2 + \left[x'_2 - x'_2(t)\right]\right\}} \times \nonumber \\
& & \times 
e^{+\frac{1}{2}D'_{1,1}\left\{\left[x_1 - x_1(t)\right] - \left[x'_1 - x'_1(t)\right]\right\}
\left\{\left[x_2 - x_2(t)\right]^2 - \left[x'_2 - x'_2(t)\right]\right\}}
e^{+i m_1 \dot x_1 (t) \left\{\left[x_1-x_1(t)\right]-\left[x'_1-x_1(t)\right]\right\}} \times \nonumber \\
& & 
\times e^{+i m_2 \dot x_2 (t) \left\{\left[x_2-x_2(t)\right]-\left[x'_2-x_2(t)\right]\right\}},
\
\eeqn
where
\beqn\label{2_RDM_12_COEFF}
& & C_{1,1} = 
\frac{(\alpha_1-\beta_1)C_{N_1,1}-(N_1-1)(C_{N_1,1}+\beta_1)\beta_1}{(\alpha_1-\beta_1)+(N_1-1)(C_{N_1,1}+\beta_1)},
\qquad
C_{N_1,1} = -\gamma^2 \frac{N_2-1}{(\alpha_2-\beta_2)+{(N_2-1)\beta_2}},
\nonumber \\
& & C'_{1,1} = 
\frac{(\alpha_2-\beta_2)C'_{1,N_2}-(N_2-1)(C'_{1,N_2}+\beta_2)\beta_2}{(\alpha_2-\beta_2)+(N_2-1)(C'_{1,N_2}+\beta_2)},
\qquad
C'_{1,N_2} = - \gamma^2 \frac{N_1-1}{(\alpha_1-\beta_1)+(N_1-1)\beta_1},
\nonumber \\
& & D_{1,1} = \gamma \frac{(\alpha_1-\beta_1)(\alpha_2-\beta_2)}
{[(\alpha_1-\beta_1)+(N_1-1)\beta_1][(\alpha_2-\beta_2)+(N_2-1)\beta_2]-\gamma^2(N_1-1)(N_2-1)}, \nonumber \\
& & D'_{1,1} = \gamma.
\eeqn
Then, the diagonal part of $\rho_{12}^{(2)}(x_1,x'_1,x_2,x'_2,t)$ is the time-dependent two-particle interspecies density and reads
\beqn
& & \rho_{12}^{(2)}(x_1,x_2,t) = N_1N_2
\left[\frac{(\alpha_1+C_{1,1})(\alpha_2+C'_{1,1})-D_{1,1}^2}{\pi^2}\right]^\frac{1}{2} \times \nonumber \\
& & \times e^{-(\alpha_1+C_{1,1})\left[x_1 - x_1(t)\right]^2}
e^{-(\alpha_2+C'_{1,1})\left[x_2 - x_2(t)\right]^2} 
e^{+2D_{1,1}\left[x_1 - x_1(t)\right]\left[x_2 - x_2(t)\right]} = \nonumber \\
& & = N_1N_2
\left[\frac{(\alpha_1+C_{1,1})(\alpha_2+C'_{1,1})-D_{1,1}^2}{\pi^2}\right]^\frac{1}{2} \times  \\
& & \times e^{-(\alpha_1+C_{1,1})\left\{x_1 - \frac{1}{M}\left[\frac{N_1f_{L,1}+N_2f_{L,2}}{\omega^2-\omega^2_L} +
\frac{m_2N_2\left(\frac{f_{L,1}}{m_1}-\frac{f_{L,2}}{m_2}\right)}{\Omega^2_{12}-\omega^2_L}
\right]\cos(\omega_L t)\right\}^2} \times \nonumber \\
& & \times e^{-(\alpha_2+C'_{1,1})\left\{x_2 - \frac{1}{M}\left[\frac{N_1f_{L,1}+N_2f_{L,2}}{\omega^2-\omega^2_L} -
\frac{m_1N_1\left(\frac{f_{L,1}}{m_1}-\frac{f_{L,2}}{m_2}\right)}{\Omega^2_{12}-\omega^2_L}
\right]\cos(\omega_L t)\right\}^2} \times \nonumber \\
& & \!\!\!\!\!\!\!\!\!\!\!
\times e^{+2D_{1,1}\left\{x_1 - \frac{1}{M}\left[\frac{N_1f_{L,1}+N_2f_{L,2}}{\omega^2-\omega^2_L} +
\frac{m_2N_2\left(\frac{f_{L,1}}{m_1}-\frac{f_{L,2}}{m_2}\right)}{\Omega^2_{12}-\omega^2_L}
\right]\cos(\omega_L t)\right\}\left\{x_2 - \frac{1}{M}\left[\frac{N_1f_{L,1}+N_2f_{L,2}}{\omega^2-\omega^2_L} -
\frac{m_1N_1\left(\frac{f_{L,1}}{m_1}-\frac{f_{L,2}}{m_2}\right)}{\Omega^2_{12}-\omega^2_L}
\right]\cos(\omega_L t)\right\}}.
\nonumber \
\eeqn
The coupling term in $\rho_{12}^{(2)}(x_1,x_2,t)$ stems from the intraspecies interaction 
and renders the time-dependent two-particle density not separable at the many-body level of theory.
Furthermore, it mixes the time-dependent oscillation amplitudes $x_1(t)$ and $x_2(t)$ of both species.
This could raise interesting questions pertaining to
the intensities of $\omega_L$-harmonics 
in many-body quantities describing driven mixtures,
which are left for investigations elsewhere. 

\subsection{Mean-field solution and the limit of an infinite number of particles}\label{subsec2}

We are now in the position to solve the driven mixture at the mean-field level,
and thereafter to compare the many-body and mean-field outcomes
at the limit of an infinite number of particles.
The related structures,
of the time-dependent reduced one-particle density matrices per particle of the mixture's ground Floquet state (\ref{1_RDM_1_2})
and of the static reduced one-particle density matrices per particle of the mixture's ground state \cite{IN10},
already tell us that at the infinite-particle-number limit
each of the driven species is $100\%$ condensed.
Furthermore and on similar grounds,
the time-dependent interspecies
reduced two-particle density matrix per particle (\ref{2_RDM_12},\ref{2_RDM_12_COEFF}) is separable and given by the product of
the time-dependent intraspecies reduced one-particle density matrices per particle at the limit of an infinite number of particles. 
To obtain further properties of the driven mixture at this limit,
we now derive the solution of the driven harmonic-interaction model for mixtures
at the mean-field level of theory.
This would allow us to compare explicitly the respective
intraspecies and interspeices time-dependent reduced density matrices per particle
and quasienergies per particle, and identify the connection 
between many-body and mean-field levels of theory
for driven interacting mixtures,
at least within the driven harmonic-interaction model for mixtures.
Beyond the above,
the Floquet solution of the coupled non-linear Schr\"odinger equations
is interesting and instrumental for itself.

The time-dependent many-body wavefunction is given at the mean-field level by
\beq\label{MF_MIX_WF}
\Phi(x_{1,1},\ldots,x_{1,N_1},x_{2,1},\ldots,x_{2,N_2},t) = \prod_{j=1}^{N_1} \phi_1(x_{1,j},t) \prod_{k=1}^{N_2} \phi_2(x_{2,k},t).
\eeq
All species $1$ bosons are described by one and the same time-dependent orbital $\phi_1(x_1,t)$
and all species $2$ bosons are similarly described by another time-dependent orbital $\phi_2(x_2,t)$
which are to be determined self-consistently.
Employing the time-dependent variational principle in any of its forms
leads to the coupled time-dependent Gross-Pitaevskii equations
\beqn\label{MIX_EQ_TDGP_1}
& &
\Bigg\{ -\frac{1}{2m_1} \frac{\partial^2}{\partial x_1^2} + \frac{1}{2} m_1 \omega^2 x_1^2 
- x_1 f_{L,1} \cos(\omega_Lt) + \nonumber \\
& & 
+ \Lambda_1 \int dx'_1 |\phi_1(x'_1,t)|^2 (x_1-x'_1)^2 
+ \Lambda_{21} \int dx_2 |\phi_2(x_2,t)|^2 (x_1-x_2)^2 \Bigg\} \phi_1(x_1,t) = 
i \frac{\partial\phi_1(x_1,t)}{\partial t},
\nonumber \\
& & 
\Bigg\{ -\frac{1}{2m_2} \frac{\partial^2}{\partial x_2^2} + \frac{1}{2} m_2 \omega^2 x_2^2 
- x_2 f_{L,2} \cos(\omega_Lt) +  \\
& &
+ \Lambda_2  \int dx'_2 |\phi_2(x'_2,t)|^2 (x_2-x'_2)^2 
+ \Lambda_{12} \int dx_1 |\phi_1(x_1,t)|^2 (x_1-x_2)^2 \Bigg\} \phi_2(x_2,t) = 
i \frac{\partial\phi_2(x_2,t)}{\partial t}, \nonumber \
\eeqn
where the mean-field interaction parameters are given by 
$\Lambda_1=\lambda_1(N_1-1)$,
$\Lambda_2=\lambda_1(N_2-1)$,
$\Lambda_{12}=\lambda_{12}N_1$, and $\Lambda_{21}=\lambda_{12}N_2$,
and satisfy $N_1 \Lambda_{21} = N_2 \Lambda_{12}$.
Only the interaction parameters appear within the mean-field treatment of the (driven) mixture. 
To compare the many-body and mean-field treatments,
the mean-field interaction parameters are held fixed while the numbers of particles
$N_1$ and $N_2$ are increased towards the infinite-particle-number limit.
In this way, the total number of particles $N=N_1+N_2$ in increased to infinity while
the ratios $\frac{N_1}{N_2} = \frac{\lambda_{12}}{\lambda_{21}}$,
$\frac{N_1}{N} = \frac{\lambda_{12}}{\lambda_{12}+\lambda_{21}}$,
$\frac{N_2}{N} = \frac{\lambda_{21}}{\lambda_{12}+\lambda_{21}}$,
defined using the (non-zero) intraspecies interaction strength $\lambda_{12}$,
are held fixed.

The solution of the time-dependent coupled non-linear Schr\"odinger equations (\ref{MIX_EQ_TDGP_1})
is somewhat involved and for brevity and completeness detailed in appendix \ref{TD_GPEs}.
The final result for the time-dependent mean-field orbitals can be written as
\beqn\label{TDGP_MIX_Orbitals}
& &
\phi_1(x_1,t) =
\left(\frac{m_1\Omega_{1,GP}}{\pi}\right)^{\frac{1}{4}} e^{-i\mu_1(t)} \times \nonumber \\
& & \times e^{-\frac{m_1\Omega_{1,GP}}{2}\left[x_1-\frac{1}{m_1\Lambda_{12}+m_2\Lambda_{21}}\left[\frac{\Lambda_{12}f_{L,1}+\Lambda_{21}f_{L,2}}{\omega^2-\omega^2_L} +
\frac{m_2\Lambda_{21}\left(\frac{f_{L,1}}{m_1}-\frac{f_{L,2}}{m_2}\right)}{\Omega^2_{12}-\omega^2_L}
\right]\cos(\omega_L t)\right]^2} \times \nonumber \\
& & \times e^{-i\frac{m_1\omega_L}{m_1\Lambda_{12}+m_2\Lambda_{21}}\left[\frac{\Lambda_{12}f_{L,1}+\Lambda_{21}f_{L,2}}{\omega^2-\omega^2_L} +
\frac{m_2\Lambda_{21}\left(\frac{f_{L,1}}{m_1}-\frac{f_{L,2}}{m_2}\right)}{\Omega^2_{12}-\omega^2_L}
\right]\sin(\omega_L t)x_1} = \nonumber \\
& & 
= \left(\frac{m_1\Omega_{1,GP}}{\pi}\right)^{\frac{1}{4}} e^{-i\mu_1(t)} e^{-\frac{m_1\Omega_{1,GP}}{2}
\left[x_1-x_1(t)\right]^2} e^{+im_1\dot x_1(t)x_1}, \nonumber \\
& &
\phi_2(x_2,t) = \left(\frac{m_2\Omega_{2,GP}}{\pi}\right)^{\frac{1}{4}} e^{-i\mu_2(t)} \times \nonumber \\
& & \times e^{-\frac{m_2\Omega_{2,GP}}{2}
\left[x_2-\frac{1}{m_1\Lambda_{12}+m_2\Lambda_{21}}\left[\frac{\Lambda_{12}f_{L,1}+\Lambda_{21}f_{L,2}}{\omega^2-\omega^2_L} -
\frac{m_1\Lambda_{12}\left(\frac{f_{L,1}}{m_1}-\frac{f_{L,2}}{m_2}\right)}{\Omega^2_{12}-\omega^2_L}
\right]\cos(\omega_L t)\right]^2} \times \nonumber \\
& & \times e^{-i\frac{m_2\omega_L}{m_1\Lambda_{12}+m_2\Lambda_{21}}\left[\frac{\Lambda_{12}f_{L,1}+\Lambda_{21}f_{L,2}}{\omega^2-\omega^2_L} -
\frac{m_1\Lambda_{12}\left(\frac{f_{L,1}}{m_1}-\frac{f_{L,2}}{m_2}\right)}{\Omega^2_{12}-\omega^2_L}
\right]\sin(\omega_L t)x_2} = \nonumber \\
& & = \left(\frac{m_2\Omega_{2,GP}}{\pi}\right)^{\frac{1}{4}} e^{-i\mu_2(t)} e^{-\frac{m_2\Omega_{2,GP}}{2}
\left[x_2-x_2(t)\right]^2} e^{+im_2\dot x_2(t)x_2},\
\eeqn
where $\Omega_{1,GP}=\sqrt{\omega^2 + \frac{2(\Lambda_1+\Lambda_{21})}{m_1}}$
and $\Omega_{2,GP}=\sqrt{\omega^2 + \frac{2(\Lambda_2+\Lambda_{12})}{m_2}}$
are interaction-dressed frequencies.
The main finding at this point, see appendix \ref{TD_GPEs} for details,
is that the time-dependent mean-field orbitals (\ref{TDGP_MIX_Orbitals})
oscillate with the same amplitudes $x_1(t)$ and $x_2(t)$ found in the many-body solution,
see (\ref{CRTZ_TRNS}),
and otherwise are similar to the ground-state mean-field orbitals of the static problem
being dressed by both the intraspecies and intersepcies interactions \cite{IN10}.
These results suggest that the respective many-body and mean-field time-dependent
reduced density matrices per particle of the driven mixture
are to coincide at the limit of an infinite number of particles, see explicitly below.
More delicate and intricate are the respective relations of the quasienergies.

As is in the driven single-species case,
the time-dependent phases $\mu_1(t)$ and $\mu_2(t)$ in (\ref{TDGP_MIX_Orbitals})
consist
of periodic and linear in $t$ parts.
The former does not enter the quasienergy
and can be read off (\ref{Phases_GP_MIX}) in appendix \ref{TD_GPEs},
and the latter are given explicitly as
\beqn\label{Phases_GP_MIX_QCP_main}
& & \mu_{1,F} = \left(\frac{\Omega_{1,GP}}{2}+\frac{\Lambda_1}{2\Omega_{1,GP}} + \frac{\Lambda_{21}}{2\Omega_{2,GP}}\right) -
\frac{1}{4(m_1\Lambda_{12}+m_2\Lambda_{21})}\Bigg\{\frac{m_1}{m_1\Lambda_{12}+m_2\Lambda_{21}}(\omega^2-\omega_L^2) \times \nonumber \\
& & \times \left[\frac{\Lambda_{12}f_{L,1}+\Lambda_{21}f_{L,2}}{\omega^2-\omega^2_L} +
\frac{m_2\Lambda_{21}\left(\frac{f_{L,1}}{m_1}-\frac{f_{L,2}}{m_2}\right)}{\Omega^2_{12}-\omega^2_L}\right]^2 + \nonumber \\
& & + 2\Lambda_{21}\left[2 \, \frac{\Lambda_{12}f_{L,1}+\Lambda_{21}f_{L,2}}{\omega^2-\omega_L^2} \,
\frac{\frac{f_{L_1}}{m_1}-\frac{f_{L_2}}{m_2}}{\Omega_{12}^2-\omega_L^2}+
(m_2\Lambda_{21}-m_1\Lambda_{12})\left(\frac{\frac{f_{L,1}}{m_1}-\frac{f_{L,2}}{m_2}}{\Omega_{12}^2-\omega_L^2}\right)^2
\right]
\Bigg\}, \nonumber \\
& & \mu_{2,F} = \left(\frac{\Omega_{2,GP}}{2}+\frac{\Lambda_2}{2\Omega_{2,GP}} + \frac{\Lambda_{12}}{2\Omega_{1,GP}}\right) -
\frac{1}{4(m_1\Lambda_{12}+m_2\Lambda_{21})}\Bigg\{\frac{m_2}{m_1\Lambda_{12}+m_2\Lambda_{21}}(\omega^2-\omega_L^2) \times \nonumber \\
& & \times \left[\frac{\Lambda_{12}f_{L,1}+\Lambda_{21}f_{L,2}}{\omega^2-\omega^2_L} -
\frac{m_1\Lambda_{12}\left(\frac{f_{L,1}}{m_1}-\frac{f_{L,2}}{m_2}\right)}{\Omega^2_{12}-\omega^2_L}
\right]^2 - \nonumber \\
& & - 2\Lambda_{12}\left[2 \, \frac{\Lambda_{12}f_{L,1}+\Lambda_{21}f_{L,2}}{\omega^2-\omega_L^2} \,
\frac{\frac{f_{L_1}}{m_1}-\frac{f_{L_2}}{m_2}}{\Omega_{12}^2-\omega_L^2}+
(m_2\Lambda_{21}-m_1\Lambda_{12})\left(\frac{\frac{f_{L,1}}{m_1}-\frac{f_{L,2}}{m_2}}{\Omega_{12}^2-\omega_L^2}\right)^2
\right]
\Bigg\}.
\
\eeqn
The quantities
$\mu_{1,F}$ and $\mu_{2,F}$ may be called quasichemical-potentials of the driven mixture,
and are simultaneous eigenvalues of the coupled non-linear Floquet equations
\beqn\label{MIX_EQ_TDGP_Floquet}
& &
\Bigg\{ -\frac{1}{2m_1} \frac{\partial^2}{\partial x_1^2} + \frac{1}{2} m_1 \omega^2 x_1^2 
- x_1 f_{L,1} \cos(\omega_Lt) + \Lambda_1 \int dx'_1 |\bar\phi_1(x'_1,t)|^2 (x_1-x'_1)^2 
+ \nonumber \\
& & 
+ \Lambda_{21} \int dx_2 |\bar\phi_2(x_2,t)|^2 (x_1-x_2)^2 - i \frac{\partial}{\partial t} \Bigg\} \bar\phi_1(x_1,t) = 
\mu_{1,F}\bar\phi_1(x_1,t),
\nonumber \\
& & 
\Bigg\{ -\frac{1}{2m_2} \frac{\partial^2}{\partial x_2^2} + \frac{1}{2} m_2 \omega^2 x_2^2 
- x_2 f_{L,2} \cos(\omega_Lt) + \Lambda_2  \int dx'_2 |\bar\phi_2(x'_2,t)|^2 (x_2-x'_2)^2 
+ \nonumber \\
& &
+ \Lambda_{12} \int dx_1 |\bar\phi_1(x_1,t)|^2 (x_1-x_2)^2 - i \frac{\partial}{\partial t} \Bigg\} \bar\phi_2(x_2,t) = 
\mu_{2,F} \bar\phi_2(x_2,t), \
\eeqn
where $\bar\phi_1(x_1,t)$ and $\bar\phi_1(x_2,t)$ are the corresponding periodic parts
of the time-dependent orbitals,
$\phi_1(x_1,t) = e^{-i\mu_{1,F}t}\bar\phi_1(x_1,t)$ and
$\phi_2(x_2,t) = e^{-i\mu_{2,F}t}\bar\phi_2(x_2,t)$.
Thus, when solved at the mean-field level of theory,
a driven mixture gives rise to a coupled non-linear Floquet equations
whose eigenvalues are analogs of the chemical potentials of a static mixture.

As can be seen from (\ref{Phases_GP_MIX_QCP_main}),
the quasichemical-potentials $\mu_{1,F}$ and $\mu_{2,F}$
are built from two parts.
The first is the contribution of the static problem to the solution
and dressed by the intraspecies and interspecies interactions,
and the second, which originates from the driving forces
and depends on these forces and the interspecies interaction only.
In particular,
the poles' structure of $\mu_{1,F}$ and $\mu_{2,F}$
would play a central role in the mean-field quasienergy
and its relation to the many-body quasienergy (\ref{E_F_MIX}).

With the mean-field wavefunction (\ref{MF_MIX_WF}) computed,
we proceed to the time-dependent reduced density matrices and thereafter to the limit of an infinite number of particles.
Thus, the lowest-order intraspecies and interspecies  
reduced density matrices are given at the mean-field level by
\beqn\label{1_2_12_density_MF_MIX}
& & \rho^{(1)}_{1,MF}(x_1,x'_1,t) = N_1 \rho^{(1)}_{1,GP}(x_1,x'_1,t), \qquad
\rho^{(1)}_{1,GP}(x_1,x'_1,t) = \left(\frac{m_1\Omega_{1,GP}}{\pi}\right)^{\frac{1}{2}} \times \nonumber \\
& & \times e^{-\frac{m_1\Omega_{1,GP}}{2}\left\{\left[x_1 - x_1(t)\right]^2 + \left[x'_1 - x_1(t)\right]^2\right\}}
e^{+i m_1 \dot{x_1}(t) \left\{\left[x_1-x_1(t)\right]-\left[x'_1-x_1(t)\right]\right\}}, \nonumber \\
& & \rho^{(1)}_{2,MF}(x_2,x'_2,t) = N_2 \rho^{(1)}_{2,GP}(x_2,x'_2,t), \qquad
\rho^{(1)}_{2,GP}(x_2,x'_2,t) = \left(\frac{m_2\Omega_{2,GP}}{\pi}\right)^{\frac{1}{2}} \times \nonumber \\
& & \times e^{-\frac{m_2\Omega_{2,GP}}{2}\left\{\left[x_2 - x_2(t)\right]^2 + \left[x'_2 - x_2(t)\right]^2\right\}}
e^{+i m_2 \dot{x_2}(t) \left\{\left[x_2-x_2(t)\right]-\left[x'_2-x_2(t)\right]\right\}}, \nonumber \\
& & \rho^{(2)}_{12,MF}(x_1,x'_1,x_2,x'_2,t) = \rho^{(1)}_{1,MF}(x_1,x'_1,t) \rho^{(1)}_{2,MF}(x_2,x'_2,t). \
\eeqn
Recall that, at the mean-field level,
$\rho^{(2)}_{12,MF}(x_1,x'_1,x_2,x'_2,t)$ is naturally built as a product of intraspecies
reduced one-particle density matrices.
Correspondingly, the diagonal parts of (\ref{1_2_12_density_MF_MIX}) are the intraspecies,
$\rho^{(1)}_{1,MF}(x_1,t) = N_1 \rho^{(1)}_{1,GP}(x_2,t)$,
$\rho^{(1)}_{1,GP}(x_1,t) = \left(\frac{m_1\Omega_{1,GP}}{\pi}\right)^{\frac{1}{2}}\times$\break\hfill
$\times e^{-m_1\Omega_{1,GP}\left[x_1 - x_1(t)\right]^2}$ and
$\rho^{(1)}_{2,MF}(x_2,t) = N_2 \rho^{(1)}_{2,GP}(x_2,t)$,
$\rho^{(1)}_{2,GP}(x_2,t) = \left(\frac{m_2\Omega_{2,GP}}{\pi}\right)^{\frac{1}{2}}\times$\break\hfill
$\times e^{-m_2\Omega_{2,GP}\left[x_2 - x_2(t)\right]^2}$,
and interspecies,
$\rho^{(2)}_{12,MF}(x_1,x_2,t) = \rho^{(1)}_{1,MF}(x_1,t) \rho^{(1)}_{2,MF}(x_2,t)$,
densities.

To proceed,
let $\bar\Phi(x_{1,1},\ldots,x_{1,N_1},x_{2,1},\ldots,x_{2,N_2},t)=
\prod_{j=1}^{N_1} \bar\phi_1(x_{1,j},t) \prod_{k=1}^{N_2} \bar\phi_2(x_{2,k},t)$
be the periodic part of (\ref{MF_MIX_WF}),
i.e., $\Phi(x_{1,1},\ldots,x_{1,N_1},x_{2,1},\ldots,x_{2,N_2},t) = e^{-i\left(N_1\mu_{1,F}+N_2\mu_{2,F}\right)t} \times$ $\times \bar\Phi(x_{1,1},\ldots,x_{1,N_1},x_{2,1},\ldots,x_{2,N_2},t)$.
The combination $N_1\mu_{1,F}+N_2\mu_{2,F}$ in the phase leads to cancellation of some terms
in the mean-field Floquet energy per particle, see appendix \ref{TD_GPEs}, 
which is defined by sandwiching the many-boson Floquet Hamiltonian with the mean-field quasienergy wavefunction,
\beqn\label{MF_Floquet_eps_N}
& & \varepsilon_F^{GP}=\frac{\langle\bar{\Phi}|\hat H - i\frac{\partial}{\partial t}|\bar{\Phi}\rangle}{N} = 
\frac{1}{2}\frac{\Lambda_{12} \Omega_{1,GP} + \Lambda_{21} \Omega_{2,GP}}{\Lambda_{12}+\Lambda_{21}}
- \frac{1}{4(\Lambda_{12}+\Lambda_{21})(m_1\Lambda_{12}+m_2\Lambda_{21})} \times \nonumber \\
& & \times
\left[\frac{m_1m_2\Lambda_{12}\Lambda_{21}
\left(\frac{f_{L,1}}{m_1}-\frac{f_{L,2}}{m_2}\right)^2\left(\omega^2-\omega_L^2\right)}{\left(\Omega_{12}^2-\omega_L^2\right)^2} +
\frac{\left(\Lambda_{12}f_{L,1}+\Lambda_{21}f_{L,2}\right)^2}{\omega^2-\omega_L^2}\right].\
\eeqn
Expression (\ref{MF_Floquet_eps_N}) is
in analogy to a mixture's ground-state in which 
the mean-field energy per particle is
obtained by sandwiching the static many-boson Hamiltonian
with the static mean-field wavefunction.

We are now in the position to examine the relations
between the many-body and mean-field solutions of the driven bosonic mixture
at the limit of an infinite number of particles.
Thus, the following relations readily hold:
\beq\label{RDM_MIX_1_limit}
\lim_{N_1 \to \infty \atop N_2 \to \infty} \frac{\rho_1^{(1)}(x_1,x'_1,t)}{N_1} = \rho^{(1)}_{1,GP}(x_1,x'_1,t), \qquad
\lim_{N_1 \to \infty \atop N_2 \to \infty} \frac{\rho_2^{(1)}(x_2,x'_2,t)}{N_2} = \rho^{(1)}_{2,GP}(x_2,x'_2,t)
\eeq
for the time-dependent reduced one-particle density matrices per particle,
with\break\hfill
$\lim_{N_1 \to \infty \atop N_2 \to \infty} \frac{\rho_1^{(1)}(x_1,t)}{N_1} = \rho^{(1)}_{1,GP}(x_1,t)$,
$\lim_{N_1 \to \infty \atop N_2 \to \infty} \frac{\rho_2^{(1)}(x_2,t)}{N_2} = \rho^{(1)}_{2,GP}(x_2,t)$ 
for the time-dependent densities per particle,
and
\beq\label{RDM_MIX_12_limit}
\lim_{N_1 \to \infty \atop N_2 \to \infty} \frac{\rho_{12}^{(2)}(x_1,x'_1,x_2,x'_2,t)}{N_1N_2} = 
\rho^{(1)}_{1,GP}(x_1,x'_1,t)\rho^{(1)}_{2,GP}(x_2,x'_2,t),
\eeq
with 
$\lim_{N_1 \to \infty \atop N_2 \to \infty} \frac{\rho_{12}^{(2)}(x_1,x_2,t)}{N_1N_2} = 
\rho^{(1)}_{1,GP}(x_1,t)\rho^{(1)}_{2,GP}(x_2,t)$
for the time-dependent interspecies two-particle density.

The above relations between the time-dependent many-body and mean-field 
reduced density matrices generalize for time-periodic (Floquet) mixtures 
literature results for the ground state of trapped mixtures and, when available, for the dynamics 
of interacting bosons with time-independent Hamiltonians.
On top of and contrary to that,
an appealing and different result emerges for the quasienergy per particle. 
Here we find
\beqn\label{QE_MIX_limit}
& & \lim_{N_1 \to \infty \atop N_2 \to \infty}\frac{\mathcal{E}_F}{N} =
\frac{1}{2}\frac{\Lambda_{12} \Omega_{1,GP} + \Lambda_{21} \Omega_{2,GP}}{\Lambda_{12}+\Lambda_{21}}
- \frac{1}{4(\Lambda_{12}+\Lambda_{21})(m_1\Lambda_{12}+m_2\Lambda_{21})} \times \nonumber \\
& & \times
\left[\frac{m_1m_2\Lambda_{12}\Lambda_{21}
\left(\frac{f_{L,1}}{m_1}-\frac{f_{L,2}}{m_2}\right)^2}{\Omega_{12}^2-\omega_L^2} +
\frac{\left(\Lambda_{12}f_{L,1}+\Lambda_{21}f_{L,2}\right)^2}{\omega^2-\omega_L^2}\right] \ne
\varepsilon_F^{GP}, \
\eeqn
implying
that the quasienergy per particle computed at the many-body level of theory (\ref{E_F_MIX}) is
different than that at mean-field level (\ref{MF_Floquet_eps_N}),
even at the infinite-particle-number limit.

Comparative-wise, the result (\ref{QE_MIX_limit}) for 
the quasienergy per particle of the driven mixture
differs from the literature results
for the ground-state energy per particle of trapped interacting bosons and mixtures
at the limit of an infinite number of particles.
Here, the difference can be seen as the additional factor $\frac{\omega^2-\omega_L^2}{\Omega_{12}^2-\omega_L^2}$,
the ratio of the two resonance frequencies' poles, 
appearing in the first term inside the square brackets of the mean-field quasienergy (\ref{MF_Floquet_eps_N}).
This additional factor
emerges because
the non-linear interaction terms in the mean-field treatment 
cannot compensate both
poles in the driven amplitudes $x_1(t)$ and $x_2(t)$.
More precisely,
the mean-field treatment renormalizes the corresponding phases only
to the center-of-mass frequency $\omega^2$ and not also to
the relative center-of-mass frequency $\Omega_{12}^2$,
see appendix \ref{TD_GPEs} for the derivation.

When are the quasienergies equal?
Trivially, if there is no interaction between the two species, implying that $\Omega_{12}=\omega$ and
therefore $\frac{\omega^2-\omega_L^2}{\Omega_{12}^2-\omega_L^2}=1$,
i.e., we have then two harmonically-trapped non-coupled driven Bose-Einstein condensates.
Otherwise, as long as the relative center-of-mass coordinate is {\it not} activated by the forces
the respective qusienergies per particle are equal,
namely,
\beqn\label{QE_MIX_limit_cut}
& & \lim_{N_1 \to \infty \atop N_2 \to \infty}\frac{\mathcal{E}_F}{N}\Bigg|_{\frac{f_{L,1}}{m_1}=\frac{f_{L,2}}{m_2}} =
\frac{1}{2}\frac{\Lambda_{12} \Omega_{1,GP} + \Lambda_{21} \Omega_{2,GP}}{\Lambda_{12}+\Lambda_{21}} - \nonumber \\
& & - \frac{f_{L,1}^2(m_1\Lambda_{12}+m_2\Lambda_{21})}{4m_1^2(\Lambda_{12}+\Lambda_{21})(\omega^2-\omega_L^2)} =
\varepsilon_F^{GP}\Bigg|_{\frac{f_{L,1}}{m_1}=\frac{f_{L,2}}{m_2}}. \
\eeqn 
The later occurs independently of the number of particles and strengths of interactions,
when the forces and masses are interrelated by $\frac{f_{L,1}}{m_1}=\frac{f_{L,2}}{m_2}$.
This brings the present section on many-body and mean-field quasienergies and
Floquet reduced density matrices to an end.

\section{Angular momentum and its fluctuations in a Bose-Einstein condensate steered by
an interacting bosonic impurity}\label{ANG_MIX}

So far,
we considered the periodically driving of a mixture in one spatial dimensions.
A natural extension and application is to consider a two-dimensional mixture and
drive it correspondingly along the $x$ and $y$ directions such that it is steered,
namely, circularly driven.
Then, angular momentum is imprinted on the bosons.
We use the harmonic-interaction model for the driven mixture 
to investigate the roles of intraspecies and interspecies 
interactions and resulting angular-momentum fluctuations \cite{IN13,IN15}
at the many-body and mean-field levels.
The availability of a solvable model is instructive.

Our strategy in this section is to begin with steering a single particle 
in two spatial dimensions and compute the angular momentum and fluctuations,
namely, the expectation value of the angular-momentum operator and
its variance.
Then, we steer the $N$-boson problem and investigate the role of interaction
at the many-body and mean-field levels of theory.
Finally, we steer species $1$ bosons, which serve
as an interacting bosonic impurity, embedded in species $2$ bosons
where the latter are not steered,
and investigate the modes of rotation,
distribution of angular momentum between the species,
and fluctuations.

Consider the circularly-driven
two-dimensional quantum harmonic oscillator [$\r=(x,y)$],
\beq\label{2D_HAM_1p}
\hat h(\r,t) = -\frac{1}{2m}\frac{\partial^2}{\partial \r^2} + 
\frac{1}{2} m \omega^2 \r^2 - f_L \left[x \cos(\omega_L t) + y \sin(\omega_L t)\right].
\eeq
The ground Floquet solution
of the time-dependent Schr\"odinger equation is given by
\beqn\label{2D_WF_1p}
& & \psi(\r,t) = 
\left(\frac{m\omega}{\pi}\right)^{\frac{1}{2}} e^{-i2\varepsilon_F t} 
e^{-\frac{m\omega}{2}\left\{\left[x - x(t)\right]^2+\left[y - y(t)\right]^2\right\}}
e^{+i m\left[\dot{x}(t) x + \dot{y}(t) y\right]}, \nonumber \\
& & x(t)=a\cos(\omega_L t), \quad y(t)=a\sin(\omega_L t), \quad a = \frac{f_L}{m(\omega^2-\omega_L^2)},
\eeqn
where $\varepsilon_F$ is given in (\ref{1P_Ef}).

The time-dependent wavefunction (\ref{2D_WF_1p}) can be seen,
apart from the time-dependent phase $e^{-i2\varepsilon_F t}$ which
is irrelevant for the properties discussed in this section,
as the ground state of the two-dimensional isotropic harmonic oscillator translated by $x(t)$ and $y(t)$ and
boosted by $m\dot x(t)$ and $m\dot y(t)$ along the $x$ and $y$ directions, respectively.
A useful way to evaluate the expectation value of the angular-momentum operator
and its variance
carried by a translated and boosted wavefunction 
is to use the transformation properties of the angular-momentum operator under translations and boosts. 
In appendix \ref{VAR_TRANS_BST}
we derive and discuss the corresponding expressions
for the most general wavefunction treated in this work, that of driven mixtures,
where each species can be translated and boosted differently.

Using (\ref{AM_trans_boost}) for $N=1$ particles we can readily evaluate the expectation value of the angular-momentum operator
$\hat l_z = \hat x \hat p_y - \hat y \hat p_x$ which reads
\beq\label{2D_AM_1p}
\langle \hat l_z \rangle\Big|_{\psi(\r,t)} = x(t) m\dot y(t) - y(t) m\dot x(t) = m \omega_L \left[\frac{f_L}{m(\omega^2-\omega_L^2)}\right]^2.
\eeq
The value (\ref{2D_AM_1p}) is that of a `classical' particle encircling the origin with angular velocity $\omega_L$ and radius
$\left|\frac{f_L}{m(\omega^2-\omega_L^2)}\right|$.
The radius is determined by the driving force and diverges as one approaches
the resonance.
On the other end,
the radius is always non-zero (for $f_L \ne 0$),
meaning that the expectation value of the angular momentum for the driven particle is non-zero (and positive). 

The Floquet solution (\ref{2D_WF_1p}) of the steered harmonic oscillator is not an eigenfunction of the angular-momentum operator.
Using the transformation properties of the latter
under translations and boosts one can readily compute the angular-momentum variance,
which is found to depend on the position and momentum 
variances along the $x$ and $y$ directions,
see appendix \ref{VAR_TRANS_BST}.
Thus, employing (\ref{Var_Lz_Spher_Symm}) for $N=1$ particles we find
\beqn\label{Var_Lz_Spher_Symm_1p}
& & \Delta^2_{\hat l_z}\Big|_{\psi(\r,t)} = \left[x^2(t) + y^2(t)\right] \frac{m\omega}{2} +
\left[m^2\dot x^2(t) + m^2\dot y^2(t)\right] \frac{1}{2m\omega} = \nonumber \\
& & = \frac{m\left(\omega^2+\omega_L^2\right)}{2\omega} \left[\frac{f_L}{m(\omega^2-\omega_L^2)}\right]^2.
\eeqn
The dependence of the variance (\ref{Var_Lz_Spher_Symm_1p})
on the radius $\left|\frac{f_L}{m(\omega^2-\omega_L^2)}\right|$ is 
in conjunction with the expectation value (\ref{2D_AM_1p}).
For instance,
for tight confinements ($\omega \gg 1$)
the radius diminishes and the variance too.
Note, however, that for $\omega \to 0$ the variance diverges whereas
the expectation value of the angular momentum does not.
This signifies that the angular-momentum variance depends on the size of the wavepacket,
and for weaker confinements this size increases.
We will return to these points when
interactions set in.

Let us move to the circularly-driven
isotropic harmonic-interaction model in two spatial dimensions,
\beqn\label{2D_Steered_HIM}
& & \hat H(\r_1,\ldots,\r_N,t) = \sum_{j=1}^N \Bigg\{ -\frac{1}{2m}\frac{\partial^2}{\partial \r_j^2} + 
\frac{1}{2}m\omega^2 \r_j^2 - \nonumber \\
& & - f_L \left[x_j \cos(\omega_L t) + y_j \sin(\omega_L t)\right] \Bigg\} + \lambda \sum_{1\le j <k}^N \left(\r_j-\r_k\right)^2. \
\eeqn
The ground Floquet solution is given by
\beqn\label{HIM_driv_crtz_WF_2D}
& & 
\Psi(\r_1,\ldots,\r_N,t) = \left(\frac{m\Omega}{\pi}\right)^{\frac{N-1}{2}} \left(\frac{m\omega}{\pi}\right)^{\frac{1}{2}}
e^{-i2\mathcal{E}_Ft}
e^{-\frac{\alpha}{2}\sum_{j=1}^N \left\{\left[x_j-x(t)\right]^2 + \left[y_j-y(t)\right]^2\right\}} \times \nonumber \\
& & \times e^{-\beta \sum_{1\le j < k}^N \left\{\left[x_j-x(t)\right]\left[x_k-x(t)\right] + \left[y_j-y(t)\right]\left[y_k-y(t)\right]\right\}}
e^{+i m \left[\dot{x}(t) \sum_{j=1}^N x_j + \dot{y}(t) \sum_{j=1}^N y_j \right]},
\eeqn
where $\mathcal{E}_F$ is given in (\ref{MB_Ef}).
Now, the expectation value of the many-particle angular-momentum operator
$\hat L_Z = \sum_{j=1}^N (\hat x_j \hat p_{y,j} - \hat y_j \hat p_{x,j})$
and the respective variance
are given by
\beqn\label{2D_AM_HIM}
& & \frac{1}{N}\langle \hat L_Z \rangle\Big|_{\Psi(\r_1,\ldots,\r_N,t)} = m \omega_L \left[\frac{f_L}{m(\omega^2-\omega_L^2)}\right]^2,
\nonumber \\
& & \frac{1}{N}\Delta^2_{\hat L_Z}\Big|_{\Psi(\r_1,\ldots,\r_N,t)} =
\frac{m\left(\omega^2+\omega_L^2\right)}{2\omega} \left[\frac{f_L}{m(\omega^2-\omega_L^2)}\right]^2, \ 
\eeqn
where (\ref{AM_trans_boost}) and (\ref{Var_Lz_Spher_Symm}) are employed.
We see that the interaction between particles 
enters neither the expectation value nor the variance of
the angular-momentum operator
at the many-body level,
and the values per particle are given by those of the steered single-particle system,
see (\ref{2D_AM_1p}) and (\ref{Var_Lz_Spher_Symm_1p}).

Let us examine the respective quantities at the mean-field level,
where the Floquet solution was derived analytically in Sec.~\ref{DRIV_SING} 
and reads in two spatial dimensions
\beqn\label{HIM_driv_crtz_WF_GP_2D}
& & 
\Phi(\r_1,\ldots,\r_N,t) = \left(\frac{m\Omega_{GP}}{\pi}\right)^{\frac{N}{2}}
e^{-i2N\mu_Ft}
e^{-\frac{m\Omega_{GP}}{2}\sum_{j=1}^N \left\{\left[x_j-x(t)\right]^2 + \left[y_j-y(t)\right]^2\right\}} \times \nonumber \\
& & \times e^{+i m \left[\dot{x}(t) \sum_{j=1}^N x_j + \dot{y}(t) \sum_{j=1}^N y_j \right]},
\eeqn
where $\mu_F$ is given in (\ref{MF_mu}).
Computing now the expectation value and variance of $\hat L_Z$ we find
\beqn\label{2D_AM_HIM_GP}
& & \frac{1}{N}\langle \hat L_Z \rangle\Big|_{\Phi(\r_1,\ldots,\r_N,t)} = m \omega_L \left[\frac{f_L}{m(\omega^2-\omega_L^2)}\right]^2,
 \\
& & \frac{1}{N}\Delta^2_{\hat L_Z}\Big|_{\Phi(\r_1,\ldots,\r_N,t)} =
\frac{m\left(\omega_{GP}^2+\omega_L^2\right)}{2\omega_{GP}} \left[\frac{f_L}{m(\omega^2-\omega_L^2)}\right]^2 =
\frac{m\left(\omega^2+\frac{2\Lambda}{m}+\omega_L^2\right)}{2\sqrt{\omega^2+\frac{2\Lambda}{m}}}\left[\frac{f_L}{m(\omega^2-\omega_L^2)}\right]^2, \nonumber \ 
\eeqn
where the variance exhibits dependence on the interaction strength.
This behavior
is because
the position and momentum variances (of the non-boosted-and-translated wavefunction $\Phi$)
depend themselves
at the mean-field level
on the interaction strength,
$\frac{1}{N} \Delta^2_{\hat X}\Big|_{\Phi} =
\frac{1}{N} \Delta^2_{\hat Y}\Big|_{\Phi} = \frac{1}{2m\sqrt{\omega^2+\frac{2\Lambda}{m}}}$
and 
$\frac{1}{N} \Delta^2_{\hat P_X}\Big|_{\Phi} = 
\frac{1}{N} \Delta^2_{\hat P_Y}\Big|_{\Phi} = \frac{m\sqrt{\omega^2+\frac{2\Lambda}{m}}}{2}$.
In particular, for repulsive [$\Lambda \to (-m\omega^2/2)^+$]
as well as for attractive ($\Lambda \gg 1$) interactions 
the angular-momentum variance (\ref{2D_AM_HIM_GP}) can diverge,
unlike the situation at the many-body level of theory (\ref{2D_AM_HIM}).
On the other end, the variance does not vanish for any interacting strength.
The situation becomes further interesting when we
steer species $1$ bosons embedded in species $2$ bosons, see below.
Finally, we note that the expectation value per particle of the angular momentum
at the many-body and mean-field levels of theory
coincide (for finite systems and)
at the limit of an infinite number of particles,
as is expected
for one-body operators \cite{IN6,IN13}.

Let us proceed to the scenario of a steered mixture.
Consider the harmonic-interaction model in two spatial dimensions for 
circularly-driven species $1$ bosons embedded in un-steered species $2$ bosons:
\beqn\label{HAM_MIX_2D}
& & \hat H(\r_{1,1},\ldots,\r_{1,N_1},\r_{2,1},\ldots,\r_{2,N_2},t) = 
 \sum_{j=1}^{N_1} \Bigg\{ -\frac{1}{2m_1} \frac{\partial^2}{\partial \r_{1,j}^2} + 
\frac{1}{2} m_1\omega^2 \r_{1,j}^2 - \nonumber \\
& & - f_{L,1} \left[x_{1,j} \cos(\omega_L t) + y_{1,j} \sin(\omega_L t)\right] \Bigg\} +
\lambda_1 \sum_{1 \le j < k}^{N_1} (\r_{1,j}-\r_{1,k})^2 + \\
& & + \sum_{j=1}^{N_2} \left( -\frac{1}{2m_2} \frac{\partial^2}{\partial \r_{2,j}^2} + 
\frac{1}{2} m_2\omega^2 \r_{2,j}^2 \right) + \lambda_2 \sum_{1 \le j < k}^{N_2} (\r_{2,j}-\r_{2,k})^2 +
\lambda_{12} \sum_{j=1}^{N_1} \sum_{k=1}^{N_2} (\r_{1,j}-\r_{2,k})^2. \nonumber \  
\eeqn
We expect the interspecies interaction $\lambda_{12}$ to mediate
the rotation from species $1$
to species $2$, and the question to answer is in what way.

The ground Floquet solution to (\ref{HAM_MIX_2D}) is given by
\beqn\label{WAVE_FUN_1_CRTZ_2D}
& & \Psi(\r_{1,1},\ldots,\r_{1,N_1},\r_{2,1},\ldots,\r_{2,N_2},t) = 
\left(\frac{m_1\Omega_1}{\pi}\right)^{\frac{N_1-1}{2}}
\left(\frac{m_2\Omega_2}{\pi}\right)^{\frac{N_2-1}{2}}
\left(\frac{M_{12}\Omega_{12}}{\pi}\right)^{\frac{1}{2}}
\left(\frac{M\omega}{\pi}\right)^{\frac{1}{2}}
\times \nonumber \\
& & \times
e^{-i2 \mathcal{E}_F t}
e^{-\frac{\alpha_1}{2} \sum_{j=1}^{N_1} \left\{\left[x_{1,j} - x_1(t)\right]^2 + \left[y_{1,j} - y_1(t)\right]^2 \right\}- 
\beta_1 \sum_{1 \le j < k}^{N_1} \left\{\left[x_{1,j} - x_1(t)\right] \left[x_{1,k} - x_1(t)\right] +
\left[y_{1,j} - y_1(t)\right] \left[y_{1,k} - y_1(t)\right]
\right\}} \times \nonumber \\
& & \times e^{-\frac{\alpha_2}{2} \sum_{j=1}^{N_2} \left\{\left[x_{2,j} - x_2(t)\right]^2 + \left[y_{2,j} - y_2(t)\right]^2\right\}
- \beta_2 \sum_{1 \le j < k}^{N_2}
\left\{\left[x_{2,j} - x_2(t)\right]\left[x_{2,k} - x_2(t)\right]+\left[y_{2,j} - y_2(t)\right]\left[y_{2,k} - y_2(t)\right]\right\}} \times \nonumber \\ 
& & \times e^{+\gamma \sum_{j=1}^{N_1} \sum_{k=1}^{N_2}
\left\{\left[x_{1,j} - x_1(t)\right]\left[x_{2,k} - x_2(t)\right] + \left[y_{1,j} - y_1(t)\right]\left[y_{2,k} - y_2(t)\right]\right\}} \times \nonumber \\
& & \times e^{+i \left\{m_1 \left[\dot x_1(t) \sum_{j=1}^{N_1} x_{1,j} + \dot y_1(t) \sum_{j=1}^{N_1} y_{1,j}\right] + 
m_2 \left[\dot x_2(t) \sum_{j=1}^{N_2} x_{2,j} + \dot y_2(t) \sum_{j=1}^{N_2} y_{2,j}\right]\right\}}, \
\
\eeqn
where the quasienergy
$\mathcal{E}_F$ is given in (\ref{E_F_MIX}),
$x_1(t) = a_1 \cos(\omega_L t)$, $y_1(t) = a_1 \sin(\omega_L t)$,
$x_2(t) = a_2 \cos(\omega_L t)$, and $y_2(t) = a_2 \sin(\omega_L t)$,
and the amplitudes $a_1$ and $a_2$,
determining the respective radii of rotations,
are:
\beqn\label{MIX_AMP_2D}
& & 
a_1 = \frac{f_{L,1} N_1}{M}\left[\frac{1}{\omega^2-\omega_L^2}+\frac{m_2N_2}{m_1N_1}\frac{1}{\Omega_{12}^2-\omega_L^2}\right] =
\frac{f_{L,1}\left[\omega^2 + \frac{2\Lambda_{12}}{m_2}-\omega^2_L\right]}{m_1\left(\omega^2-\omega_L^2\right)
\left[\omega^2 + 2\left(\frac{\Lambda_{12}}{m_2}+\frac{\Lambda_{21}}{m_1}\right)-\omega_L^2\right]}, \nonumber \\
& &
a_2 = \frac{f_{L,1} N_1}{M}\left[\frac{1}{\omega^2-\omega_L^2}-\frac{1}{\Omega_{12}^2-\omega_L^2}\right] =
\frac{f_{L,1} \frac{2\Lambda_{12}}{m_2}}{m_1\left(\omega^2-\omega_L^2\right)
\left[\omega^2 + 2\left(\frac{\Lambda_{12}}{m_2}+\frac{\Lambda_{21}}{m_1}\right)-\omega_L^2\right]}.
\eeqn
The resulting amplitudes $a_1$ and $a_2$ deserve a discussion.
Despite only species $1$ is being steered,
there are contributions from
two distinct poles, at $\omega^2$ and at $\Omega_{12}^2=
\omega^2 + 2\left(\frac{\Lambda_{12}}{m_2}+\frac{\Lambda_{21}}{m_1}\right)$,
signifying the activation of the center-of-mass $Q_N$ and 
relative center-of-mass $Q_{N-1}$ coordinates in the presence
of coupling between the two species.
The radius of rotation of species $2$, $a_2$, 
is linear in the intraspecies interacting parameter, $\Lambda_{12}=N_1\lambda_{12}$,
and vanishes as expected for diminishing coupling between
the steered species $1$ and un-steered species $2$ 
Bose-Einstein condensates.

It is appealing to see that there is also a zero for $a_1$,
when
\beq\label{w_L_0}
\omega^0_L = \sqrt{\omega^2 + \frac{2\Lambda_{12}}{m_2}} > 0 \qquad \Longrightarrow \qquad
a_1\Big|_{\omega_L=\omega_L^0}=0, \quad
a_2\Big|_{\omega_L=\omega_L^0}=-\frac{f_{L,1}}{\Lambda_{21}},
\eeq
which is `born' only due to the interaction between species $1$ and species $2$ bosons.
Physically, a vanishing value for $a_1$ means that species $1$ does not rotate or move,
despite being driven by force $f_{L,1}$ at the angular frequency $\omega^0_L$.
Alternatively, we may say that
the effect of the driving force $f_{L,1}$ exerted on species $1$ is then
completely transfered to species $2$.
The value of $a_2$ at $\omega^0_L$, see (\ref{w_L_0}),
is inversely proportional to the intraspecies
interaction parameter $\Lambda_{21} = N_2 \lambda_{12}$.
Further analysis of $a_1$ and $a_2$ is given below 
when analyzing the distribution of angular momentum in 
between the two species.

The mean-field Floquet wavefunction in two spatial dimensions is needed to compute respective properties 
and 
is given by
\beqn\label{WAVE_FUN_1_CRTZ_GP_2D}
& & \Phi(\r_{1,1},\ldots,\r_{1,N_1},\r_{2,1},\ldots,\r_{2,N_2},t) = 
\left(\frac{m_1\Omega_{1,GP}}{\pi}\right)^{\frac{N_1}{2}}
\left(\frac{m_2\Omega_{2,GP}}{\pi}\right)^{\frac{N_2}{2}}
e^{-i2 \left(N_1\mu_{1,F}+N_2\mu_{2,F}\right) t} \times \nonumber \\
& & \times
e^{-\frac{m_1\Omega_{1,GP}}{2}\sum_{j=1}^{N_1} \left\{\left[x_{1,j} - x_1(t)\right]^2 + \left[y_{1,j} - y_1(t)\right]^2 \right\}}
e^{-\frac{m_2\Omega_{2,GP}}{2} \sum_{j=1}^{N_2} \left\{\left[x_{2,j} - x_2(t)\right]^2 + \left[y_{2,j} - y_2(t)\right]^2\right\}} \times \nonumber \\
& & \times e^{+i \left\{m_1 \left[\dot x_1(t) \sum_{j=1}^{N_1} x_{1,j} + \dot y_1(t) \sum_{j=1}^{N_1} y_{1,j}\right] + 
m_2 \left[\dot x_2(t) \sum_{j=1}^{N_2} x_{2,j} + \dot y_2(t) \sum_{j=1}^{N_2} y_{2,j}\right]\right\}}, \
\
\eeqn
where the quasichemical-potentials $\mu_{1,F}$ and $\mu_{2,F}$
are given in (\ref{Phases_GP_MIX_QCP_main}).

We are now in the position to compute the expectation value of the angular-momentum operator and its variance,
and compare the many-body and mean-field results.
For a mixture $\hat L_Z = \hat L_{Z_1} + \hat L_{Z_2}$,
hence we shall also investigate the connection
between the mixture's and individual species' angular momenta.
We start with the expectation values at the many-body level of theory
which take on the form:
\beqn\label{2D_AM_HIM_MIX_MB_GP_Lz}
& & \!\!\!\!\!\! \frac{1}{N}\langle \hat L_Z \rangle\Big|_{\Psi(\r_{1,1},\ldots,\r_{2,N_2},t)} =
\frac{\Lambda_{12}}{\Lambda_{12}+\Lambda_{21}} \cdot \frac{1}{N_1}\langle \hat L_{Z_1} \rangle\Big|_{\Psi(\r_{1,1},\ldots,\r_{2,N_2},t)} +
\frac{\Lambda_{21}}{\Lambda_{12}+\Lambda_{21}} \cdot \frac{1}{N_2}\langle \hat L_{Z_2} \rangle\Big|_{\Psi(\r_{1,1},\ldots,\r_{2,N_2},t)} = \nonumber \\
& & = \frac{\Lambda_{12}^2 f_{L,1}^2 \omega_L}{(\Lambda_{12}+\Lambda_{21})(m_1\Lambda_{12}+m_2\Lambda_{21})}
\left\{\frac{1}{(\omega^2-\omega_L^2)^2} +
\frac{m_2\Lambda_{21}}{m_1\Lambda_{12}}\frac{1}{\left[\omega^2 + 2\left(\frac{\Lambda_{12}}{m_2}+\frac{\Lambda_{21}}{m_1}\right)-\omega_L^2\right]^2}\right\}, \nonumber \\
& & \frac{1}{N_1}\langle \hat L_{Z_1} \rangle\Big|_{\Psi(\r_{1,1},\ldots,\r_{2,N_2},t)} = 
\frac{f_{L,1}^2 \omega_L \left(\omega^2 + \frac{2\Lambda_{12}}{m_2}-\omega^2_L\right)^2}{m_1\left(\omega^2-\omega_L^2\right)^2
\left[\omega^2 + 2\left(\frac{\Lambda_{12}}{m_2}+\frac{\Lambda_{21}}{m_1}\right)-\omega_L^2\right]^2}, \nonumber \\
& & \frac{1}{N_2}\langle \hat L_{Z_2} \rangle\Big|_{\Psi(\r_{1,1},\ldots,\r_{2,N_2},t)} =
\frac{f_{L,1}^2 \omega_L \left(\frac{2\Lambda_{12}}{m_2}\right)^2\frac{m_2}{m_1}}{m_1\left(\omega^2-\omega_L^2\right)^2
\left[\omega^2 + 2\left(\frac{\Lambda_{12}}{m_2}+\frac{\Lambda_{21}}{m_1}\right)-\omega_L^2\right]^2}. \
\eeqn
The amount of angular momentum carried by each species is proportional to the square of the radios of its rotation.
The poles and zeros of the radii govern the sizes of angular momenta,
yet only the zeros dictates the relative distribution of angular momentum between the two species;
recall that only species $1$ is steered.
Explicitly,
the ratio of the angular momenta per particle of the two species is
$\frac{\frac{1}{N_1}\langle \hat L_{Z_1} \rangle\big|_{\Psi(\r_{1,1},\ldots,\r_{2,N_2},t)}}
{\frac{1}{N_2}\langle \hat L_{Z_2} \rangle\big|_{\Psi(\r_{1,1},\ldots,\r_{2,N_2},t)}}=
\frac{m_1}{m_2}\frac{\left(\omega^2 + \frac{2\Lambda_{12}}{m_2}-\omega^2_L\right)^2}
{\left(\frac{2\Lambda_{12}}{m_2}\right)^2}$
and can vary from arbitrary large for small intraspecies interaction parameters $|\Lambda_{12}|$
to arbitrary small due to the zero of $a_1$ at $\omega^0_L = \sqrt{\omega^2 + \frac{2\Lambda_{12}}{m_2}}$, see (\ref{w_L_0}).
Another interesting regime occurs in the vicinity of the center-of-mass pole
$\frac{1}{\left(\omega^2-\omega_L^2\right)^2}$, see (\ref{2D_AM_HIM_MIX_MB_GP_Lz}),
where the ratio of the angular momenta per particle approaches the mass ratio $\frac{m_1}{m_2}$
and is independent of other parameters.
Of course, in the vicinity of any of the poles the rotation radii increase ad infinitum.
 
For the mean-field expectation values,
$\frac{1}{N_1}\langle \hat L_{Z_1} \rangle\Big|_{\Phi(\r_{1,1},\ldots,\r_{2,N_2},t)}$,
$\frac{1}{N_2}\langle \hat L_{Z_2} \rangle\Big|_{\Phi(\r_{1,1},\ldots,\r_{2,N_2},t)}$,
and therefore for
$\frac{1}{N}\langle \hat L_Z \rangle\Big|_{\Phi(\r_{1,1},\ldots,\r_{2,N_2},t)}$,
we find explicitly from (\ref{WAVE_FUN_1_CRTZ_GP_2D}), as expected, the same results as the many-body ones,
see (\ref{2D_AM_HIM_MIX_MB_GP_Lz}).
On the other hand, the variances computed at the many-body and mean-field levels of theory are different and intricate,
as we now explicitly show.

To compute the variances of $\hat L_Z$, $\hat L_{Z_1}$, and $\hat L_{Z_2}$
of the time-dependent mixture
we need according to appendix \ref{VAR_TRANS_BST} 
the respective position, momentum, and angular-momentum variances 
of the individual species,
computed at the many-body and mean-field levels with respect to the wavefunctions $\Psi$ and $\Phi$
without translations and boosts,
see (\ref{Var_Lz1_Lz2_Spher_Symm}) and (\ref{Var_Lz_Spher_Symm_MIX}).
Using the definition of the center-of-mass and relative center-of-mass Jacoby coordinates
in terms of the individual species' center-of-mass coordinates,
the final result
at the many-body level reads \cite{HM17}:
\beqn\label{VAR_X1_PX2_MB}
& & \frac{1}{N_1}\Delta^2_{\hat X_1}\Big|_{\Psi} =
\frac{1}{N_1}\Delta^2_{\hat Y_1}\Big|_{\Psi} = 
\frac{\Lambda_{12}}{2(m_1\Lambda_{12}+m_2\Lambda_{21})}
\left[\frac{1}{\omega} + \frac{m_2\Lambda_{21}}{m_1\Lambda_{12}}\frac{1}{\sqrt{\omega^2 + 2\left(\frac{\Lambda_{12}}{m_2}+\frac{\Lambda_{21}}{m_1}\right)}}\right], \nonumber \\
& & \frac{1}{N_2}\Delta^2_{\hat X_2}\Big|_{\Psi} =
\frac{1}{N_2}\Delta^2_{\hat Y_2}\Big|_{\Psi} = 
\frac{\Lambda_{21}}{2(m_1\Lambda_{12}+m_2\Lambda_{21})}
\left[\frac{1}{\omega} + \frac{m_1\Lambda_{12}}{m_2\Lambda_{21}}\frac{1}{\sqrt{\omega^2 + 2\left(\frac{\Lambda_{12}}{m_2}+\frac{\Lambda_{21}}{m_1}\right)}}\right], \nonumber \\
& & \frac{1}{N_1}\Delta^2_{\hat P_{X_1}}\Big|_{\Psi} =
\frac{1}{N_1}\Delta^2_{\hat P_{Y_1}}\Big|_{\Psi} = 
\frac{m_1^2\Lambda_{12}}{2(m_1\Lambda_{12}+m_2\Lambda_{21})}
\left[\omega + \frac{m_2\Lambda_{21}}{m_1\Lambda_{12}}
\sqrt{\omega^2 + 2\left(\frac{\Lambda_{12}}{m_2}+\frac{\Lambda_{21}}{m_1}\right)}\right], \nonumber \\
& & \frac{1}{N_2}\Delta^2_{\hat P_{X_2}}\Big|_{\Psi} =
\frac{1}{N_2}\Delta^2_{\hat P_{Y_2}}\Big|_{\Psi} = 
\frac{m_2^2\Lambda_{21}}{2(m_1\Lambda_{12}+m_2\Lambda_{21})}
\left[\omega + \frac{m_1\Lambda_{12}}{m_2\Lambda_{21}}
\sqrt{\omega^2 + 2\left(\frac{\Lambda_{12}}{m_2}+\frac{\Lambda_{21}}{m_1}\right)}\right], \nonumber \\
& & \Delta^2_{\hat L_{Z_1}}\Big|_{\Psi} = \Delta^2_{\hat L_{Z_2}}\Big|_{\Psi} = 
\frac{\frac{\Lambda_{12}}{m_2}\frac{\Lambda_{21}}{m_1}}
{2\left(\frac{\Lambda_{12}}{m_2}+\frac{\Lambda_{21}}{m_1}\right)^2}
\frac{\left[\sqrt{\omega^2 + 2\left(\frac{\Lambda_{12}}{m_2}+\frac{\Lambda_{21}}{m_1}\right)}-\omega\right]^2}
{\omega\sqrt{\omega^2 + 2\left(\frac{\Lambda_{12}}{m_2}+\frac{\Lambda_{21}}{m_1}\right)}}. \
\eeqn
The respective quantities at the mean-field level are simpler to evaluate since $\Phi$ is a product state,
and the final result reads \cite{HM17}:
\beqn\label{VAR_X1_PX2_GP}
& & \frac{1}{N_1}\Delta^2_{\hat X_1}\Big|_{\Phi} =
\frac{1}{N_1}\Delta^2_{\hat Y_1}\Big|_{\Phi} = 
\frac{1}{2m_1 \sqrt{\omega^2 + \frac{2}{m_1}\left(\Lambda_1+\Lambda_{21}\right)}}, \nonumber \\
& & \frac{1}{N_2}\Delta^2_{\hat X_2}\Big|_{\Phi} =
\frac{1}{N_2}\Delta^2_{\hat Y_2}\Big|_{\Phi} = 
\frac{1}{2m_2 \sqrt{\omega^2 + \frac{2}{m_2}\left(\Lambda_2+\Lambda_{12}\right)}}, \nonumber \\
& & \frac{1}{N_1}\Delta^2_{\hat P_{X_1}}\Big|_{\Phi} =
\frac{1}{N_1}\Delta^2_{\hat P_{Y_1}}\Big|_{\Phi} = 
\frac{m_1}{2}\sqrt{\omega^2 + \frac{2}{m_1}\left(\Lambda_1+\Lambda_{21}\right)}, \nonumber \\
& & \frac{1}{N_2}\Delta^2_{\hat P_{X_2}}\Big|_{\Phi} =
\frac{1}{N_2}\Delta^2_{\hat P_{Y_2}}\Big|_{\Phi} = 
\frac{m_2}{2}\sqrt{\omega^2 + \frac{2}{m_2}\left(\Lambda_2+\Lambda_{12}\right)}, \nonumber \\
& & \Delta^2_{\hat L_{Z_1}}\Big|_{\Phi} = \Delta^2_{\hat L_{Z_2}}\Big|_{\Phi} = 0. \
\eeqn
It is worthwhile to list the main properties of and differences
between the variances (\ref{VAR_X1_PX2_MB}) and (\ref{VAR_X1_PX2_GP}),
starting with the position and momentum variances,
as far as they are needed for our investigations:
(i) The individual species' variances depend only on the interspecies interaction parameters $\Lambda_{12}$ and $\Lambda_{21}$
at the many-body level whereas at the mean-field level they additionally depend on the intraspecies interaction parameters $\Lambda_1$
and $\Lambda_2$.
(ii) For attractive interactions the momentum variances can diverge and for repulsive interactions the position variances can diverge.
Combining (i) and (ii) one can have situations where a position or momentum variance at the many-body level does not diverge 
whereas it does diverge at the mean-field level, and vice versa,
i.e., mean-field variances tending to zero
whereas many-body variances remaining finite.
Furthermore, for the individual species' angular-momentum variances:
(iii) There is a non-vanishing value (\ref{VAR_X1_PX2_MB}) at the many-body level of theory 
which solely originates
from the interaction between the species,
and which can diverge for both repulsive and attractive interspecies interaction.
Interestingly, it is marginal in the number of particles,
that is, the angular-momentum variance per, unlike the position and momentum variances per particle,
diminishes for large and at the limit of an infinite number of particles.
Finally, at the mean-field level of theory,
the angular-momentum variance (\ref{VAR_X1_PX2_GP}) is zero due to separability of $\Phi$.
We are now in the position to put the pieces together.

We find for the angular-momentum variance at the many-body level
\begin{subequations}\label{2D_AM_HIM_MIX_VAR_MB_Lz}
\beqn\label{2D_AM_HIM_MIX_VAR_MB_Lz_a}
& & \frac{1}{N}\Delta^2_{\hat L_Z}\Big|_{\Psi(\r_{1,1},\ldots,\r_{2,N_2},t)} =
\frac{\Lambda_{12}^2 f_{L,1}^2}{2(\Lambda_{12}+\Lambda_{21})(m_1\Lambda_{12}+m_2\Lambda_{21})} \times \nonumber \\
& & \times
\left\{\frac{\omega^2+\omega_L^2}{\omega(\omega^2-\omega_L^2)^2} +
\frac{m_2\Lambda_{21}}{m_1\Lambda_{12}}
\frac{\omega^2 + 2\left(\frac{\Lambda_{12}}{m_2}+\frac{\Lambda_{21}}{m_1}\right)+\omega_L^2}
{\sqrt{\omega^2 + 2\left(\frac{\Lambda_{12}}{m_2}+\frac{\Lambda_{21}}{m_1}\right)}\left[\omega^2 + 2\left(\frac{\Lambda_{12}}{m_2}+\frac{\Lambda_{21}}{m_1}\right)-\omega_L^2\right]^2}\right\}, \nonumber \\
\eeqn
where the individual species' values are
\beqn\label{2D_AM_HIM_MIX_VAR_MB_Lz_b}
& & \frac{1}{N_1}\Delta^2_{\hat L_{Z_1}}\Big|_{\Psi(\r_{1,1},\ldots,\r_{2,N_2},t)} = 
\frac{1}{N_1} \cdot \frac{\frac{\Lambda_{12}}{m_2}\frac{\Lambda_{21}}{m_1}}
{2\left(\frac{\Lambda_{12}}{m_2}+\frac{\Lambda_{21}}{m_1}\right)^2}
\frac{\left[\sqrt{\omega^2 + 2\left(\frac{\Lambda_{12}}{m_2}+\frac{\Lambda_{21}}{m_1}\right)}-\omega\right]^2}
{\omega\sqrt{\omega^2 + 2\left(\frac{\Lambda_{12}}{m_2}+\frac{\Lambda_{21}}{m_1}\right)}} + \nonumber \\
& & +
\frac{f_{L,1}^2 \left[\omega^2 + \frac{2\Lambda_{12}}{m_2}-\omega^2_L\right]^2}{m_1\left(\omega^2-\omega_L^2\right)^2
\left[\omega^2 + 2\left(\frac{\Lambda_{12}}{m_2}+\frac{\Lambda_{21}}{m_1}\right)-\omega_L^2\right]^2} \times \nonumber \\
& & \times \frac{m_1\Lambda_{12}}{2(m_1\Lambda_{12}+m_2\Lambda_{21})}
\left[\frac{\omega^2+\omega_L^2}{\omega} +
\frac{m_2\Lambda_{21}}{m_1\Lambda_{12}}\frac{\omega^2 +
2\left(\frac{\Lambda_{12}}{m_2}+\frac{\Lambda_{21}}{m_1}\right) + \omega_L^2}{\sqrt{\omega^2 + 2\left(\frac{\Lambda_{12}}{m_2}+\frac{\Lambda_{21}}{m_1}\right)}}\right], \nonumber \\ \nonumber \\
& & \frac{1}{N_2} \Delta^2_{\hat L_{Z_2}}\Big|_{\Psi(\r_{1,1},\ldots,\r_{2,N_2},t)} =
\frac{1}{N_2} \cdot \frac{\frac{\Lambda_{12}}{m_2}\frac{\Lambda_{21}}{m_1}}
{2\left(\frac{\Lambda_{12}}{m_2}+\frac{\Lambda_{21}}{m_1}\right)^2}
\frac{\left[\sqrt{\omega^2 + 2\left(\frac{\Lambda_{12}}{m_2}+\frac{\Lambda_{21}}{m_1}\right)}-\omega\right]^2}
{\omega\sqrt{\omega^2 + 2\left(\frac{\Lambda_{12}}{m_2}+\frac{\Lambda_{21}}{m_1}\right)}} + \nonumber \\
& & +
\frac{f_{L,1}^2 \left(\frac{2\Lambda_{12}}{m_2}\right)^2\frac{m_2}{m_1}}{m_1\left(\omega^2-\omega_L^2\right)^2
\left[\omega^2 + 2\left(\frac{\Lambda_{12}}{m_2}+\frac{\Lambda_{21}}{m_1}\right)-\omega_L^2\right]^2} \times \nonumber \\
& & \times
\frac{m_2\Lambda_{21}}{2(m_1\Lambda_{12}+m_2\Lambda_{21})}
\left[\frac{\omega^2+\omega_L^2}{\omega} + \frac{m_1\Lambda_{12}}{m_2\Lambda_{21}}
\frac{\omega^2 + 2\left(\frac{\Lambda_{12}}{m_2}+\frac{\Lambda_{21}}{m_1}\right)+\omega_L^2}{\sqrt{\omega^2 + 2\left(\frac{\Lambda_{12}}{m_2}+\frac{\Lambda_{21}}{m_1}\right)}}\right]. \
\eeqn
\end{subequations}
Examining the expressions (\ref{2D_AM_HIM_MIX_VAR_MB_Lz}) for the angular-momentum variances 
the first point to note
is that the individual species' variances (\ref{2D_AM_HIM_MIX_VAR_MB_Lz_b})
do not sum up to the
mixture's variance (\ref{2D_AM_HIM_MIX_VAR_MB_Lz_a}).
This is due to the position
$\langle\Psi| \hat X_1 \hat X_2 |\Psi\rangle = \langle\Psi| \hat Y_1 \hat Y_2 |\Psi\rangle$
and momentum
$\langle\Psi| \hat P_{X_1} \hat P_{X_2} |\Psi\rangle = \langle\Psi| \hat P_{Y_1} \hat P_{Y_2} |\Psi\rangle$ 
cross terms
appearing in (\ref{Var_Lz_Spher_Symm_MIX}).
For the individual species, see (\ref{Var_Lz1_Lz2_Spher_Symm}),
one of course 
does not have contributions from any cross term.
On the other hand for the mixture's angular-momentum variance,  
the angular-momentum cross term
$\langle\Psi| \hat L_{Z_1} \hat L_{Z_2} |\Psi\rangle$ cancels
the individual species' contributions
$\langle\Psi| \hat L^2_{Z_1} |\Psi\rangle$ and $\langle\Psi| \hat L^2_{Z_2} |\Psi\rangle$
due to conservation of angular momentum, $\left(\hat L_{Z_1} + \hat L_{Z_2}\right)|\Psi\rangle = 0$,
and hence this cross term
does not appear in (\ref{Var_Lz_Spher_Symm_MIX}).
All in all,
the angular-momentum variance of the driven mixture 
(\ref{2D_AM_HIM_MIX_VAR_MB_Lz_a}) does not vanish,
depends on the interspecies interaction parameters $\Lambda_{12}$ and $\Lambda_{21}$ only,
can diverge either due to repulsive or attractive interspecies interaction, 
and is governed by the poles of the center-of-mass and relative center-of-mass coordinates.

To proceed, the angular-momentum variances at the
mean-field level of theory are evaluated and read
\beqn\label{2D_AM_HIM_MIX_VAR_GP_Lz}
& & \frac{1}{N}\Delta^2_{\hat L_Z}\Big|_{\Phi(\r_{1,1},\ldots,\r_{2,N_2},t)} =
\frac{\Lambda_{12}}{\Lambda_{12}+\Lambda_{21}} \cdot \frac{1}{N_1}\Delta^2_{\hat L_{Z_1}}\Big|_{\Phi(\r_{1,1},\ldots,\r_{2,N_2},t)} +
\frac{\Lambda_{21}}{\Lambda_{12}+\Lambda_{21}} \cdot \frac{1}{N_2}\Delta^2_{\hat L_{Z_2}}\Big|_{\Phi(\r_{1,1},\ldots,\r_{2,N_2},t)}, \nonumber \\
& & \frac{1}{N_1}\Delta^2_{\hat L_{Z_1}}\Big|_{\Phi(\r_{1,1},\ldots,\r_{2,N_2},t)} = 
\frac{f_{L,1}^2 \left[\omega^2 + \frac{2\Lambda_{12}}{m_2}-\omega^2_L\right]^2}{m_1\left(\omega^2-\omega_L^2\right)^2
\left[\omega^2 + 2\left(\frac{\Lambda_{12}}{m_2}+\frac{\Lambda_{21}}{m_1}\right)-\omega_L^2\right]^2}
\frac{\omega^2 + \frac{2}{m_1}\left(\Lambda_1+\Lambda_{21}\right)+\omega_L^2}
{\sqrt{\omega^2 + \frac{2}{m_1}\left(\Lambda_1+\Lambda_{21}\right)}}, \nonumber \\
& & \frac{1}{N_2}\Delta^2_{\hat L_{Z_2}}\Big|_{\Phi(\r_{1,1},\ldots,\r_{2,N_2},t)} =
\frac{f_{L,1}^2 \left(\frac{2\Lambda_{12}}{m_2}\right)^2\frac{m_2}{m_1}}{m_1\left(\omega^2-\omega_L^2\right)^2
\left[\omega^2 + 2\left(\frac{\Lambda_{12}}{m_2}+\frac{\Lambda_{21}}{m_1}\right)-\omega_L^2\right]^2}
\frac{\omega^2 + \frac{2}{m_2}\left(\Lambda_2+\Lambda_{12}\right)+\omega_L^2}
{\sqrt{\omega^2 + \frac{2}{m_2}\left(\Lambda_2+\Lambda_{12}\right)}}, \nonumber \\
\eeqn
where (\ref{Var_Lz1_Lz2_Spher_Symm}) and (\ref{Var_Lz_Spher_Symm_MIX}) are employed.
At the mean-field level,
in contrast with the outcome (\ref{2D_AM_HIM_MIX_VAR_MB_Lz}) at the many-body level,
the individual species' angular-momentum variances
add up to the mixture's angular-momentum variance,
since all cross terms 
$\langle\Phi| \hat X_1 \hat X_2 |\Phi\rangle$, $\langle\Phi| \hat Y_1 \hat Y_2 |\Phi\rangle$,
$\langle\Phi| \hat P_{X_1} \hat P_{X_2} |\Phi\rangle$, $\langle\Phi| \hat P_{Y_1} \hat P_{Y_2} |\Phi\rangle$,
and
$\langle\Phi| \hat L_{Z_1} \hat L_{Z_2} |\Phi\rangle$
vanish.
On top of that,
the mean-field quantities
depend on all interaction parameters,
intraspecies and interspecies,
not only on the latter ones.
All together,
the angular-momentum variances per particle
at the mean-field level
(\ref{2D_AM_HIM_MIX_VAR_GP_Lz})
are different
from those at the many-body level
(\ref{2D_AM_HIM_MIX_VAR_MB_Lz})
for finite systems as well as at the limit
of an infinite number of particles
where each of the species is $100\%$ condensed, i.e.,
$\lim_{N_1 \to \infty \atop N_2 \to \infty}\frac{1}{N}\Delta^2_{\hat L_Z}\big|_{\Psi(\r_{1,1},\ldots,\r_{2,N_2},t)} \ne
\lim_{N_1 \to \infty \atop N_2 \to \infty}\frac{1}{N}\Delta^2_{\hat L_Z}\big|_{\Phi(\r_{1,1},\ldots,\r_{2,N_2},t)}$
for the mixture's angular-momentum variance
and
$\lim_{N_1 \to \infty \atop N_2 \to \infty}\frac{1}{N_1}\Delta^2_{\hat L_{Z_1}}\big|_{\Psi(\r_{1,1},\ldots,\r_{2,N_2},t)} \ne
\lim_{N_1 \to \infty \atop N_2 \to \infty}\frac{1}{N_1}\Delta^2_{\hat L_{Z_1}}\big|_{\Phi(\r_{1,1},\ldots,\r_{2,N_2},t)}$,
$\lim_{N_1 \to \infty \atop N_2 \to \infty}\frac{1}{N_2}\Delta^2_{\hat L_{Z_2}}\big|_{\Psi(\r_{1,1},\ldots,\r_{2,N_2},t)} \ne
\lim_{N_1 \to \infty \atop N_2 \to \infty}\frac{1}{N_2}\Delta^2_{\hat L_{Z_2}}\big|_{\Phi(\r_{1,1},\ldots,\r_{2,N_2},t)}$
for the individual species' angular-momentum variances.

A final point.
What happens to the angular-momentum expectation values and variances
at the frequency $\omega_L=\omega_L^0$,
when the effect of the
steering force exerted on species $1$ is completely transfered to species $2$?
First, since $a_1\big|_{\omega_L=\omega_L^0} = 0$
the expectation value of the mixture's angular momentum and its fluctuations,
at the many-body and mean-field levels,
are now given by the respective expressions of the un-steered species $2$ only.
Second, one gets for the angular-momentum variance of species $1$
that the only contribution at the many-body level is the marginal part,
which diminishes to zero, the mean-field value,
as the number of particles increases.
And third, the respective many-body and mean-field expressions for the mixture's
and species $2$ variances remain different,
even at the limit of an infinite number of particles
when both species are $100\%$ condensed.
This wraps up the present work.

\section{Summary and Outlook}\label{SUM_OUT}

We have presented in this work a solvable time-dependent model of a many-particle system
and used it to amalgamate together several themes:
(i) Mixtures of trapped Bose-Einstein condensates;
(ii) Periodically-driven many-particle systems;
(iv) Many-body versus mean-field theory
of Floquet bosonic systems at the limit of an infinite number of particles, and;
(iv) The imprinting and fluctuations of angular momentum  
when a Bose-Einstein condensate is steered by interacting bosons consisting of a different species.

We introduced the driven harmonic-interaction model for mixtures,
namely,
$N_1$ bosons of mass $m_1$ interacting by harmonic interparticle interaction of strength $\lambda_1$
and driven by a time-periodic force of amplitude $f_{L,1}$ and frequency $\omega_L$,
and
$N_2$ bosons of mass $m_2$ interacting by harmonic interparticle interaction of strength $\lambda_2$
and driven by a time-periodic force of amplitude $f_{L,1}$ and the same frequency $\omega_L$.
All bosons are trapped in an harmonic potential of frequency $\omega$
and, furthermore, bosons of species $1$ and species $2$ interact with each other
by harmonic interparticle interaction of strength $\lambda_{12}$.
The model generalizes the harmonic-interaction model for mixtures to the time-dependent domain.

Using Jacoby coordinates the time-dependent model is diagonalized,
and governed by two time-dependent many-particle oscillators,
associate with the center-of-mass and relative center-of-mass of the driven mixture,
and two poles of respective frequencies.
The frequency of the center-of-mass motion does not depend on any interaction strength
whereas the frequency of the relative center-of-mass depends on the interspecies interaction strength. 
We concentrated in this work on the ground Floquet solution of the driven mixture,
and begun by computing its quasienergy,
the eigenvalue of the time-periodic many-particle Floquet Hamiltonain,
as a function of the above parameters.

Back at the laboratory frame,
all Cartesian coordinates in the time-dependent
many-particle quasienergy state are coupled,
and the coordinates of each species oscillate with a different amplitude combining the two poles
associated with the center-of-mass and relative center-of-mass motions.
This structure of the time-dependent many-boson wavefunction allows one to
obtain explicitly the intraspecies and interspecies time-dependent reduced
one-particle and two-particle density matrices of the driven mixture,
which possess the same coherence properties as the corresponding static quantities 
computed for the ground state of the trapped mixture.

One of the main goals of the present work is to ask and answer 
questions on the relation between many-body and mean-field theories of (trapped) interacting bosons.
These have been dealt with in the literature for the ground state or for the dynamics with time-independent Hamiltonians,
and we would like to extend the topic to the realm of Floquet many-boson systems.
To discuss properties of the driven mixture at the limit of an infinite number of particles,
explicitly the connection between the respective
many-body and mean-field results for 
the time-dependent reduced density matrices, densities,
the quasienergies which are unique for Floquet systems,
and the variances,
the solution of the coupled Floquet non-linear Schr\"odinger equations of the driven mixture has
been explicitly derived.

With closed-form expressions at the many-body and mean-field levels of theory,
we have proven explicitly that, at the limit of an infinite number of particles,
the intraspecies time-dependent reduced one-particle and two-particle density matrices per particle
are $100\%$ condensed and given by the respective mean-field quantities.
Furthermore in this limit we have proven
that the
interspecies time-dependent reduced two-particle density matrix per particle
is separable, i.e., given by the product of the time-dependent reduced one-particle density matrices per particle
of species $1$ and species $2$ bosons,
and is given by the respective mean-field quantity.
These results generalize literature results for the ground state and the dynamics with time-independent Hamiltonians
to a Floquet bosonic system.

We have found that the relation between the many-body and mean-field quasienergies per particle is intriguing,
namely, that the many-body and mean-field results {\it need not} coincide.
This is unlike the situation for the many-body and mean-field energies per particle of the ground state of trapped bosons and mixtures.
Explicitly,
the coupled non-linear Schr\"odinger equations cannot renormalize appropriately the intensity
of both poles of the oscillating mixture in the time-dependent mean-field wavefucntion,
and hence in the mean-field quasienergy,
in comparison with the many-body treatment.
Thus, the many-body quasienergy per particle coincides with the mean-field value 
at the infinite-particle-number limit in the specific case where {\it only} the center-of-mass
motion of the driven mixture is activated by the driving forces. 
In the general case when the relative center-of-mass motion is
also activated,
the many-body quasienergy per particle differs from the mean-field one
even at the limit of an infinite number of particles.

We then investigated an application in two spatial dimensions,
in which only species $1$ bosons are circularly driven by a time-periodic force,
and through their interspecies interaction with the un-steered species $2$ bosons,
induce rotation and imprint angular momentum on the latter.
The dynamics is solved analytically and analyzed within many-body and mean-field theories as presented above.
We first concentrated on the average of the angular-momentum operators of each of the species.
Despite steering species $1$ bosons only,
we show that the ratio of the angular momenta per particle of species $1$ and species $2$ bosons
can be marginally small or arbitrary large,
because of the interspecies interaction.
Explicitly, the latter induces a zero in the radius of rotation of species $1$ bosons
at a frequency just between the center-of-mass and relative center-of-mass resonances.
We have also shown that
the expectation values per particle of the angular-momentum operators
for the many-body and mean-field solutions of the time-periodic mixture
coincide (for finite systems and)
at the limit of an infinite number of particles.

Finally, we investigated the variances of the angular-momentum operators of each of the
species and of the mixture.
To this end,
it is useful to express the Floquet quasienergy state of the steered mixture
using translation and boost operators along the $x$ and $y$ directions.
Consequently, we can express the angular-momentum variance computed
from a translated and boosted wavefunction using the
position, momentum, and angular-momentum variances of the
un-translated-and-boosted wavefunction.
The angular-momentum variances of species $1$ and species $2$ bosons
and of the mixture were computed explicitly and compared at the many-body and mean-field levels of theory.
The mean-field variances depend on the resonances and all interactions,
and can diverge due to either the attractive or repulsive intraspecies and interspecies interactions.
Interestingly, at the mean-field level of theory the individual species' angular-momentum variances are additive,
and sum up to the mixture's angular-momentum variance.
On the other hand,
the many-body quantities depend on the resonances and interspecies interaction only,
can diverge due to either the attractive or repulsive interspecies interaction,
and are non-additive because of the correlation term in the time-dependent many-boson wavefunction.
These differences persist at the limit of an infinite number of particles,
where each of the species comprising the Floquet time-dependent mixture is $100\%$ condensed.

Last but not least, a brief outlook.
There are several directions worth pursing of which we list four.
We have discussed in the present work the driving of a bosonic mixture by a monochromatic force,
and can envision that using a polychromatic force would be instructive.
We have studied the driving of structureless bosons,
and foresee that solvable models for driving structured targets embedded in a Bose-Einstein condensate would be interesting.
We have investigated the imprinting of angular-momentum in a steered two-species mixture in two spatial dimensions,
and suggest that exploring three-species mixtures would be valuable.
Finally, the present model can serve to benchmark numerical algorithms developed to compute
real-space many-boson Floquet states,
which would exhibit both fundamental and applicative intriguing facets. 

\section*{Acknowledgements}

I thank Kaspar Sakmann for discussions. 
This research was supported by the Israel Science Foundation 
(Grants No. 600/15 and No. 1516/19). 

\appendix

\section{Time-dependent intraspecies two-particle reduced density matrices and
their limit at an infinite number of particles}\label{TD_2RDMs}

In this appendix we prescribe explicitly the many-body and mean-field
time-dependent intraspecies two-particle reduced density matrices
and establish their relation at the limit of an infinite number of particles.
All wavefunctions appearing below are normalized to unity.
We start from the single-species driven harmonic-interaction model
and proceed to the harmonic-interaction model for driven mixtures.

The time-dependent two-particle reduced density matrix
is defined as $\rho^{(2)}(x,x',x'',x''',t) = N(N-1) \int dx_3 \cdots dx_N \Psi(x,x',x_3,\ldots,x_N,t)\Psi^\ast(x'',x''',x_3,\ldots,x_N,t)$
and given for the ground Floquet solution of the driven harmonic-interaction model (\ref{HIM_driv_crtz_WF}) by
\beqn\label{2_RDM}
& & 
\rho^{(2)}(x,x',x'',x''',t) = 
N(N-1) \left(\frac{\alpha+C_1}{\pi}\right)^{\frac{1}{2}} \left(\frac{\alpha+C_2}{\pi}\right)^{\frac{1}{2}} \times \nonumber \\
& & 
\times e^{-\frac{\alpha}{2}\left\{\left[x-x(t)\right]^2 + \left[x'-x(t)\right]^2 +
\left[x''-x(t)\right]^2 + \left[x'''-x(t)\right]^2 \right\}} \times \nonumber \\
& & 
\times e^{- \beta \left\{\left[x-x(t)\right]\left[x'-x(t)\right] + \left[x''-x(t)\right]\left[x'''-x(t)\right] \right\}} \times \nonumber \\
& & 
\times e^{- \frac{1}{4} C_2 \left\{\left[x-x(t)\right]+\left[x'-x(t)\right]+\left[x''-x(t)\right]+\left[x'''-x(t)\right]\right\}^2} \times \nonumber \\
& &
\times e^{+i m \dot{x}(t) \left\{\left[x-x(t)\right]+\left[x'-x(t)\right]-\left[x''-x(t)\right]-\left[x'''-x(t)\right]\right\}},
\qquad C_2 = - \beta^2 \frac{N-2}{(\alpha-\beta)+(N-2)\beta},
\eeqn
where $C_1$ can be read from (\ref{1_RDM}).
The diagonal of (\ref{2_RDM}) is the time-dependent two-particle density and reads
\beqn\label{2_density}
& & 
\rho^{(2)}(x,x',t) = 
N(N-1) \left(\frac{\alpha+C_1}{\pi}\right)^{\frac{1}{2}} \left(\frac{\alpha+C_2}{\pi}\right)^{\frac{1}{2}} \times \nonumber \\
& & 
\times e^{-(\alpha+C_2)\left\{\left[x-x(t)\right]^2 + \left[x'-x(t)\right]^2\right\}}
e^{- 2(\beta+C_2)\left[x-x(t)\right]\left[x'-x(t)\right]} = \nonumber \\
& & = N(N-1) \left(\frac{\alpha+C_1}{\pi}\right)^{\frac{1}{2}} \left(\frac{\alpha+C_2}{\pi}\right)^{\frac{1}{2}} \times \nonumber \\
& & \times e^{-(\alpha+C_2)\left[x-\frac{f_L}{m(\omega^2-\omega_L^2)}\cos(\omega_Lt)\right]^2}
e^{-(\alpha+C_2)\left[x'-\frac{f_L}{m(\omega^2-\omega_L^2)}\cos(\omega_Lt)\right]^2} \times \nonumber \\
& & \times e^{-2(\beta+C_2)\left[x-\frac{f_L}{m(\omega^2-\omega_L^2)}\cos(\omega_Lt)\right]\left[x'-\frac{f_L}{m(\omega^2-\omega_L^2)}\cos(\omega_Lt)\right]}.
\eeqn
From the mean-field wavefunction (\ref{MF_ansatz}), 
the reduced two-particle density matrix at the mean-field level reads
\beq\label{2_density_MF}
\rho^{(2)}_{MF}(x,x',x'',x''',t) = N(N-1) \rho^{(1)}_{GP}(x,x'',t)\rho^{(1)}_{GP}(x',x''',t),
\eeq
where
$\rho^{(1)}_{GP}(x,x',t) = \left(\frac{m\Omega_{GP}}{\pi}\right)^{\frac{1}{2}} 
e^{-\frac{m\Omega_{GP}}{2}\left\{\left[x - x(t)\right]^2 + \left[x' - x(t)\right]^2\right\}}
 e^{+i m \dot{x}(t) \left\{\left[x-x(t)\right]-\left[x'-x(t)\right]\right\}}$ [Eq.~(\ref{1_density_MF})].
The two-particle mean-field density reads
$\rho^{(2)}_{MF}(x,x',t) = N(N-1) \rho^{(1)}_{GP}(x,t) \rho^{(1)}_{GP}(x',t)$,
where $ \rho^{(1)}_{GP}(x,t) = \left(\frac{m\Omega_{GP}}{\pi}\right)^{\frac{1}{2}} 
e^{-m\Omega_{GP}\left[x - x(t)\right]^2}$.

Thus,
at the limit of an infinite number of particles one finds that
\beq\label{RDM_2_limit}
 \lim_{N \to \infty} \frac{\rho^{(2)}(x,x',x'',x''',t)}{N(N-1)} = \rho^{(1)}_{GP}(x,x'',t) \rho^{(1)}_{GP}(x',x''',t) 
\eeq
for the time-dependent reduced two-particle density matrix per particle
and
$\lim_{N \to \infty} \frac{\rho^{(2)}(x,x',t)}{N(N-1)} = \rho^{(1)}_{GP}(x,t) \rho^{(1)}_{GP}(x',t)$
for the time-dependent
two-particle density per particle.

For the driven mixture we have 
for the ground Floquet solution (\ref{WAVE_FUN_1_CRTZ},\ref{WAVE_FUN_3})
the following closed-form expressions:
\beqn\label{2_RDM_1}
& & 
\rho_1^{(2)}(x_1,x'_1,x''_1,x'''_1,t) = 
N_1(N_1-1) \left(\frac{\alpha_1+C_{1,0}}{\pi}\right)^{\frac{1}{2}} \left(\frac{\alpha_1+C_{2,0}}{\pi}\right)^{\frac{1}{2}} \times \nonumber \\
& & 
\times e^{-\frac{\alpha_1}{2}\left\{\left[x_1-x_1(t)\right]^2 + \left[x'_1-x_1(t)\right]^2 +
\left[x''_1-x_1(t)\right]^2 + \left[x'''_1-x_1(t)\right]^2 \right\}} \times \nonumber \\
& & 
\times e^{- \beta_1 \left\{\left[x_1-x_1(t)\right]\left[x'_1-x_1(t)\right] +
\left[x''_1-x_1(t)\right]\left[x'''_1-x_1(t)\right] \right\}} \times \nonumber \\
& & 
\times e^{- \frac{1}{4} C_{2,0} \left\{\left[x_1-x_1(t)\right]+\left[x'_1-x_1(t)\right]+\left[x''_1-x_1(t)\right]+\left[x'''_1-x_1(t)\right]\right\}^2} \times \nonumber \\
& &
\times e^{+i m_1 \dot x_1(t) \left\{\left[x_1-x_1(t)\right]+\left[x'_1-x_1(t)\right]-\left[x''_1-x_1(t)\right]-\left[x'''_1-x_1(t)\right]\right\}},
\nonumber \\
& & \quad C_{2,0} =
\frac{(\alpha_1-\beta_1)C_{N_1,0}-(N_1-2)(C_{N_1,0}+\beta_1)\beta_1}{(\alpha_1-\beta_1)+(N_1-2)(C_{N_1,0}+\beta_1)} \
\eeqn
and
\beqn\label{2_RDM_2}
& & 
\rho_2^{(2)}(x_2,x'_2,x''_2,x'''_2,t) = 
N_2(N_2-1) \left(\frac{\alpha_2+C'_{0,1}}{\pi}\right)^{\frac{1}{2}} \left(\frac{\alpha_2+C'_{0,2}}{\pi}\right)^{\frac{1}{2}} \times \nonumber \\
& & 
\times e^{-\frac{\alpha_2}{2}\left\{\left[x_2-x_2(t)\right]^2 + \left[x'_2-x_2(t)\right]^2 +
\left[x''_2-x_2(t)\right]^2 + \left[x'''_2-x_2(t)\right]^2 \right\}} \times \nonumber \\
& & 
\times e^{- \beta_2 \left\{\left[x_2-x_2(t)\right]\left[x'_2-x_2(t)\right] +
\left[x''_2-x_2(t)\right]\left[x'''_2-x_2(t)\right] \right\}} \times \nonumber \\
& & 
\times e^{- \frac{1}{4} C'_{0,2} \left\{\left[x_2-x_2(t)\right]+\left[x'_2-x_2(t)\right]+\left[x''_2-x_2(t)\right]+\left[x'''_2-x_2(t)\right]\right\}^2} \times \nonumber \\
& &
\times e^{+i m_2 \dot x_2(t) \left\{\left[x_2-x_2(t)\right]+\left[x'_2-x_2(t)\right]-\left[x''_2-x_2(t)\right]-\left[x'''_2-x_2(t)\right]\right\}},
\nonumber \\
& & \quad C'_{0,2} =  
\frac{(\alpha_2-\beta_2)C'_{0,N_2}-(N_2-2)(C'_{0,N_2}+\beta_2)\beta_2}{(\alpha_2-\beta_2)+(N_2-2)(C'_{0,N_2}+\beta_2)},
\eeqn
where the time-dependent
two-body densities are given explicitly by
\beqn\label{2_density_1}
& & \!\!\!\!\!\!\!\!\!\!\!\!\!\!\! 
\rho_1^{(2)}(x_1,x'_1,t) = 
N_1(N_1-1) \left(\frac{\alpha_1+C_{1,0}}{\pi}\right)^{\frac{1}{2}} \left(\frac{\alpha_1+C_{2,0}}{\pi}\right)^{\frac{1}{2}} \times  \\
& & \!\!\!\!\!\!\!\!\!\!\!\!\!\!\! 
\times e^{-(\alpha_1+C_{2,0})\left\{\left[x_1-x_1(t)\right]^2 + \left[x'_1-x_1(t)\right]^2\right\}}
e^{- 2(\beta_1+C_{2,0})\left[x_1-x_1(t)\right]\left[x'_1-x_1(t)\right]} = \nonumber \\
& & \!\!\!\!\!\!\!\!\!\!\!\!\!\!\! = N_1(N_1-1) \left(\frac{\alpha_1+C_{1,0}}{\pi}\right)^{\frac{1}{2}} \left(\frac{\alpha_1+C_{2,0}}{\pi}\right)^{\frac{1}{2}} \times \nonumber \\
& & \!\!\!\!\!\!\!\!\!\!\!\!\!\!\! \times
e^{-(\alpha_1+C_{2,0})\left\{x_1-\frac{1}{M}\left[\frac{N_1f_{L,1}+N_2f_{L,2}}{\omega^2-\omega^2_L} +
\frac{m_2N_2\left(\frac{f_{L,1}}{m_1}-\frac{f_{L,2}}{m_2}\right)}{\Omega^2_{12}-\omega^2_L}
\right]\cos(\omega_L t)\right\}^2} \times \nonumber \\
& & \!\!\!\!\!\!\!\!\!\!\!\!\!\!\! \times e^{-(\alpha_1+C_{2,0})\left\{x'_1-\frac{1}{M}\left[\frac{N_1f_{L,1}+N_2f_{L,2}}{\omega^2-\omega^2_L} +
\frac{m_2N_2\left(\frac{f_{L,1}}{m_1}-\frac{f_{L,2}}{m_2}\right)}{\Omega^2_{12}-\omega^2_L}
\right]\cos(\omega_L t)\right\}^2} \times \nonumber \\
& & \!\!\!\!\!\!\!\!\!\!\!\!\!\!\! \times 
e^{-2(\beta_1+C_{2,0})\left\{x_1-\frac{1}{M}\left[\frac{N_1f_{L,1}+N_2f_{L,2}}{\omega^2-\omega^2_L} +
\frac{m_2N_2\left(\frac{f_{L,1}}{m_1}-\frac{f_{L,2}}{m_2}\right)}{\Omega^2_{12}-\omega^2_L}
\right]\cos(\omega_L t)\right\}
\left\{x'_1-\frac{1}{M}\left[\frac{N_1f_{L,1}+N_2f_{L,2}}{\omega^2-\omega^2_L} +
\frac{m_2N_2\left(\frac{f_{L,1}}{m_1}-\frac{f_{L,2}}{m_2}\right)}{\Omega^2_{12}-\omega^2_L}
\right]\cos(\omega_L t)\right\}}, \nonumber \
\eeqn
and
\beqn\label{2_density_2}
& & \!\!\!\!\!\!\!\!\!\!\!\!\!\!\!
\rho_1^{(2)}(x_2,x'_2,t) = 
N_2(N_2-1) \left(\frac{\alpha_2+C'_{0,1}}{\pi}\right)^{\frac{1}{2}} \left(\frac{\alpha_2+C'_{0,2}}{\pi}\right)^{\frac{1}{2}} \times \\
& & \!\!\!\!\!\!\!\!\!\!\!\!\!\!\!
\times e^{-(\alpha_2+C'_{0,2})\left\{\left[x_2-x_2(t)\right]^2 + \left[x'_2-x_2(t)\right]^2\right\}}
e^{- 2(\beta_2+C'_{0,2})\left[x_2-x_2(t)\right]\left[x'_2-x_2(t)\right]} = \nonumber \\
& & \!\!\!\!\!\!\!\!\!\!\!\!\!\!\! = N_2(N_2-1) \left(\frac{\alpha_2+C'_{0,1}}{\pi}\right)^{\frac{1}{2}} \left(\frac{\alpha_2+C'_{0,2}}{\pi}\right)^{\frac{1}{2}} \times \nonumber \\
& & \!\!\!\!\!\!\!\!\!\!\!\!\!\!\! \times
e^{-(\alpha_2+C'_{0,2})\left\{x_2-\frac{1}{M}\left[\frac{N_1f_{L,1}+N_2f_{L,2}}{\omega^2-\omega^2_L} -
\frac{m_1N_1\left(\frac{f_{L,1}}{m_1}-\frac{f_{L,2}}{m_2}\right)}{\Omega^2_{12}-\omega^2_L}
\right]\cos(\omega_L t)\right\}^2} \times \nonumber \\
& & \!\!\!\!\!\!\!\!\!\!\!\!\!\!\! \times e^{-(\alpha_2+C'_{0,2})\left\{x'_2-\frac{1}{M}
\left[\frac{N_1f_{L,1}+N_2f_{L,2}}{\omega^2-\omega^2_L} -
\frac{m_1N_1\left(\frac{f_{L,1}}{m_1}-\frac{f_{L,2}}{m_2}\right)}{\Omega^2_{12}-\omega^2_L}
\right]\cos(\omega_L t)\right\}^2} \times \nonumber \\
& & \!\!\!\!\!\!\!\!\!\!\!\!\!\!\! \times 
e^{-2(\beta_2+C'_{0,2})\left\{x_2-\frac{1}{M}\left[\frac{N_1f_{L,1}+N_2f_{L,2}}{\omega^2-\omega^2_L} -
\frac{m_1N_1\left(\frac{f_{L,1}}{m_1}-\frac{f_{L,2}}{m_2}\right)}{\Omega^2_{12}-\omega^2_L}
\right]\cos(\omega_L t)\right\}
\left\{x'_2-\frac{1}{M}\left[\frac{N_1f_{L,1}+N_2f_{L,2}}{\omega^2-\omega^2_L} -
\frac{m_1N_1\left(\frac{f_{L,1}}{m_1}-\frac{f_{L,2}}{m_2}\right)}{\Omega^2_{12}-\omega^2_L}
\right]\cos(\omega_L t)\right\}}. \nonumber \
\eeqn
From the mean-field wavefunction (\ref{MF_MIX_WF}), 
the time-dependent
intraspecies reduced two-particle density matrices are given at the mean-field level of theory by
\beqn\label{2_density_MF_MIX}
& & \rho^{(2)}_{1,MF}(x_1,x'_1,x''_1,x'''_1,t) = N_1(N_1-1) \rho^{(1)}_{1,GP}(x_1,x''_1,t)\rho^{(1)}_{1,GP}(x'_1,x'''_1,t), \nonumber \\
& & \rho^{(2)}_{2,MF}(x_2,x'_2,x''_2,x'''_2,t) = N_2(N_2-1) \rho^{(1)}_{2,GP}(x_2,x''_2,t)\rho^{(1)}_{2,GP}(x'_2,x'''_2,t), \
\eeqn
with
$\rho^{(1)}_{1,GP}(x_1,x'_1,t) = \left(\frac{m_1\Omega_{1,GP}}{\pi}\right)^{\frac{1}{2}} 
e^{-\frac{m_1\Omega_{1,GP}}{2}\left\{\left[x_1- x_1(t)\right]^2 + \left[x'_1 - x_1(t)\right]^2\right\}}
e^{+i m_1 \dot x_1(t) \left\{\left[x_1-x_1(t)\right]-\left[x'_1-x_1(t)\right]\right\}}$ 
and
$\rho^{(1)}_{2,GP}(x_2,x'_2,t) = \left(\frac{m_2\Omega_{2,GP}}{\pi}\right)^{\frac{1}{2}} 
e^{-\frac{m_2\Omega_{2,GP}}{2}\left\{\left[x_2- x_2(t)\right]^2 + \left[x'_2 - x_2(t)\right]^2\right\}}
e^{+i m_2 \dot x_2(t) \left\{\left[x_2-x_2(t)\right]-\left[x'_2-x_2(t)\right]\right\}}$ 
[Eq.~(\ref{1_2_12_density_MF_MIX})].
The two-particle mean-field intraspecies densities read
$\rho^{(2)}_{1,MF}(x_1,x'_1,t) =$\break\hfill $= N_1(N_1-1) \rho^{(1)}_{1,GP}(x_1,t) \rho^{(1)}_{1,GP}(x'_1,t)$
and
$\rho^{(2)}_{2,MF}(x_2,x'_2,t) = N_2(N_2-1) \rho^{(1)}_{2,GP}(x_2,t) \rho^{(1)}_{2,GP}(x'_2,t)$,
where
$\rho^{(1)}_{1,GP}(x_1,t) = \left(\frac{m_1\Omega_{1,GP}}{\pi}\right)^{\frac{1}{2}} 
e^{-m_1\Omega_{1,GP}\left[x_1 - x_1(t)\right]^2}$
and
$\rho^{(1)}_{2,GP}(x_2,t) = \left(\frac{m_2\Omega_{2,GP}}{\pi}\right)^{\frac{1}{2}} \times$\break\hfill
$\times e^{-m_2\Omega_{2,GP}\left[x_2 - x_2(t)\right]^2}$.

Therefore, with the mean-field quantities explicitly prescribed,
one obtains at the limit of an infinite number of particles that
\beqn\label{RDM_2_limit_MIX}
& & \lim_{N_1 \to \infty \atop N_2 \to \infty} \frac{\rho_1^{(2)}(x_1,x'_1,x''_1,x'''_1,t)}{N_1(N_1-1)} =
\rho^{(1)}_{1,GP}(x_1,x''_1,t) \rho^{(1)}_{1,GP}(x'_1,x'''_1,t), \nonumber \\
& & \lim_{N_1 \to \infty \atop N_2 \to \infty} \frac{\rho_2^{(2)}(x_2,x'_2,x''_2,x'''_2,t)}{N_2(N_2-1)} =
\rho^{(1)}_{2,GP}(x_2,x''_2,t) \rho^{(1)}_{2,GP}(x'_2,x'''_2,t) \
\eeqn
for the reduced two-particle density matrices per particle
and
$\lim_{N_1 \to \infty \atop N_2 \to \infty} \frac{\rho_1^{(2)}(x_1,x'_1,t)}{N_1(N_1-1)} =$\break\hfill
$= \rho^{(1)}_{1,GP}(x_1,t) \rho^{(1)}_{1,GP}(x'_1,t)$,
$\lim_{N_1 \to \infty \atop N_2 \to \infty} \frac{\rho_2^{(2)}(x_2,x'_2,t)}{N_2(N_2-1)} = \rho^{(1)}_{2,GP}(x_2,t) \rho^{(1)}_{2,GP}(x'_2,t)$
for the time-dependent
two-particle intraspecies densities per particle themselves.

\section{Further details of solving the time-dependent coupled Gross-Pitaevskii mean-field equations}\label{TD_GPEs} 

In this appendix we derive the solution of the Floquet non-linear Schr\"odinger equations of the driven mixture.
To this end, we first obtain the respective solution of the driven single-species harmonic-interaction model
and thereafter generalize the techniques used.

The Gross-Pitaevskii equation for the driven harmonic-interaction model, see (\ref{TDGP}), is
$\left[-\frac{1}{2m}\frac{\partial^2}{\partial x^2} + \frac{1}{2} m \omega^2 x^2 - x f_L \cos(\omega_L t)
+ \Lambda \int dx' |\phi(x',t)|^2(x-x')^2\right]\phi(x,t)
=i\frac{\partial \phi(x,t)}{\partial t}$.
We suspect that and examine whether the solution can be cast as
\beq\label{TDGP_Shape}
\phi(x,t) = \left(\frac{m\Omega_{GP}}{\pi}\right)^{\frac{1}{4}} e^{-i\mu(t)} e^{-\frac{m\Omega_{GP}}{2}
\left[x-a\cos(\omega_L t)\right]^2} e^{-im\omega_L a \sin(\omega_L t)x},
\eeq
where the interaction-dressed frequency, $\Omega_{GP}$, time-dependent phase, $\mu(t)$,
and amplitude of oscillations, $a$, are to be determined self-consistently.
Substituting the function (\ref{TDGP_Shape}) into the square brackets on the left-hand-side of
the time-dependent Gross-Pitaevskii equation,
expanding the interaction term using the driven oscillations,
\beqn
& &
(x-x')^2 = x^2 - 2ax\cos(\omega_L t) +a^2 \cos^2(\omega_L t) - 2[x-a\cos(\omega_L t)][x'-a\cos(\omega_L t)] + \nonumber \\
& & 
\quad + [x'-a\cos(\omega_L t)]^2, \
\eeqn
and collecting
terms
we find for the square brackets
$\big[\!\!-\frac{1}{2m}\frac{\partial^2}{\partial x^2} + \frac{1}{2}m\left(\omega^2 + \frac{2\Lambda}{m}\right)x^2 - x(f_L+2a\Lambda)\cos(\omega_L t) 
+ \Lambda a^2\cos^2(\omega_L t) + \frac{\Lambda}{2\Omega_{GP}}\big]$.
This implies $\Omega_{GP}=\sqrt{\omega^2 + \frac{2\Lambda}{m}}$ to hold,
just like in the static problem.
Next, being the solution of a driven oscillator with the interaction-dressed frequency $\Omega_{GP}$, 
the amplitude $a$ must satisfy
\beq\label{GP_Amplitude}
a = \frac{f_L+2a\Lambda}{m\left(\Omega^2_{GP}-\omega^2_L\right)} \quad \Longrightarrow \quad 
a = \frac{f_L}{m\left(\omega^2-\omega^2_L\right)}, 
\eeq
i.e.,
resulting in the same amplitude of oscillations as for
the bare driven harmonic oscillator and for the many-body solution in the laboratory frame.
To verify consistency,
we re-expand 
$\frac{1}{2}m\Omega_{GP}^2 x^2 =
\frac{1}{2}m\Omega_{GP}^2\left[x-a\cos(\omega_L t)\right]^2 + m\Omega_{GP}^2 x a\cos(\omega_L t) -
\frac{1}{2}m\Omega_{GP}^2 a^2 \cos^2(\omega_L t)$
and indeed find
$m\Omega_{GP}^2 a - (f_L+2a\Lambda) = m\omega_L^2 a = f_L \frac{\omega^2_L}{\omega^2-\omega_L^2}$
for the linear-in-$x$ term
and $- \frac{1}{2}m\Omega_{GP}^2 a^2 \cos^2(\omega_L t) + \Lambda a^2 \cos^2(\omega_L t) =
-\frac{1}{2}m\omega^2a^2 \cos^2(\omega_L t) =
-\frac{f_L^2 \omega^2}{2m\left(\omega^2-\omega^2_L\right)^2}\cos^2(\omega_L t)$ 
for the contribution to the phase, as is the case with the bare system.
Finally, when operating from the left-hand-side of the Gross-Pitaevskii equation
with the kinetic-energy operator
and from the right-hand-side with the derivative in time on (\ref{TDGP_Shape}),
the remaining pieces of the time-dependent phase are collected,
$\mu(t) = \left(\frac{\Omega_{GP}}{2}+\frac{\Lambda}{2\Omega_{GP}}\right) t +
\int^t dt' \Big[\frac{f_L^2\omega_L^2}{2m(\omega^2-\omega_L^2)^2} \sin^2(\omega_L t') -
\frac{f_L^2\omega^2}{2m(\omega^2-\omega_L^2)^2} \cos^2(\omega_L t') \Big] = 
\left(\frac{\Omega_{GP}}{2}+\frac{\Lambda}{2\Omega_{GP}}\right) t
+ \int^t dt' \left[\frac{1}{2}m\dot{x}^2(t) - \frac{1}{2}m \omega^2 x^2(t)\right]$,
determining the time-dependent mean-field solution (\ref{TDGP_SOL_1}).
Summing up,
despite the interaction dressing of the frequency of the (driven) oscillator from $\omega$ to $\Omega_{GP}$,
the second and third contributions (see main text) to the time-dependent phase of
$\phi(x,t)$ are renormalized `back' to those of the bare system.

We employ the same strategy to derive the more involved solution of the
coupled time-dependent Gross-Pitaevskii
equations for the driven-mixture system.
The coupled Gross-Pitaevskii equations for the harmonic-interaction model for a driven mixture are\break\hfill
$\Big\{ -\frac{1}{2m_1} \frac{\partial^2}{\partial x_1^2} + \frac{1}{2} m_1 \omega^2 x_1^2 
- x_1 f_{L,1} \cos(\omega_Lt) + \Lambda_1 \int dx'_1 |\phi_1(x'_1,t)|^2 (x_1-x'_1)^2 
+ \Lambda_{21} \int dx_2 |\phi_2(x_2,t)|^2 (x_1-x_2)^2 \Big\} \phi_1(x_1,t) = 
i \frac{\partial\phi_1(x_1,t)}{\partial t}$,
$\Big\{ -\frac{1}{2m_2} \frac{\partial^2}{\partial x_2^2} + \frac{1}{2} m_2 \omega^2 x_2^2 
- x_2 f_{L,2} \cos(\omega_Lt) + \Lambda_2  \int dx'_2 |\phi_2(x'_2,t)|^2 (x_2-x'_2)^2 
+ \Lambda_{12} \int dx_1 |\phi_1(x_1,t)|^2 (x_1-x_2)^2 \Big\} \phi_2(x_2,t) = 
i \frac{\partial\phi_2(x,t)}{\partial t}$,
see (\ref{MIX_EQ_TDGP_1}).
We guess that and check whether the solution can be cast as
\beqn\label{TDGP_MIX_Shape}
& & 
\phi_1(x_1,t) = \left(\frac{m_1\Omega_{1,GP}}{\pi}\right)^{\frac{1}{4}} e^{-i\mu_1(t)} e^{-\frac{m_1\Omega_{GP,1}}{2}
\left[x_1-a_1\cos(\omega_L t)\right]^2} e^{-im_1\omega_L a_1 \sin(\omega_L t)x_1}, \nonumber \\
& &  
\phi_2(x_2,t) = \left(\frac{m_2\Omega_{2,GP}}{\pi}\right)^{\frac{1}{4}} e^{-i\mu_2(t)} e^{-\frac{m_2\Omega_{GP,2}}{2}
\left[x_2-a_2\cos(\omega_L t)\right]^2} e^{-im_2\omega_L a_2 \sin(\omega_L t)x_2},\
\eeqn
where the interaction-dressed frequencies, $\Omega_{1,GP}$ and $\Omega_{2,GP}$,
time-dependent phases, $\mu_1(t)$ and $\mu_2(t)$,
and amplitudes of oscillations, $a_1$ and $a_2$, are to be determined self-consistently.
Substituting the functions (\ref{TDGP_MIX_Shape}) into the curly
brackets on the left-hand-sides of
the two coupled time-dependent Gross-Pitaevskii equations,
expanding the interaction terms with the respective oscillation
of each orbital,
and collecting
the terms in each curly brackets
we find
\beqn
& &
\Bigg\{-\frac{1}{2m_1}\frac{\partial^2}{\partial x_1^2} + 
\frac{1}{2}m_1\left[\omega^2 + \frac{2(\Lambda_1+\Lambda_{21})}{m_1}\right]x_1^2 - 
x_1(f_{L,1}+2a_1\Lambda_1+2a_2\Lambda_{21})\cos(\omega_L t) + \nonumber \\
& & 
+ (\Lambda_1 a_1^2 + \Lambda_{21} a_2^2) \cos^2(\omega_L t) + 
\frac{\Lambda_1}{2\Omega_{1,GP}} + \frac{\Lambda_{21}}{2\Omega_{2,GP}}\Bigg\}, \nonumber \\
& & 
\Bigg\{-\frac{1}{2m_2}\frac{\partial^2}{\partial x_2^2} + 
\frac{1}{2}m_2\left[\omega^2 + \frac{2(\Lambda_2+\Lambda_{12})}{m_2}\right]x_2^2 - 
x_2(f_{L,2}+2a_2\Lambda_2+2a_1\Lambda_{12})\cos(\omega_L t) + \nonumber \\
& & 
+ (\Lambda_2 a_2^2 + \Lambda_{12} a_1^2) \cos^2(\omega_L t) + 
\frac{\Lambda_2}{2\Omega_{2,GP}} + \frac{\Lambda_{12}}{2\Omega_{1,GP}}\Bigg\}.
\eeqn
These imply that $\Omega_{1,GP}=\sqrt{\omega^2 + \frac{2(\Lambda_1+\Lambda_{21})}{m_1}}$
and $\Omega_{2,GP}=\sqrt{\omega^2 + \frac{2(\Lambda_2+\Lambda_{12})}{m_2}}$ hold,
just as in the static mixture's problem.

Next, being the solutions of driven coupled (many-particle) oscillators with the interaction-dressed frequencies 
$\Omega_{1,GP}$ and $\Omega_{2,GP}$, 
the amplitudes $a_1$ and $a_2$ have to satisfy the coupled linear relations
\beq\label{GP_MIX_Amplitudes}
a_1 = \frac{f_{L,1}+2a_1\Lambda_1+2a_2\Lambda_{21}}{m_1\left(\Omega^2_{1,GP}-\omega^2_L\right)}, \quad
a_2 = \frac{f_{L,2}+2a_2\Lambda_2+2a_1\Lambda_{12}}{m_2\left(\Omega^2_{2,GP}-\omega^2_L\right)},
\eeq
the solution of which is readily found
\beqn\label{GP_MIX_Amplitudes_SOL}
& &
a_1 = 
\frac{1}{m_1\Lambda_{12}+m_2\Lambda_{21}}\left[\frac{\Lambda_{12}f_{L,1}+\Lambda_{21}f_{L,2}}{\omega^2-\omega^2_L} +
\frac{m_2\Lambda_{21}\left(\frac{f_{L,1}}{m_1}-\frac{f_{L,2}}{m_2}\right)}{\Omega^2_{12}-\omega^2_L}
\right],
\nonumber \\
& &
a_2 = 
\frac{1}{m_1\Lambda_{12}+m_2\Lambda_{21}}\left[\frac{\Lambda_{12}f_{L,1}+\Lambda_{21}f_{L,2}}{\omega^2-\omega^2_L} -
\frac{m_1\Lambda_{12}\left(\frac{f_{L,1}}{m_1}-\frac{f_{L,2}}{m_2}\right)}{\Omega^2_{12}-\omega^2_L}
\right]. \
\eeqn
Interestingly, these are the same amplitudes of oscillations, with the same poles,
as for the the many-body solution, see (\ref{CRTZ_TRNS}).
We may say that the non-linear mean-field terms renormalize the poles 
from the values of the interaction-dressed oscillators
$\Omega^2_{1,GP}$ and $\Omega^2_{2,GP}$, see (\ref{GP_MIX_Amplitudes}), to the oscillators' frequencies of the many-body problem,
$\omega^2$ and $\Omega^2_{12}$, see (\ref{GP_MIX_Amplitudes_SOL}).

To proceed for the terms linear in $x_1$ and $x_2$,
after re-expanding
$\frac{1}{2}m_1\Omega_{1,GP}^2 x_1^2=$\break\hfill
$=\frac{1}{2}m_1\Omega_{1,GP}^2\left[x_1-a_1\cos(\omega_L t)\right]^2 + m_1\Omega_{1,GP}^2 x_1 a_1\cos(\omega_L t) -
\frac{1}{2}m_1\Omega_{1,GP}^2 a_1^2 \cos^2(\omega_L t)$
and likewise for $\frac{1}{2}m_2\Omega_{2,GP}^2 x_2^2$,
we find
\beqn\label{MIX_LIN_RENORM}
& &
m_1\Omega_{1,GP}^2 a_1 - (f_{L,1}+2a_1\Lambda_1+2a_2\Lambda_{21}) = \nonumber \\
& & = \frac{m_1 \omega_L^2}{m_1\Lambda_{12}+m_2\Lambda_{21}}\left[\frac{\Lambda_{12}f_{L,1}+\Lambda_{21}f_{L,2}}{\omega^2-\omega^2_L} +
\frac{m_2\Lambda_{21}\left(\frac{f_{L,1}}{m_1}-\frac{f_{L,2}}{m_2}\right)}{\Omega^2_{12}-\omega^2_L}
\right] = m_1 \omega_L^2 a_1, \nonumber \\
& &
m_2\Omega_{2,GP}^2 a_2 - (f_{L,2}+2a_2\Lambda_2+2a_1\Lambda_{12}) = \nonumber \\
& & = \frac{m_2 \omega_L^2}{m_1\Lambda_{12}+m_2\Lambda_{21}}\left[\frac{\Lambda_{12}f_{L,1}+\Lambda_{21}f_{L,2}}{\omega^2-\omega^2_L} -
\frac{m_1\Lambda_{12}\left(\frac{f_{L,1}}{m_1}-\frac{f_{L,2}}{m_2}\right)}{\Omega^2_{12}-\omega^2_L}
\right] = m_2 \omega_L^2 a_2, \
\eeqn
and for the contributions to the phases 
\beqn\label{MIX_PHASE_RENORM}
& & 
- \frac{1}{2}m_1\Omega_{1,GP}^2 a_1^2 \cos^2(\omega_L t) + 
\left(\Lambda_1 a_1^2 + \Lambda_{21} a_2^2\right) \cos^2(\omega_L t) = \nonumber \\
& & =
- \left[\frac{1}{2}m_1\omega^2 a_1^2 + \Lambda_{21}(a_1^2-a_2^2) \right] \cos^2(\omega_L t),
\nonumber \\
& & 
- \frac{1}{2}m_2\Omega_{2,GP}^2 a_2^2 \cos^2(\omega_L t) + 
\left(\Lambda_2 a_2^2 + \Lambda_{12} a_1^2\right) \cos^2(\omega_L t) = \nonumber \\
& & = - \left[\frac{1}{2}m_2\omega^2 a_2^2 - \Lambda_{12}(a_1^2-a_2^2) \right] \cos^2(\omega_L t).
\
\eeqn
The meaning of the above is that
the linear terms (\ref{MIX_LIN_RENORM}) are appropriately renormalized in comparison with the many-body treatment 
by the non-linear interaction terms whereas
the phases (\ref{MIX_PHASE_RENORM}) are not, also see below.
In summery, the mean-field treatment captures correctly the amplitudes of oscillations,
contributions of the linear terms (associated with operation of the kinetic-energy operator) to the wavefunction,
but not the contribution of the time-dependent phases.
These contributions do not match {\it both} poles' frequencies $\omega^2$ and $\Omega_{12}^2$
appearing in the oscillations' terms at the mean-field level.

Adding up all terms,
we have for the time-dependent phases in (\ref{TDGP_MIX_Shape})
\beqn\label{Phases_GP_MIX}
& & 
\mu_1(t) = \left(\frac{\Omega_{1,GP}}{2}+\frac{\Lambda_1}{2\Omega_{1,GP}} + \frac{\Lambda_{21}}{2\Omega_{2,GP}}\right) t +
\nonumber \\
& & 
+ \int^t dt' \left\{\frac{1}{2}m_1 \omega_L^2 a_1^2 \sin^2(\omega_L t') 
- \left[\frac{1}{2}m_1\omega^2 a_1^2 + \Lambda_{21}(a_1^2-a_2^2) \right] \cos^2(\omega_L t') \right\} =
\nonumber \\
& & = \mu_{1,F} - \frac{1}{8}\left[m_1(\omega^2+\omega_L^2)a_1^2 + 2\Lambda_{21}(a_1^2-a_2^2)\right]\sin(2\omega_Lt),  \nonumber \\
& & 
\mu_2(t) = \left(\frac{\Omega_{2,GP}}{2}+\frac{\Lambda_2}{2\Omega_{2,GP}} + \frac{\Lambda_{12}}{2\Omega_{1,GP}}\right) t +
\nonumber \\
& & 
+ \int^t dt' \left\{\frac{1}{2}m_2 \omega_L^2 a_2^2 \sin^2(\omega_L t') 
- \left[\frac{1}{2}m_2\omega^2 a_2^2 - \Lambda_{12}(a_1^2-a_2^2) \right] \cos^2(\omega_L t') \right\} =
\nonumber \\
& & = \mu_{2,F} - \frac{1}{8}\left[m_2(\omega^2+\omega_L^2)a_2^2 - 2\Lambda_{12}(a_1^2-a_2^2)\right]\sin(2\omega_Lt), \
\eeqn
where
\beqn\label{Phases_GP_MIX_QCP}
& & \mu_{1,F} = \left(\frac{\Omega_{1,GP}}{2}+\frac{\Lambda_1}{2\Omega_{1,GP}} + \frac{\Lambda_{21}}{2\Omega_{2,GP}}\right) -
\frac{1}{4}\left[m_1(\omega^2-\omega_L^2)a_1^2 + 2\Lambda_{21}(a_1^2-a_2^2)\right], \nonumber \\
& & \mu_{2,F} = \left(\frac{\Omega_{2,GP}}{2}+\frac{\Lambda_2}{2\Omega_{2,GP}} + \frac{\Lambda_{12}}{2\Omega_{1,GP}}\right) -
\frac{1}{4}\left[m_2(\omega^2-\omega_L^2)a_2^2 - 2\Lambda_{12}(a_1^2-a_2^2)\right]
\
\eeqn
are the quasichemical-potentials,
determining the time-dependent mean-field solution of the driven mixture
in the main text,
see (\ref{TDGP_MIX_Orbitals}).
After substitution of the amplitudes $a_1$ and $a_2$
the final expressions for $\mu_{1,F}$ and $\mu_{2,F}$
are obtained, see (\ref{Phases_GP_MIX_QCP_main}).

The final expression for the quasienergy per particle at the mean-field level of theory 
simplifies, since the interspecies interaction terms in (\ref{Phases_GP_MIX_QCP})
cancel each other upon the addition $N_1 \mu_1(t) + N_2 \mu_2(t)$ within
the time-dependent phase of the mean-field 
Floquet wavefunction,  
and reads
\beqn\label{MIX_eps_N_general}
& &
\varepsilon_F^{GP} =
\frac{N_1}{N}\frac{\Omega_{1,GP}}{2} + \frac{N_2}{N}\frac{\Omega_{2,GP}}{2} -
\frac{1}{4}(\omega^2-\omega_L^2) \left(\frac{N_1}{N} m_1  a_1^2 + \frac{N_2}{N} m_2  a_2^2\right). 
\eeqn 
Upon substitution of the oscillations' amplitudes $a_1$ and $a_2$ the expression for the mixture's
quasienergy given in the main text is obtained,
see (\ref{MF_Floquet_eps_N}).

\section{Angular-momentum variances in presence of translations and boosts}\label{VAR_TRANS_BST}

In this appendix we derive expressions for the angular-momentum variance
of translated and boosted many-particle wavefunctions,
in terms of quantities computed from the un-translated and un-boosted wavefunctions.
We start with the analysis for single-species bosons
(for the case of translations only see \cite{IN15}),
then generalize it to the two-species-mixture or interacting-impurity system. 

Consider the translated--boosted many-particle wavefunction in two spatial dimensions,
$e^{-i(\hat P_X a + \hat P_Y b)} e^{+i(\hat X \alpha + \hat Y \beta)} \Psi(X,Y) 
\equiv \Psi(a,b;\alpha,\beta)$,
where $\hat P_X=\sum_{j=1}^N \hat p_{x,j}, \hat P_Y=\sum_{j=1}^N \hat p_{y,j}$
and $\hat X=\sum_{j=1}^N \hat x_j, \hat Y=\sum_{j=1}^N \hat y_j$
and the shorthand collective notation for the coordinates $\Psi(X,Y) \equiv \Psi(x_1,y_1,\ldots,x_N,y_N)$ is implied.
The order of the translation and boost operators does not matter for the computation of matrix elements.
The wavefunction (before translations and boosts) as well as the translation and boost operators themselves can be time dependent.
What are the implications on the variances of observables
when computed with respect to the translated--boosted wavefunction $\Psi(a,b;\alpha,\beta)$?

For the position operator
$\hat X$ (and equivalently for $\hat Y$) we have
$e^{-i(\hat X \alpha + \hat Y \beta)} e^{+i(\hat P_X a + \hat P_Y b)} \hat X \times$
$\times e^{-i(\hat P_X a + \hat P_Y b)} e^{+i(\hat X \alpha + \hat Y \beta)} = \hat X + Na$,
and using its square one has
\beq
\frac{1}{N} \Delta^2_{\hat X}\Big|_{\Psi(a,b;\alpha,\beta)} = \frac{1}{N} \Delta^2_{\hat X}\Big|_{\Psi}.
\eeq
Similarly, for the momentum operator
$\hat P_X$ (and equivalently for $\hat P_Y$) we have
$e^{-i(\hat X \alpha + \hat Y \beta)} \times$ $\times e^{+i(\hat P_X a + \hat P_Y b)} \hat P_X 
e^{-i(\hat P_X a + \hat P_Y b)} e^{+i(\hat X \alpha + \hat Y \beta)} = \hat P_X + N\alpha$
and using its square leads to 
\beq
\frac{1}{N} \Delta^2_{\hat P_X}\Big|_{\Psi(a,b;\alpha,\beta)} = \frac{1}{N} \Delta^2_{\hat P_X}\Big|_{\Psi},
\eeq
i.e., both the position and momentum variances are invariant under translations and boosts and their combined operation.

For the angular-momentum operator
$\hat L_Z=\sum_{j=1}^N \left(\hat x_j \hat p_{y,j} - \hat y_j \hat p_{x,j}\right)$ the situation is more intricate,
because it consists of the position and momentum operators which do not commute with each other.
From
\beqn\label{AM_trans_boost}
& & e^{-i(\hat X \alpha + \hat Y \beta)} e^{+i(\hat P_X a + \hat P_Y b)} \hat L_Z
e^{-i(\hat P_X a + \hat P_Y b)} e^{+i(\hat X \alpha + \hat Y \beta)} = \nonumber \\
& & = \hat L_Z + a \hat P_Y - b \hat P_X + \beta \hat X - \alpha \hat Y + N(a\beta-b\alpha) \
\eeqn
and its square
we have
\beqn\label{Var_Lz_a_b}
& & 
\frac{1}{N} \Delta^2_{\hat L_Z}\Big|_{\Psi(a,b;\alpha,\beta)} = 
\frac{1}{N} \Delta^2_{\hat L_Z}\Big|_{\Psi} + 
a^2 \frac{1}{N} \Delta^2_{\hat P_Y}\Big|_{\Psi} + 
b^2 \frac{1}{N} \Delta^2_{\hat P_X}\Big|_{\Psi} + 
\alpha^2 \frac{1}{N} \Delta^2_{\hat Y}\Big|_{\Psi} + 
\beta^2 \frac{1}{N} \Delta^2_{\hat X}\Big|_{\Psi} + 
\nonumber \\
& & \qquad + \frac{1}{N} \Bigg[ a \left( \langle\Psi| \hat L_Z \hat P_Y + \hat P_Y \hat L_Z |\Psi\rangle - 
2 \langle\Psi| \hat L_Z |\Psi\rangle \langle\Psi| \hat P_Y |\Psi\rangle \right) - \nonumber \\
& & \qquad - b \left( \langle\Psi| \hat L_Z \hat P_X + \hat P_X \hat L_Z |\Psi\rangle - 
2 \langle\Psi| \hat L_Z |\Psi\rangle \langle\Psi| \hat P_X |\Psi\rangle \right) - \nonumber \\
& & \qquad - 2ab \left( \langle\Psi| \hat P_Y \hat P_X |\Psi\rangle - 
\langle\Psi| \hat P_Y |\Psi\rangle \langle\Psi| \hat P_X |\Psi\rangle \right) + \nonumber \\
& & \qquad + \beta \left( \langle\Psi| \hat L_Z \hat X + \hat X \hat L_Z |\Psi\rangle - 
2 \langle\Psi| \hat L_Z |\Psi\rangle \langle\Psi| \hat X |\Psi\rangle \right) - \nonumber \\
& & \qquad - \alpha \left( \langle\Psi| \hat L_Z \hat Y + \hat Y \hat L_Z |\Psi\rangle - 
2 \langle\Psi| \hat L_Z |\Psi\rangle \langle\Psi| \hat Y |\Psi\rangle \right) - \nonumber \\
& & \qquad - 2\alpha\beta \left( \langle\Psi| \hat X \hat Y |\Psi\rangle - 
\langle\Psi| \hat X |\Psi\rangle \langle\Psi| \hat Y |\Psi\rangle \right) - \nonumber \\
& & \qquad - a\alpha \left( \langle\Psi| \hat P_Y \hat Y + \hat Y \hat P_Y |\Psi\rangle - 
2 \langle\Psi| \hat P_Y |\Psi\rangle \langle\Psi| \hat Y |\Psi\rangle \right) - \nonumber \\
& & \qquad - b\beta \left( \langle\Psi| \hat P_X \hat X + \hat X \hat P_X |\Psi\rangle - 
2 \langle\Psi| \hat P_X |\Psi\rangle \langle\Psi| \hat X |\Psi\rangle \right) + \nonumber \\
& & \qquad + 2 a\beta \left( \langle\Psi| \hat P_Y \hat X |\Psi\rangle - 
\langle\Psi| \hat P_Y |\Psi\rangle \langle\Psi| \hat X |\Psi\rangle \right) + \nonumber \\
& & \qquad + 2 b\alpha \left( \langle\Psi| \hat P_X \hat Y |\Psi\rangle - 
\langle\Psi| \hat P_X |\Psi\rangle \langle\Psi| \hat Y |\Psi\rangle \right) \Bigg]. \
\eeqn
Eq.~(\ref{Var_Lz_a_b}) warrants a discussion.
We see that the angular-momentum variance of a translated--boosted wavefunction 
can be decomposed into
various terms of physical meaning computed from the un-translated-and-boosted wavefunction.
These include, for a start, the angular-momentum variance of the wavefunction
and the momentum and position variances in each direction, see the first line in (\ref{Var_Lz_a_b}).
Then, there are terms which originate from translations only, and `interference' between them,
followed by terms which originate from boosts only, and `interference' between them,
and, finally, `interference' terms between translations and boosts.
For the most general wavefunction, all involved expectation values contribute.
When symmetries set in, see below, or when the wavefunction is real (excluding a trivial phase factor),
corresponding terms fall.

For our needs, let us examine translated and boosted rotationally-symmetric systems.
Consider the ground state $\Psi$
of an interacting many-boson system in a rotationally-symmetric trap
(to be more precise,
we do not consider the ground-state of a rotationally-symmetric system in the rotating frame).
Clearly, $\frac{1}{N}\Delta^2_{\hat L_Z}\Big|_{\Psi}=0$ holds.
Now, when the ground state is translated and boosted,
the angular-momentum variance can be written as
\beq\label{Var_Lz_Spher_Symm}
\frac{1}{N} \Delta^2_{\hat L_Z}\Big|_{\Psi(a,b;\alpha,\beta)} = 
a^2 \frac{1}{N} \Delta^2_{\hat P_Y}\Big|_{\Psi} + 
b^2 \frac{1}{N} \Delta^2_{\hat P_X}\Big|_{\Psi} +
\alpha^2 \frac{1}{N} \Delta^2_{\hat Y}\Big|_{\Psi} + 
\beta^2 \frac{1}{N} \Delta^2_{\hat X}\Big|_{\Psi},
\eeq
and seen to solely be described by the momentum and position
variances along the $y$ and $x$ directions 
multiplied by the respective values of the translations and boosts along the $x$ and $y$ directions.

Interestingly, the angular-momentum variance of a translated-boosted many-boson
system $\frac{1}{N} \Delta^2_{\hat L_Z}\Big|_{\Psi(a,b;\alpha,\beta)}$
differs at the many-body and mean-field levels of theory,
i.e.,
when $a, b \ne 0$, $\alpha, \beta \ne 0$, and $\lambda \ne 0$. 
This, as can be seen in (\ref{Var_Lz_Spher_Symm}),
is because of the respective many-body and mean-field
momentum, $\frac{1}{N}\Delta^2_{\hat P_X}\Big|_{\Psi}$ and $\frac{1}{N}\Delta^2_{\hat P_Y}\Big|_{\Psi}$,
and position,
$\frac{1}{N}\Delta^2_{\hat X}\Big|_{\Psi}$ and $\frac{1}{N}\Delta^2_{\hat Y}\Big|_{\Psi}$,
variances.
For instance, in harmonic traps these variances do not dependent on the
interaction strength at the many-body level of theory
but are dressed by the interaction at the mean-field level of theory,
see in this context \cite{IN6}.
The analytical result (\ref{Var_Lz_Spher_Symm})
is employed to compute and analyze the findings for the
angular-momentum variance of a many-boson Floquet state in the main text.
Generally, as mentioned above, in the absence of spatial symmetries
more terms contribute to the translated and boosted
angular-momentum variance, see (\ref{Var_Lz_a_b}).

We now generalize the above analysis to the mixture or impurity system.
The notation used hereafter is a straightforward generalization of the notation employed above
in the single-species case.
Starting from the many-particle wavefunction $\Psi(X_1,Y_1,X_2,Y_2)$,
we note that each species can be translated and boosted differently,
$e^{-i(\hat P_{X_1} a_1 + \hat P_{Y_1} b_1)} e^{+i(\hat X_1 \alpha_1 + \hat Y_1 \beta_1)} \times$\break\hfill
$\times e^{-i(\hat P_{X_2} a_2 + \hat P_{Y_2} b_2)} e^{+i(\hat X_2 \alpha_2 + \hat Y_2 \beta_1)}
\Psi(X_1,Y_1,X_2,Y_2) \equiv \Psi(\vec a,\vec b;\vec \alpha,\vec \beta)$,
as is, for instance, the result in the main text for the steered species $1$ bosons
embedded in un-steered species $2$ bosons, see Sec.~\ref{ANG_MIX}.

With $\Psi(\vec a,\vec b;\vec \alpha,\vec \beta)$,
it is straightforward to see that the variance of the (mass-weighted) position operator,
$\frac{N}{M}\left(m_1 \hat X_1 + m_2 \hat X_2\right)$, total momentum operator, $\hat P_1 + \hat P_2$,
or any linear combination of $\hat X_1$ and $\hat X_2$ or of $\hat P_1$ and $\hat P_2$
is invariant
under translations and boosts and combinations thereof.

For the variance of the angular-momentum operator of either species $1$ or $2$ just use (\ref{Var_Lz_a_b})
with the corresponding notations,
since
$\frac{1}{N_1} \Delta^2_{\hat L_{Z_1}}\Big|_{\Psi(\vec a,\vec b;\vec \alpha,\vec \beta)} = 
\frac{1}{N_1} \Delta^2_{\hat L_{Z_1}}\Big|_{\Psi(a_1,b_1;\alpha_1,\beta_1)}$ 
and 
$\frac{1}{N_2} \Delta^2_{\hat L_{Z_2}}\Big|_{\Psi(\vec a,\vec b;\vec \alpha,\vec \beta)} =
\frac{1}{N_2} \Delta^2_{\hat L_{Z_2}}\Big|_{\Psi(a_2,b_2;\alpha_2,\beta_2)}$.
Symmetry can be of assistance as well,
yet note that
$\frac{1}{N_1} \Delta^2_{\hat L_{Z_1}}\Big|_{\Psi}$ and
$\frac{1}{N_2} \Delta^2_{\hat L_{Z_2}}\Big|_{\Psi}$
do generally not vanish at the many-body level
(and differences between the mean-field and many-body treatments can thus occur),
even for the ground state of a mixture in spherically symmetric traps,
due to the coupling introduced by the interspecies interaction $\lambda_{12}$ \cite{HM17}.
All in all in our case we have
\beqn\label{Var_Lz1_Lz2_Spher_Symm}
& &
\frac{1}{N_1} \Delta^2_{\hat L_{Z_1}}\Big|_{\Psi(\vec a,\vec b;\vec \alpha,\vec \beta)} = 
\frac{1}{N_1} \Delta^2_{\hat L_{Z_1}}\Big|_{\Psi} +
a_1^2 \frac{1}{N_1} \Delta^2_{\hat P_{Y_1}}\Big|_{\Psi} + 
b_1^2 \frac{1}{N_1} \Delta^2_{\hat P_{X_1}}\Big|_{\Psi} +
\alpha_1^2 \frac{1}{N_1} \Delta^2_{\hat Y_1}\Big|_{\Psi} + 
\beta_1^2 \frac{1}{N_1} \Delta^2_{\hat X_1}\Big|_{\Psi}, \nonumber \\
& &
\frac{1}{N_2} \Delta^2_{\hat L_{Z_2}}\Big|_{\Psi(\vec a,\vec b;\vec \alpha,\vec \beta)} = 
\frac{1}{N_2} \Delta^2_{\hat L_{Z_2}}\Big|_{\Psi} +
a_2^2 \frac{1}{N_2} \Delta^2_{\hat P_{Y_2}}\Big|_{\Psi} + 
b_2^2 \frac{1}{N_2} \Delta^2_{\hat P_{X_2}}\Big|_{\Psi} +
\alpha_2^2 \frac{1}{N_2} \Delta^2_{\hat Y_2}\Big|_{\Psi} + 
\beta_2^2 \frac{1}{N_2} \Delta^2_{\hat X_2}\Big|_{\Psi} \nonumber \\ 
& & \
\eeqn
for the angular-momentum variances of the individual species in the driven mixture.

To derive the working expression for the variance of the total angular-momentum operator
$\hat L_Z = \hat L_{Z_1} + \hat L_{Z_2}$ of the mixture,
we begin from the transformed quantity 
\beqn\label{Lz_Mix_transform}
& & e^{-i(\hat X_2 \alpha_2 + \hat Y_2 \beta_1)} e^{+i(\hat P_{X_2} a_2 + \hat P_{Y_2} b_2)}
e^{-i(\hat X_1 \alpha_1 + \hat Y_1 \beta_1)} e^{+i(\hat P_{X_1} a_1 + \hat P_{Y_1} b_1)}
\hat L_Z \times \nonumber \\
& & \times e^{-i(\hat P_{X_1} a_1 + \hat P_{Y_1} b_1)} e^{+i(\hat X_1 \alpha_1 + \hat Y_1 \beta_1)}
e^{-i(\hat P_{X_2} a_2 + \hat P_{Y_2} b_2)} e^{+i(\hat X_2 \alpha_2 + \hat Y_2 \beta_1)} = \nonumber \\
& & = \hat L_Z
+ a_1 \hat P_{Y_1} + a_2 \hat P_{Y_2}
-  b_1 \hat P_{X_1} - b_2 \hat P_{X_2}
+ \beta_1 \hat X_1 + \beta_2 \hat X_2
- \alpha_1 \hat Y_1 - \alpha_2 \hat Y_2 + \nonumber \\
& & + N_1(a_1\beta_1-b_1\alpha_1)
+ N_2(a_2\beta_2-b_2\alpha_2) \
\eeqn
and its square.
The final result is lengthier than (\ref{Var_Lz_a_b}) but otherwise straightforward.
For spherically-symmetric traps many terms fall,
and the final result can be written as
\beqn\label{Var_Lz_Spher_Symm_MIX}
& &
\frac{1}{N} \Delta^2_{\hat L_Z}\Big|_{\Psi(\vec a,\vec b;\vec\alpha,\vec\beta)} = \nonumber \\
& & = \frac{1}{N} \Delta^2_{a_1 \hat P_{Y_1} + a_2 \hat P_{Y_2}}\Big|_{\Psi} +
\frac{1}{N} \Delta^2_{b_1 \hat P_{X_1} + b_2 \hat P_{X_2}}\Big|_{\Psi} +
\frac{1}{N} \Delta^2_{\beta_1 \hat X_1 + \beta_2 \hat X_2}\Big|_{\Psi} +
\frac{1}{N} \Delta^2_{\alpha_1 \hat Y_1 + \alpha_2 \hat Y_2}\Big|_{\Psi} +
\nonumber \\ 
& & = \frac{N_1}{N} \left(
a_1^2 \frac{1}{N_1} \Delta^2_{\hat P_{Y_1}}\Big|_{\Psi} + 
b_1^2 \frac{1}{N_1} \Delta^2_{\hat P_{X_1}}\Big|_{\Psi} +
\alpha_1^2 \frac{1}{N_1} \Delta^2_{\hat Y_1}\Big|_{\Psi} +
\beta_1^2 \frac{1}{N_1} \Delta^2_{\hat X_1}\Big|_{\Psi} \right) + \nonumber \\
& & +
\frac{N_2}{N} \left(
a_2^2 \frac{1}{N_2} \Delta^2_{\hat P_{Y_2}}\Big|_{\Psi} + 
b_2^2 \frac{1}{N_2} \Delta^2_{\hat P_{X_2}}\Big|_{\Psi} +
\alpha_2^2 \frac{1}{N_2} \Delta^2_{\hat Y_2}\Big|_{\Psi} +
\beta_2^2 \frac{1}{N_2} \Delta^2_{\hat X_2}\Big|_{\Psi} \right) +  \\
& & + 2 \frac{1}{N} \Big(
a_1 a_2\langle\Psi| \hat P_{Y_1} \hat P_{Y_2} |\Psi\rangle +
b_1 b_2\langle\Psi| \hat P_{X_1} \hat P_{X_2} |\Psi\rangle + \alpha_1 \alpha_2\langle\Psi| \hat Y_1 \hat Y_2 |\Psi\rangle +
\beta_1 \beta_2\langle\Psi| \hat X_1 \hat X_2 |\Psi\rangle
\Big). \nonumber \
\eeqn
When applied to the driven mixture in the main text,
the angular-momentum variance (\ref{Var_Lz_Spher_Symm_MIX}) 
can be expressed using the variances of time-dependent interspecies-weighted momenta,
$a_1 \hat P_{Y_1} + a_2 \hat P_{Y_2}$ and 
$b_1 \hat P_{X_1} + b_2 \hat P_{X_2}$,
and
position,
$\beta_1 \hat X_1 + \beta_2 \hat X_2$ and
$\alpha_1 \hat Y_1 + \alpha_2 \hat Y_2$,
operators.
Alternatively,
it can be expressed using the intraspecies momentum and position variances
and respective interspecies cross terms. 
The contributions of the various intraspecies and interspecies terms to the mixture's angular-momentum variance is 
different at the many-body and mean-field levels of theory,
and computed and discussed in the main text.
In particular,
the momentum and position interspecies
cross terms in (\ref{Var_Lz_Spher_Symm_MIX}) vanish at the
mean-field treatment of the (driven) mixture.

\end{document}